\documentclass[11pt,oneside,english]{article}
\usepackage[a4paper, total={7.0in, 9.5in}]{geometry}
\usepackage[utf8]{inputenc}
\pdfoutput=1

\usepackage{rotating}
\usepackage{graphicx}
\usepackage{amssymb}
\usepackage[table]{xcolor}
\usepackage{xcolor}
\usepackage[T1]{fontenc}
\usepackage{lmodern}
\usepackage{pdflscape}

\usepackage{enumitem}

\usepackage{multirow,booktabs,array,longtable,caption}

\usepackage{pdflscape}

\definecolor{verylightgray}{gray}{0.9}

\newcommand{\changes}[1]{#1}

\newcommand{\changess}[1]{#1}

\usepackage{enumitem}

\newlist{itemizelessVspace}{itemize}{1}
\setlist[itemizelessVspace]{label=\textbullet, itemsep=1pt,
                    parsep=2pt, topsep=1pt, partopsep=2pt}
  
\newlist{inneritemizelessVspace}{itemize}{1}
\setlist[inneritemizelessVspace]{label=\textendash, itemsep=1pt,
                    parsep=1pt, topsep=1pt, partopsep=1pt}

\newlist{compactimize}{itemize}{1}
\setlist[compactimize]{label=\textbullet,leftmargin=*,topsep=0ex, itemsep=0pt,
parsep=0pt,after=\vspace{-\baselineskip},before=\vspace{-0.80\baselineskip}}

\newcommand{\myparagraph}[1]{\paragraph{#1}\mbox{}\\}

\usepackage{caption}

\usepackage{courier}  
\usepackage[bottom, multiple]{footmisc}  
\usepackage[normalem]{ulem}    
\usepackage{anyfontsize}

\PassOptionsToPackage{hyphens}{url}\usepackage[hidelinks, hyperfootnotes=false]{hyperref}
\usepackage{fancyhdr}                           
\fancypagestyle{firstpage} 
{
   \footskip=10mm
   \fancyhf{}
   \fancyfoot[C]{\vspace{-6mm}\thepage}

   \fancyfoot[L]{\vspace{-0mm}\noindent\rule{7.0in}{0.4pt}
     \scriptsize{
         \textcopyright \hspace{0.5mm} \changes{Barclays 2024} \\
       This work is licensed under a \href{https://creativecommons.org/licenses/by/4.0/}{Creative Commons Attribution 4.0 International License (CC
       BY).}  \\ Provided you adhere to the CC BY license, including as to attribution, you are
       free to copy and redistribute this work in any medium or format and remix, transform,
       and build upon the work for any purpose, even commercially.                  \\
       \changes{BARCLAYS is a registered trademark, all rights are reserved.} \\
     }
 }
}
\title{\vspace{-2cm} 
  Payments Use Cases and Design Options for \\
  Interoperability and Funds Locking across \\
  Digital Pounds and Commercial Bank Money \\
}

\author{%
  \begin{tabular}{c} {\fontsize{10.75}{1cm}\selectfont Lee Braine,
    Shreepad Shukla, Piyush Agrawal, Shrirang Khedekar and Aishwarya Nair} \\ 
    {\fontsize{10.75}{1cm}\selectfont Chief Technology Office} \\
    {\fontsize{10.75}{1cm}\selectfont Barclays} \\ \hskip 1em \end{tabular} }

\date{\vspace{-0.5cm}\changess{September 13, 2024}}

\begin{document}
\maketitle
\thispagestyle{firstpage} 
\vspace{-0.75cm}


\begin{abstract}
\noindent
Central banks are actively exploring retail central bank digital
currencies (CBDCs),
with the Bank of England currently in the design phase for a
potential UK retail CBDC, the digital pound.
In a previous paper, we defined and explored the important
concept of functional consistency 
(which is the principle that different forms of money have the
same operational characteristics) and
evaluated design options to support functional consistency
across digital pounds and commercial bank money,
based on a set of key capabilities.
In this paper, we continue to analyse the design options for
supporting functional consistency and, in order to perform a
detailed analysis, we focus on three key
capabilities: 
communication between digital pound ecosystem participants,
funds locking, and interoperability across digital pounds
and commercial bank money.
We explore these key capabilities via three payments \linebreak
use cases:
person-to-person push payment,
merchant-initiated request to pay, and 
lock funds and pay on physical delivery.
We then present and evaluate the suitability of design options to 
provide the specific capabilities for each use case and draw
initial insights.
We conclude that a financial market infrastructure (FMI)
providing specific capabilities could
simplify the experience of ecosystem participants,
simplify the operating platforms for both the Bank of England
and digital pound Payment Interface Providers (PIPs), and 
facilitate the creation of innovative services.
We also identify potential next steps.

\end{abstract}

\vspace{0.2cm}


\section{Introduction}
\label{sec:introduction}

\noindent
A central bank digital currency (CBDC) is a digital payment
instrument, denominated in a national unit of account, that is
a direct liability of a central bank \cite{bis-aer2021-cbdcs}.
Many central banks are actively exploring retail CBDCs 
\cite{boe-cbdc-mainpage, bis-cbdc-survey-2024, ecb-digeuro-stocktake,
rbi-cbdc-note, pboc-ecny-paper}.
\changes{In the United Kingdom (UK), the Bank of England and HM Treasury are 
exploring a retail CBDC, known as a digital pound. 
The digital pound project has moved into a design phase which is supported by three
external engagement groups, the CBDC Engagement Forum, the CBDC Technology Forum and 
the CBDC Academic Advisory Group} to gather input on various
aspects of a potential UK retail CBDC, the
digital pound \cite{boe-cbdc-mainpage}.
The Bank of England and HM Treasury have also published
a Consultation Paper (CP) \cite{boe-cbdc-cons-paper}, 
a Technology Working Paper (TWP) \cite{boe-cbdc-twp},
the response \cite{boe-cbdc-cp-resp} to the CP,
and the response \cite{boe-cbdc-twp-resp} to the TWP.
The CP and the TWP describe the Bank of England's `platform model'
for the digital pound, which comprises the Bank of England
operating a digital pound core ledger and providing access via
application programming interfaces (APIs) 
to authorised and regulated Payment Interface Providers (PIPs)
and External Service Interface Providers (ESIPs) that provide
user access to the digital pound.
\changes{This paper uses publicly available information on a digital 
pound and does not provide insight to any future policy decisions 
the Bank of England and HM Treasury make.}

In our first paper \cite{barc-cbdc-iia-doi} on the digital pound,
we:
  (i) identified the risk of fragmentation in payments
  markets and retail deposits if digital pounds and commercial bank
  money do not have common operational characteristics, and
  (ii) presented an illustrative industry architecture intended to
  mitigate this risk by placing the digital pound and commercial bank
  money on a similar footing.
We then developed prototypes of this industry
architecture as part of Barclays CBDC Hackathon 2022
\cite{barclays-cbdchack-site}
and Project Rosalind \cite{bis-rosalind-site} conducted by
the Bank for International Settlements' Innovation Hub London Centre and
the Bank of England.
In our second paper \cite{barc-cbdc-func-cons} on the digital pound, we:
  (i) defined and explored the important concept of functional consistency
  (which is the principle that different forms of
  money have the same operational characteristics)
  as a means to mitigate the risk of fragmentation, and 
  (ii) evaluated design options to support functional consistency
  across digital pounds and commercial bank money based on a
  set of key capabilities.
We also participated in the UK Regulated Liability Network
(UK RLN) discovery phase \cite{ukrln-disc-report} and experimentation
phase \cite{ukrln-exp-release}, which \changes{included a
prototype} common `platform for innovation' \changes{that
supported} functional consistency across digital pounds and
commercial bank money.

In this paper, we continue to analyse the design options for supporting
functional consistency, but
we now focus on three key capabilities in order to perform a
detailed analysis: 
(i) communication between PIPs and other ecosystem participants such as
merchants, acquirers and FMIs,
(ii) funds locking, and \linebreak
(iii) interoperability between the digital pound and
commercial bank money.
We explore these key capabilities via three payments use
cases\footnote{
  ``A use case describes a function that a system performs to
  achieve the user's goal.
  A use case must yield an observable result that is of value
  to the user of the system.
  Use cases contain detailed information about the system,
  the system's users, relationships between the system and
  the users, and the required behavior of the system.
  Use cases do not describe the details of how the system is
  implemented.'' \cite{ibm-rational-use-case}
}:
(i) `Person-to-person push payment with interoperability across 
the digital pound and commercial bank money',
(ii) `Merchant initiated request to pay with interoperability 
across the digital pound and commercial bank money', and
(iii) `Lock digital pounds and pay on physical delivery from 
digital pounds to commercial bank money'.
We identify potential variants for each use case but
focus on a standard case and its normal flow in the use
case details.
In order to simplify our analysis, we do not explore all
potential variants and alternate flows for each use case. 
We present design options
\changes{(with sequence diagrams)}
that are based on the
platform model for the digital pound, and 
that support the specific capabilities for each use case\footnote{
  We use a hierarchical numbering scheme for use cases (e.g. U1, U2, ...),
  their specific capabilities (e.g. U1.S1, U1.S2, ...), and the
  design options that support them (e.g. U1.S1.D1, U1.S1.D2, ...)
 to aid readability.
}.
We evaluate the suitability of the design options and
draw initial insights.
Finally, we summarise our findings, draw initial
conclusions and identify potential next steps.

This paper is structured with Sections 2, 3 and 4 each
presenting the description, design options, evaluation
and insights for a use case,
Section 5 summarising our findings and initial
conclusions, and 
Section 6 identifying potential further work.
\changes{The appendices contain a common component model and
a sequence diagram for each design option.}

The contributions of this paper include:
  (i) detailed use case descriptions that focus on specific capabilities
  that support functional consistency,
  (ii) design options that provide the specific capabilities for each
  use case, and
  (iii) a preliminary evaluation of the suitability of the
  design options.
We hope the insights presented in this paper will aid the
design and experimentation for the digital pound, and look forward
to ongoing industry engagement.



\begingroup
\let\clearpage\relax


\section{Use case for `Person-to-person push payment with interoperability across 
the digital pound and commercial bank money' (U1)}
\label{sec:p2p-push-payment}


In this section, we present the use case for person-to-person (P2P) push 
payment, with interoperability across the digital pound and commercial bank money.
We then describe and evaluate design options for the specific capabilities for 
this use case.
\subsection{Use case details}
\label{subsec:p2p-details}

\subsubsection{Use case description}
\label{subsubsec:p2p-description}
A parent (payer) holding a commercial bank account wishes to top-up  
his/her child's (payee) digital pound wallet for low value expenses.
The parent uses his/her bank's mobile app to initiate the payment. 
Once the transfer is complete, both the parent and the child are notified of the 
transaction status in their commercial bank account / digital pound wallet respectively.

Note that we focus on the standard case to  
simplify our preliminary evaluation.

\subsubsection{Use case variants}
\label{subsubsec:p2p-variants}
We first identify a number of criteria that could potentially be used to 
develop variants for this use case. 
For each criterion, we highlight the standard case.
\begin{itemizelessVspace}
    \item Source and target account types:
    \begin{inneritemizelessVspace}
        \item From payer's commercial bank account to payee's digital 
        pound wallet (standard case).
        \item From payer's digital pound wallet to payee's commercial bank 
        account.
    \end{inneritemizelessVspace}
    \item Financial entity used by payer and payee:
    \begin{inneritemizelessVspace}
        \item The payer's and payee's account/wallet are managed by two 
        different financial entities (standard case).
        \item The payer's and payee's account/wallet are managed by the
        same financial entity.
    \end{inneritemizelessVspace}
    \item Payee's digital pound wallet details (digital pound wallet alias) 
    entered by the payer on his/her commercial bank smartphone app (only applicable for use case 
    variant where payer pays using his/her commercial bank account):
    \begin{inneritemizelessVspace}
        \item Payee's mobile phone number alias (standard case).
        \item Payee wallet's account number and sort code assigned by his/her PIP. 
    \end{inneritemizelessVspace}
    \item Funds availability in payer's digital pound wallet (only 
    applicable for use case variant where payer pays using his/her 
    digital pound wallet):
    \begin{inneritemizelessVspace}
        \item The payer has sufficient funds in his/her digital pound wallet.
        \item The payer does not have sufficient funds in his/her digital 
        pound wallet, resulting in sweeping of funds from a linked commercial 
        bank account to his/her digital pound wallet (reverse waterfall 
        approach\footnote{
            Similar to European Central Bank's notions of waterfall approach and 
            reverse waterfall approach for the digital euro \cite{ecb-digeuro-stocktake}.\label{fn-ecb-wf-rwf}}).
        \item The payer does not have sufficient funds \changes{across} his/her digital pound 
        wallet and commercial bank account, resulting in failure of the 
        transaction.
    \end{inneritemizelessVspace}
    \item Payee's digital pound holding threshold  
    (only applicable for use case variant where the payee receives funds in 
    his/her digital pound wallet):
    \begin{inneritemizelessVspace}
        \item The payee does not exceed his/her defined digital pound 
        holding threshold after receiving the funds (standard case).
        \item The payee would exceed his/her defined digital pound 
        holding threshold, resulting in a transfer of funds that 
        exceed the threshold into a linked commercial bank account 
        (waterfall approach\footref{fn-ecb-wf-rwf}).
    \end{inneritemizelessVspace}
    \item Alternative methods to capture payee payment details:
    \begin{inneritemizelessVspace}
        \item The payer enters the payee's digital pound wallet alias / 
        commercial bank details on the 
        smartphone app provided by the PIP / commercial bank (standard case).
        \item The payer scans a QR code presented by the payee.
    \end{inneritemizelessVspace}
    \item Payment status:
    \begin{inneritemizelessVspace}
        \item The payment is successful (standard case).
        \item The payment fails (potential reasons include failure at payment 
        scheme, invalid payee details provided, etc).
    \end{inneritemizelessVspace}
\end{itemizelessVspace}

\pagebreak

\subsubsection{Actors\protect\footnote{An actor is an entity that initiates or participates
in the use case.}}
\label{subsubsec:p2p-actors}
\begin{itemizelessVspace}
    \item \emph{Payer}: Person initiating funds transfer.
    \item \emph{Payee}: Beneficiary of funds transfer.
    \item \emph{Commercial bank}: Commercial bank holding the payer's account and 
    providing a user interface to the payer.
    \item \emph{Digital pound wallet provider (PIP)}: Authorised and regulated firm 
    providing user interface between the user and the digital pound ledger.
    \item \emph{Technical Service Provider (TSP)}: A TSP provides 
    technical services such as communication, technical onboarding, 
    and information processing/storage to support an authorised 
    financial services provider. 
    It does not have a direct relationship with, or provide services to, 
    end users.
    \item \emph{Financial Market Infrastructures (FMI)}: An FMI is a multilateral system 
    among participating institutions, including the operator of the system, used for clearing, 
    settling or recording payments, securities, derivatives or other financial transactions \cite{bis-cpmi-glossary}. 

    \item \emph{Faster Payment System (FPS)}: UK payment system facilitating real-time payments \cite{abt-fps}.
    \item \emph{New Payment Architecture (NPA)}: The UK payments industry's proposed new way of 
    organising the clearing and settlement of interbank payments made from one payment account to another \cite{payuk-npa}.
    \item \emph{CBDC core ledger}: The CBDC core ledger would record issuance and transfer
    of digital pounds. 
    \item \emph{RTGS}: The Real-Time Gross Settlement (RTGS) system is the infrastructure 
    operated by the Bank of England that holds accounts 
    for commercial banks, building societies and other institutions. 
    The sterling balances held in these accounts are used to settle the 
    sterling payment obligations which arise from that institution's 
    direct participation in one or more UK payment or securities 
    settlement systems \cite{boe-rtgs}.
\end{itemizelessVspace}

\subsubsection{Preconditions\protect\footnote{A precondition of a use case explains the state 
that the system must be in for the use case to start.}}
\label{subsubsec:p2p-precond}
\begin{itemizelessVspace}
    \item Payer holds a commercial bank account and payee holds a digital pound wallet.
    \item Payer has a smartphone app with network connectivity to access his/her 
    commercial bank account, and has successfully authenticated on the app.
\end{itemizelessVspace}

\subsubsection{Normal flow\protect\footnote{A normal flow provides a detailed description of 
the user actions and system responses that will take place during execution of the use case under 
normal conditions.}}
\label{subsubsec:p2p-flows}
\begin{enumerate}[itemsep=1pt, parsep=2pt, topsep=1pt, partopsep=2pt]
    \item The use case begins with a parent (payer) using his/her bank's smartphone app to access 
    his/her commercial bank account.
    \item The parent (payer) initiates the payment by entering the
    amount and digital pound wallet 
    alias (such as mobile phone number) of his/her child (payee).
    \item The payer's bank performs the confirmation of payee check
    using the payee's digital pound wallet alias.
    \item The payer validates the payee's name and confirms initiation of the payment to the 
    payee.
    \item The payer authorises the payment using a pre-configured authorisation method, 
    such as biometric or one-time-password (OTP).
    \item Compliance checks, such as AML and fraud checks are performed by payer's commercial bank, the transfer amount is debited 
    from the payer's commercial bank account and is credited to the payee's digital pound wallet.
    \item The payer and the payee are notified of successful transfer
    of funds.
    \item The use case ends successfully.    
\end{enumerate}

\subsubsection{Subflows\protect\footnote{
    In some cases the normal flow of events should be decomposed into a 
    set of subflows that has a clear purpose.
}}
\label{subsubsec:p2p-subflows}
The following subflows are referenced in this use case but are not 
elaborated further within this document:
\begin{itemizelessVspace}
    \item Validate a \changes{user's} digital pound alias.
\end{itemizelessVspace}
\begin{itemizelessVspace}
    \item Perform AML and fraud check by payer's commercial bank and payee's PIP.
\end{itemizelessVspace}

\subsubsection{Postconditions\protect\footnote{Postconditions describe 
the state of the system at the conclusion of the use case. 
Postconditions may include conditions for both successful and unsuccessful 
execution of the use case.}}
\label{subsubsec:p2p-postcond}
\begin{itemizelessVspace}
    \item Successful completion:
    \begin{inneritemizelessVspace}
        \item Payer's commercial bank account is debited with the funds 
        transferred to payee.
        \item Payee's digital pound wallet is credited with the funds 
        \changes{received} from the payer.
        \item Funds are settled between the settlement ledger and digital 
        pound core ledger operated by the Bank of England.  
    \end{inneritemizelessVspace}
    \item Failure:
    \begin{inneritemizelessVspace}
        \item No funds are moved, and no change are made to the payer's 
        commercial bank account balance and the payee's digital pound 
        wallet balance.

    \end{inneritemizelessVspace}
\end{itemizelessVspace}

\subsubsection{Additional requirements\protect\footnote{
  Additional requirements 
  for the use case that may need to be addressed 
  during design or \changes{build}.}
}
\label{subsubsec:p2p-addReq}
\begin{itemizelessVspace}
    \item Personal data\footnote{
        ``Personal data means any information relating to an identified or 
        identifiable natural person (`data subject'); 
        an identifiable natural person is one who can be identified, 
        directly or indirectly, in particular by reference to an identifier 
        such as a name, an identification number, location data, an 
        online identifier or to one or more factors specific to the physical, 
        physiological, genetic, mental, economic, cultural or social 
        identity of that natural person;'' (Article 4(1) of \cite{uk-gdpr})
    } should not be available 
    to the Bank of England and appropriate privacy controls 
    should be applied to protect user privacy.
\end{itemizelessVspace}

\subsection{Design options for specific capabilities}
\label{subsec:p2p-design-opts}
In this subsection, we list our design assumptions and present design options
for two specific capabilities for this use case: 
\begin{enumerate}[label=(\roman*), itemsep=1pt, parsep=2pt, topsep=1pt, partopsep=2pt]
    \item U1.S1. Confirmation of payee check for digital pound wallets, and 
    \item U1.S2. Clearing and settlement of funds transfer from commercial bank
    money to digital pounds. 
\end{enumerate}

\noindent
\\
Note that we consider only the standard case in our preliminary evaluation.


\subsubsection{Assumptions}
\label{subsubsec:p2p-assumptions}
We make the following assumptions for the design options:
\begin{itemizelessVspace}
    \item The design options must align with the Bank of England's
    platform model.

    \item The payee's PIP could be potentially identified by using an alias 
    and PIP lookup service, which would be accessible to PIPs, ESIPs, commercial 
    banks and other participants of the digital pound ecosystem. 

    \item The UK FPS/NPA would be used to clear and settle funds transfers from/to 
    commercial bank accounts, as it provides immediate clearing 
    for low value payments.

    \item The Bank of England would integrate the new digital pound core 
    ledger with their existing reserve/settlement account ledger to enable 
    funds \changes{transfers} between the two ledgers.

    \item Any credit to users' digital pound wallets would require 
    confirmation by their PIPs to ensure that PIPs perform compliance checks 
    (such as AML and fraud checks) before the funds are made available to their customers.

    \item Any credit to an intermediary organisation's\footnote{
        Intermediary organisations (such as FMI, partner PIP) that have 
        access to both digital pounds and commercial bank money, can provide
        clearing and settlement across the two forms of money. \label{fn-interm-dg-cr}
    } digital pound wallet 
    must be notified in near real-time to the intermediary organisation. 
    This would enable it to perform further necessary  
    timely actions (such as credit their customer's digital pound wallet).
    
    \item Holding limits would not be imposed on digital pound 
    wallets of regulated financial institutions (such as PIPs, FMI and 
    commercial banks).

\end{itemizelessVspace}

\subsubsection{Design options for `Confirmation of payee check for digital 
pound wallets' (U1.S1) }
\label{subsubsec:p2p-design-CoP}
This use case requires the specific capability of performing confirmation of 
payee checks on the payee's digital pound wallet as the payer's commercial 
bank would need a mechanism to help the payer check the name of the digital pound
wallet holder.
The confirmation of payee service would also provide the 
target account number and sort code to which the payer's commercial bank 
would initiate the payment. \linebreak
Note this service could also be provided as an additional 
service by the PIP, or a digital pound alias service provider.

We present \changes{four} design options for this specific capability. 
Note we do not explore the design option where confirmation of payee service for 
digital pounds is provided by the CBDC system (e.g. as part of alias service) 
because this would allow the 
CBDC system to access personal data of the payee.

\myparagraph{Design option `Provided by PIPs over a peer-to-peer 
    channel using a TSP' (U1.S1.D1)}
    In this design option, the communication between the payer's 
    commercial bank and the payee's 
    PIP is over a peer-to-peer channel facilitated by a TSP, similar to the 
    Open Banking model \cite{open-banking}. 
    PIPs would need to expose standard, secure and reliable 
    APIs for the registration of participants (such as commercial banks) in 
    an automated manner, with registrations processed 
    and API responses provided in near real-time.
    
    The commercial bank would connect with the alias and PIP lookup service to 
    validate the digital pound alias and fetch the payee's PIP details. 
    The commercial bank would onboard onto the TSP and use a dynamic 
    client registration service to directly access the confirmation of payee 
    service provided by the payee's PIP. 
    The payee's PIP would respond with the name 
    of the digital pound wallet owner along with an account number and sort 
    code in which the payee's PIP could receive payment from 
    commercial banks.

\myparagraph{Design option `Provided by PIPs via a TSP' (U1.S1.D2)}
    In this design option, the communication between the commercial 
    bank and the PIPs is via a TSP. 
    PIPs would need to expose 
    APIs to the TSP, which in turn provides access to these APIs to other 
    entities (such as commercial banks, PIPs). 
    In this model, the TSP would act as an aggregator, connecting multiple 
    backend PIPs. 
    This model is similar to the new aggregator model for the  
    Confirmation of Payee (CoP) service \changes{\cite{abt-cop}} for commercial bank accounts.
    
    The commercial bank would onboard onto the TSP and access the 
    confirmation of payee service. 
    The TSP would in turn, connect with the alias and PIP lookup 
    service to validate the digital pound alias and fetch the payee's PIP 
    details. 
    The TSP would then invoke the API exposed by the payee's PIP 
    and pass the response received to the commercial bank.
    This response would include the name of the digital pound wallet owner, 
    and a commercial bank account number and sort code in which the payee's 
    PIP could receive payments from commercial banks.

\myparagraph{Design option `Provided by the digital pound alias 
    service provider' (U1.S1.D3)}
    In this design option, the digital pound alias and PIP lookup service 
    provider would extend the alias service to also provide confirmation of
    payee for digital pound wallets. 
    This service would therefore need to acquire and maintain digital pound
    users' personal data (such as name) and target commercial bank account 
    numbers and sort codes for digital pound wallets from all the PIPs. 

    This service would be used by commercial banks / PIPs for checking 
    the name of the a digital pound wallet owner and retrieving the 
    target account number and sort code 
    mapped against a digital pound wallet to initiate fund transfers. 
    
    \myparagraph{\changes{Design option `Provided by PIPs via the CBDC 
    system' (U1.S1.D4)}}
    \changes{In this design option, confirmation of payee checks 
    are routed via the CBDC system to the payee's PIP. 
    The CBDC system would provide key management and retrieval 
    capabilities to PIPs to ensure that personal data in
    the requests is not available to the Bank of England.
    Entities such as commercial banks that are not PIPs or ESIPs
    would need to partner with a PIP or an ESIP in order to access
    the confirmation of payee service for digital pound
    wallets.}

    \changes{The commercial bank would connect with the 
    alias and PIP lookup service to validate the digital pound alias and 
    fetch the payee's PIP details. 
    The commercial bank would then use its partner PIP's service to request 
    confirmation of payee via the CBDC system. 
    The partner PIP would encrypt confidential information
    (such as personal data of the payee) before submitting the 
    request to the CBDC system.}

    \changes{On receiving the request, the CBDC system would forward the 
    confirmation of payee request to the payee's PIP.
    The payee's PIP would decrypt the confidential information, 
    lookup payee details and respond with an encrypted message
    containing the name of the digital pound 
    wallet owner, and a commercial bank account number and sort code to 
    receive payments from commercial banks.
    The CBDC system would forward the encrypted response to the partner PIP which 
    would, in turn, forward it to the commercial bank.
    } \\
    \noindent
    \newline
    Appendix \ref{app:uc1-sequence-diagram-a} presents the sequence diagrams 
    for each design option above.

\subsubsection{Design options for `Clearing and settlement of funds transfer from 
commercial bank money to digital pounds' (U1.S2)} 
\label{subsubsec:p2p-design-SoF}

This use case requires the specific capability of clearing and 
settlement of funds transfer from commercial bank money to digital pounds so that the payer can transfer funds 
from his/her commercial bank account to payee's digital pound wallet.

We present six design options for this specific capability.

    \myparagraph{Design option `Provided by the CBDC system' (U1.S2.D1)}
    In this design option, the CBDC system operated by the Bank of England 
    provides the capability to transfer funds from the payer's commercial bank account 
    to the payee's digital pound wallet. 
    The Bank of England would have an FPS-reachable sort code and settlement account 
    in the reserve/settlement ledger (operated by the Bank of England).
    
    In this use case, after performing the 
    confirmation of payee check, the payer's commercial bank would have the 
    target account number and sort code (the Bank of England's dedicated sort 
    code and settlement account for digital pounds) to which it would transfer 
    the funds. 
    The payer's commercial bank would initiate an FPS/NPA instant payment 
    from payer's commercial bank account to the Bank of England's settlement account 
    (for digital pounds). 
    \changes{The payee's digital pound wallet identifier would be included 
    in the payment instruction.}
    On receiving the payment instruction, the Bank of England would request 
    the payee's PIP to approve the credit to the payee's digital pound wallet. 
    On approval by the payee's PIP, the Bank of England would fund an 
    equivalent amount of digital pounds in the payee's digital pound wallet.

    \myparagraph{Design option `Provided by payer's commercial bank which is 
    also a PIP' (U1.S2.D2)}
    In this design option, the payer's commercial bank is also a PIP and 
    has access to the digital pound core ledger and has its own digital pound 
    wallet. The payer's commercial bank would transfer funds from 
    the payer's commercial bank account to an internal account.
    The commercial bank, being a PIP, would have access to the CBDC system and would initiate 
    transfer of funds from its digital pound wallet to the payee's digital pound 
    wallet. 
    On receiving the payment instruction, the Bank of England would 
    request the payee's PIP to approve the transfer to the payee's digital pound 
    wallet. 
    The payee's PIP would retrieve confidential payment information 
    (such as payer's personal data) from the payer's PIP 
    to perform compliance checks, and would confirm 
    the transfer to the CBDC system. 
    On confirmation by the payee's PIP, the Bank of England would 
    transfer funds from the payer PIP's digital pound 
    wallet to the payee's digital pound wallet.
    
    The payer's commercial bank would need to maintain sufficient funds
    in its digital pound wallet\changes{\footnote{
        Note that, if settlement account funds are needed to fund digital 
        pound wallets, then sufficient liquidity would need to be 
        provisioned during RTGS operating hours to enable near real-time 
        settlement.
    }} for instant settlement of funds from the payer's 
    bank account to the payee's digital pound wallet.

    \myparagraph{Design option `Provided by payee's PIP, which is either a 
    commercial bank or partners with a commercial bank which is a PIP'
    (U1.S2.D3)}
    In this design option, the settlement of funds from a commercial bank 
    account to a digital pound wallet would be provided by the payee's PIP either
    directly, if it is also a commercial bank, or by partnering with a commercial bank 
    which is a PIP. 
    For the analysis below, we assume the payee's PIP partners 
    with a commercial bank that is also a PIP, as it would be the more
    general case.
    In order to settle payments from commercial bank money to digital pounds, 
    each PIP would rely on its partner commercial bank which is both a direct 
    participant of UK payment systems and has direct access to the digital 
    pound core ledger. 
    In this use case, after performing the alias validation and confirmation 
    of payee check, the payer's commercial bank \changes{would have} the dedicated target 
    account number and sort code of the payee PIP's partner commercial bank to  
    which it would transfer the funds. 
    The payer's commercial bank would initiate an FPS/NPA instant 
    payment from the payer's commercial bank account to the target account. 
    \changes{The payee's digital pound wallet identifier would be included 
    in the payment instruction.}
    
    After receiving funds from the payer's commercial bank, the partner 
    bank would then initiate a transfer of funds from its digital pound 
    wallet to the payee's digital pound wallet. On receiving the payment 
    instruction, the Bank of England would request the payee's PIP to 
    approve the transfer to the payee's digital pound wallet. The payee's 
    PIP would retrieve confidential payment information (such as payer's  
    personal data) from the payer's PIP to perform compliance 
    checks, and would confirm the transfer to the CBDC system. 
    On confirmation by the payee's PIP, the Bank of England would transfer 
    funds from the partner bank's digital pound 
    wallet to the payee's digital pound wallet.

    The partner commercial bank, which is also a PIP in this design 
    option, would need to maintain sufficient funds in its digital pound 
    wallet for instant settlement of funds from the payer's 
    bank account to the payee's digital pound wallet.

    \myparagraph{Design option `Provided by PIPs which are either \changes{Directly Connected 
    Non-Settling Participants} (DCNSP) or indirect FPS scheme participants' (U1.S2.D4)}
    In this design option, the payee's PIP would either be a DCNSP or an 
    indirect participant of the UK payment scheme (FPS) \cite{fps-principles}. 
    This would allow PIPs to participate in the 
    UK payment schemes \changes{in order to} send
    and receive commercial bank money, using 
    a \changes{Directly Connected Settling Participant (DCSP)} sponsor bank.
    
    In this use case, after performing 
    the confirmation of payee check, the payer's commercial bank would have 
    the target account number and sort code at the payee's PIP to which 
    it would transfer the funds. 
    The payer's commercial bank would initiate an FPS/NPA instant payment 
    from the payer's commercial bank account to the payee PIP target 
    sort code and account number. 
    \changes{The payee's digital pound wallet identifier would be included 
    in the payment instruction.}

    The FPS scheme would transfer funds to the payee PIP's sponsor bank account
    and notify the payee PIP of the incoming funds transfer.
    The payee's PIP would then initiate transfer of funds from its prefunded 
    digital pound wallet to the payee's digital pound wallet. Payee's PIP 
    would need to maintain sufficient funds in \changes{its} prefunded digital pound 
    wallet
    \changes{in order to settle the transfer of funds immediately.}
    
    If the payee's PIP is an indirect participant of the FPS scheme, 
    after the funds are transferred to the sponsor bank, then the sponsor
    bank would notify the payee's PIP of the incoming funds transfer.
    
    \myparagraph{Design option `Provided by an FMI' (U1.S2.D5)}
    In this design option, the clearing and settlement of funds transfers 
    from commercial bank accounts to digital pound wallets would be 
    provided by an FMI. 
    The FMI would be a participant of both the UK payment systems and the CBDC 
    system in order to clear and settle payments between digital pounds and 
    commercial bank money.
    It would hold (and manage liquidity across) technical settlement accounts on
    both the Bank of England's reserve/settlement ledger and the digital pound core 
    ledger. 
    PIPs would onboard onto the FMI, potentially including an exchange 
    of cryptographic keys.
 
    In this use case, after performing the 
    confirmation of payee check, the payer's commercial bank would initiate 
    an FPS/NPA instant payment from payer's commercial bank account to the 
    FMI's technical settlement account.
    \changes{The payee's digital pound wallet identifier would be included 
    in the payment instruction.}
    
    After receiving the funds transfer from the payment scheme, the FMI 
    would then initiate a transfer of funds from its 
    technical digital pound wallet to the payee's digital pound wallet. 
    The payment instruction could include encrypted personal data (such as payer's and payee's 
    personal data) to allow the payee's PIP to perform compliance checks.
    In another approach, the FMI could expose an API for PIPs to 
    retrieve confidential information for a payment instruction.
    
    On receiving the payment instruction from the FMI, 
    the Bank of England would request 
    the payee's PIP to approve the transfer to the payee's digital pound wallet. 
    The payee's PIP would retrieve the confidential payment information,
    perform compliance checks and confirm the transfer to the CBDC system. \linebreak 
    On confirmation by the payee's PIP, the Bank of England would transfer 
    funds from the FMI's technical digital pound wallet to the payee's 
    digital pound wallet.
    
    \myparagraph{Design option `Provided by an enhanced payment system' (U1.S2.D6)}
    In this design option, the clearing and settlement of funds transfer 
    from commercial bank money to digital pounds would be provided by an enhanced 
    payment system (such as NPA) potentially operated by a third party. 
    This payment system will provide settlement across commercial bank accounts and 
    digital pound wallets using the reserve/settlement account ledger operated by 
    the Bank of England.
    PIPs and commercial banks would onboard onto the enhanced payment system, including 
    an exchange of cryptographic keys.

    In this use case,
    after performing the confirmation of payee check,
    the payer's commercial bank will initiate a payment 
    from the payer's commercial bank account to the payee's digital pound wallet using 
    the enhanced payment system. 
    The payment system would debit the payer bank's settlement 
    account at the Bank of England and initiate an
    instruction to the CBDC system to credit an equivalent amount of digital pounds 
    in the payee's digital pound wallet. 
    The payment system would encrypt confidential information (such as
    personal data of the payer and payee) before submitting the payment 
    to the CBDC system. 
    
    On receiving the payment instruction, the Bank of England would request the 
    payee's PIP to approve the transfer to the payee's digital pound wallet. 
    The payee's PIP would decrypt the confidential payment information, 
    perform compliance checks, and confirm the transfer to the CBDC system. 
    On confirmation by the payee's PIP, the Bank of England would fund an 
    equivalent amount of digital pounds in the payee's digital pound wallet.
    
    \changes{Note, the settlement of funds between the Bank of 
    England's dedicated digital pound settlement account and 
    commercial banks' settlement accounts may be deferred and 
    use batching and netting because the RTGS system may not 
    support retail payment volumes and 24x7 operations. 
    This deferred settlement would require the Bank of England and commercial 
    banks to manage the resulting settlement risk.}
    \\
    \noindent
    \newline
    Appendix \ref{app:uc1-sequence-diagram-b} presents the sequence diagrams 
    for each design option above.
    \changes{Note the sequence diagram for design option U1.S2.D4
    presents the DCNSP model and not the indirect participant
    model.}


\subsection{Evaluation of design options}
\label{subsec:p2p-eval-designOpts}
In this section, we present our preliminary evaluation of each 
design option for the two specific capabilities of this use case.
Our preliminary evaluation comprises pros, cons and a suitability rating 
for each design option.
\subsubsection{Preliminary evaluation of design options for `Confirmation of 
payee for digital pound wallets' (U1.S1)}
\label{subsubsec:p2p-eval-designOpts-CoP}
Table \ref{table:table-p2p-designOpts-CoP} presents our preliminary 
evaluation of the design options to support confirmation of payee 
for digital pound wallets.

\begin{center}
    \begingroup
    \renewcommand{\arraystretch}{1.75}     
    \captionsetup{width=15cm}
      
      \begin{longtable}{ >{\raggedright}p{0.14\textwidth} >{\raggedright}p{0.11\textwidth} >{\raggedright}p{0.32\textwidth} >{\raggedright\arraybackslash}p{0.34\textwidth} }
       
       \toprule
       \textbf{Design option} & \textbf{Suitability rating} & \textbf{Pros} & \textbf{Cons} \\  [0.5ex] 
       \midrule
       \endfirsthead
  
       \multicolumn{4}{r}{\textit{Continued from previous page}} \\
       \toprule
       \textbf{Design option} & \textbf{Suitability rating} & \textbf{Pros} & \textbf{Cons} \\  [0.5ex] 
       \midrule
       \endhead
  
       \bottomrule
       \multicolumn{4}{r}{\textit{Continued on next page}} \\
       \endfoot
  
       \bottomrule
       \caption{Preliminary evaluation of design options to
       \changes{support} confirmation of payee for
       digital pound wallets (U1.S1).}
       \label{table:table-p2p-designOpts-CoP}
       \endlastfoot

        Provided by PIPs over peer-to-peer channel using 
        a TSP (U1.S1.D1)& 

        Partially suitable & 
        
        \begin{compactimize}
            \item Similar to the proven Confirmation of Payee 
            implementation for commercial bank accounts.
            \item Users' personal data is not available to the
            Bank of England.
        \end{compactimize}  &

        \begin{compactimize}
            \item APIs exposed by all PIPs may not be secure,
             reliable and as per standard.
            \item Commercial banks / PIPs would need to maintain multiple 
            peer-to-peer integrations with every PIPs.
            \item Overhead and complexity of operating a TSP
            service.
            \item Stringent security measures
            are required to mitigate cyber risks caused by
            use of the public internet.  
        \end{compactimize} \\ 
  
        Provided by PIPs via a TSP (U1.S1.D2) & 
  
        Suitable & 

        \begin{compactimize}
            \item Similar to the new aggregator model of Confirmation of 
            Payee for commercial bank accounts.
            \item Users' personal data is not available to the
            Bank of England.
        \end{compactimize} &
  
        \begin{compactimize}
          \item Would potentially add a point of failure (i.e. TSP) in the 
          overall process of transferring funds from commercial bank money to 
          digital pounds.
          \item Overhead and complexity of operating a TSP
          service.
        \item Introduces additional complexity
          (e.g. PIP key management and retrieval) 
          to ensure that personal data included in confirmation of payee check is 
          not available to the TSP.
        \end{compactimize} \\   

        Provided by the digital pound alias service provider (U1.S1.D3)& 
  
        Unsuitable & 

        \begin{compactimize}
            \item Simple communication design as PIPs and commercial banks 
            would integrate with a single service for both alias validation 
            and confirmation of payee check.
        \end{compactimize} &
  
        \begin{compactimize}
          \item Acquiring and maintaining data from PIPs to support confirmation of 
          payee checks would greatly increase complexity for the alias service and PIPs.
          \item If the service is operated by a TSP then users' personal data would 
          be available with the TSP, creating a risk of exposure to other 
          third parties.
        \end{compactimize} \\

        \changes{Provided by PIPs via the CBDC system (U1.S1.D4)}& 
  
        \changes{Partially suitable} & 

        \begin{compactimize}
            \item \changes{Simple design for PIPs because they would rely on
            the CBDC system and not require multiple point-to-point integrations.}
        \end{compactimize} &
  
        \begin{compactimize}
            \item \changes{Introduces additional complexity at the CBDC 
            system (e.g. PIP key management and retrieval) to ensure that 
            personal data in the request is not available to the Bank of England.
            \item Increases traffic and dependency on the CBDC system 
            for ancillary operations (e.g. confirmation of payee).
            \item Non-PIP ecosystem participants 
            would depend on PIPs
            to confirm digital pound payees.
            }
        \end{compactimize} \\

      \end{longtable}
  
    \endgroup
        
  \end{center}


\subsubsection{Preliminary evaluation of design options for `Clearing and 
settlement of funds transfer from commercial bank money to digital pounds' (U1.S2)}
\label{subsubsec:p2p-eval-designOpts-Settlement}
Table \ref{table:table-p2p-designOpts-settlement} presents our preliminary
evaluation of the design options to provide clearing and settlement of funds 
from commercial bank money to digital pounds.

\begin{center}
    \begingroup
    \renewcommand{\arraystretch}{1.75}     
    \captionsetup{width=15cm}
      
      \begin{longtable}{ >{\raggedright}p{0.14\textwidth} >{\raggedright}p{0.11\textwidth} >{\raggedright}p{0.32\textwidth} >{\raggedright\arraybackslash}p{0.34\textwidth} }
       
       \toprule
       \textbf{Design option} & \textbf{Suitability rating} & \textbf{Pros} & \textbf{Cons} \\  [0.5ex] 
       \midrule
       \endfirsthead
  
       \multicolumn{4}{r}{\textit{Continued from previous page}} \\
       \toprule
       \textbf{Design option} & \textbf{Suitability rating} & \textbf{Pros} & \textbf{Cons} \\  [0.5ex] 
       \midrule
       \endhead
  
       \bottomrule
       \multicolumn{4}{r}{\textit{Continued on next page}} \\
       \endfoot
  
       \bottomrule
       \caption{Preliminary evaluation of design options to
       provide 
       clearing and settlement of funds transfer from
       commercial bank money to digital pounds (U1.S2).}
       \label{table:table-p2p-designOpts-settlement}
       \endlastfoot

        Provided by the CBDC system (U1.S2.D1)& 

        Unsuitable & 
        
        \begin{compactimize}
            \item Simple design for PIPs because they can rely on the CBDC 
            system to settle payments across digital pounds and commercial bank money.
        \end{compactimize}  &

        \begin{compactimize}
            \item Users' personal data  (such as payer and 
            payee details) contained in the message formats of most payment 
            systems would be available to the Bank of England\footnote{ 
            If the CBDC system were to integrate directly
            with payment systems without processing personal data, then the 
            payment systems would need to be enhanced to support
            encrypted fields in messages and 
            a participant key exchange mechanism that supports field
            encryption.
            It is not clear whether, in the future, novel 
            technologies (such as privacy enhancing technologies) could 
            potentially allow the CBDC system operated by the 
            Bank of England to provide clearing and settlement of funds between 
            the digital pound and commercial bank money without compromising user 
            privacy. \label{ft-pet-cbdc-privacy}}.
            \item Introduces higher complexity in the CBDC system to build 
            and maintain integrations with UK payment systems.
        \end{compactimize} \\ 
  
        Provided by payer's commercial bank which is also a PIP (U1.S2.D2)&

        \changess{Partially suitable} & 

        \begin{compactimize}
            \item Commercial banks could perform instant settlement across 
            digital pounds and commercial bank money on their own via a
            shorter end-to-end process.
            \item Existing UK payment schemes would not be impacted by 
            transactions between digital pounds and commercial bank money.
            \item Users' personal data is not available to the
            Bank of England.
        \end{compactimize} &
  
        \begin{compactimize}
            \item Payers' commercial banks would need to be PIPs.
            \item \changess{The CP states that financial firms' access to the
            digital pound could be restricted, which may prevent
            payers' commercial banks from using funds in their own
            digital pound wallets to support interoperability.}
        \end{compactimize} \\   

        Provided by payee's PIP, which is either a 
        commercial bank or partners with a commercial bank
        which is a PIP (U1.S2.D3)& 
  
        Partially suitable & 

        \begin{compactimize}
            \item PIPs that are commercial banks and are 
            existing participants of UK payment systems 
            could provide interoperability on their own via a
            shorter end-to-end process.
            \item Users' personal data is not available to the
            Bank of England.
        \end{compactimize} &
  
        \begin{compactimize}
            \item PIPs that are not existing participants of UK payment 
            systems may incur higher costs and would be dependent on their
             partner commercial banks.
            \item Users' personal data would be available to the
            PIP's partner commercial bank.
            \item \changess{The CP states that financial firms'
            access to the digital pound could be restricted,
            which may prevent the PIP or its partner bank from
            using funds in their own digital pound wallets to
            support interoperability.}
        \end{compactimize} \\
        
        Provided by PIPs which are either DCNSPs or indirect FPS scheme participants (U1.S2.D4)&
  
        Unsuitable & 

        \begin{compactimize}
            \item Users' personal data is not available to the
            Bank of England.
            \item PIP's partner bank does not need to be a PIP.
            \item If the PIP is a DCNSP then it can apply incoming
            credits to payee's digital pound wallet in near real-time.
        \end{compactimize} &
  
        \begin{compactimize}
            \item PIPs would have to bear the cost and complexities of being 
            a DCNSP or an indirect participant.
            \item \changess{The CP states that financial firms'
            access to the digital pound could be restricted,
            which may prevent PIPs from using funds in their
            own digital pound wallets to support interoperability.}
            \item PIPs would need to depend on another intermediary 
            for their participation in UK payment systems and to 
            funds from its commercial bank account to its digital pound wallet.
            \item In the indirect participation model, the transfer of 
            funds would not be immediate as it would be intermediated
            by the partner institutions.
            \item Payment reconciliation complexities will be high.
        \end{compactimize} \\
  
        Provided by an FMI (U1.S2.D5)&
  
        \changess{Partially suitable} & 

        \begin{compactimize}
            \item Provides a common interoperability service that could be used by 
            PIPs and commercial banks.
            \item Users' personal data is not available to the
            Bank of England.
        \end{compactimize} &
  
        \begin{compactimize}
            \item Cost and complexity of establishing and operating the FMI service.
            \item \changess{The CP states that financial firms' 
            access to the digital pound could be restricted,
            which may prevent an FMI from using funds in its own
            technical digital pound wallet to support
            interoperability.}
        \end{compactimize} \\


        Provided by an enhanced payment system (U1.S2.D6)&
  
        Partially suitable & 

        \begin{compactimize}
            \item Simple design for PIPs and commercial banks because 
            can rely on the enhanced payment system to settle payments across 
            digital pounds and commercial bank money.
            \item Users' personal data is not available to the
            Bank of England.
        \end{compactimize} &
  
        \begin{compactimize}
            \item \changess{The Bank of England may need to manage the settlement
            risk arising from deferred net settlement of funds transfers
            from commercial bank money to digital pounds
            (similar to how commercial banks manage deferred net settlement
            risk for FPS payments today).}
            \item Cost and complexities of enhancing the payment system.
        \end{compactimize} \\

      \end{longtable}
  
    \endgroup
        
  \end{center}



\subsection{Initial insights}
\label{subsec:p2p-key-thoughts}
Our initial insights from the preliminary evaluation include:
\begin{itemizelessVspace}
    \item Is there a need for an additional digital pound wallet alias, 
    such as a unique sort code and account number, to enable payments to and 
    from commercial bank accounts?
    Or would a well-known user alias (such as mobile phone number) be sufficient?
    Our preliminary evaluation suggests that a well-known user alias could be 
    used to send and receive payments from commercial bank accounts.

    \item \changess{Several design options that provide
    clearing and settlement of funds transfers from
    commercial bank money to digital pounds depend on
    financial intermediaries (such as PIPs, commercial banks
    or an FMI) holding digital pounds (either for themselves
    or temporarily on behalf of users) in order to support
    interoperability.
    These financial intermediaries would also need to forecast,
    provision and monitor liquidity in their digital pound wallets.
    Note, however, the CP states that financial firms'
    access to the digital pound could be restricted,
    which may prevent them from using these funds to
    support interoperability.}

    \item Some commercial banks may wish to provide interoperability 
    with digital pounds to their customers, as described in design option 
    U1.S2.D2 (Provided by payer's commercial bank which 
    is also a PIP), but they may not wish to become PIPs.
    To support this, commercial banks could be given `PIP lite' access to 
    the CBDC system where they have their own digital pound wallet and 
    access to core ledger APIs, but would not onboard digital pound users.
    Further analysis would be needed to explore the feasibility of extending
    the platform model to include such a `PIP lite' role.

    \item Design option U1.S1.D2 (Provided by the PIPs via a TSP) could be 
    suitable for providing confirmation of payee for digital 
    pound wallets. 
    Further analysis could explore extending the existing 
    CoP service for commercial bank accounts to also include digital pound 
    wallets.

    \item Design option U1.S2.D5 (Provided by an FMI) 
    could be the most suitable option for providing
    interoperability, \changess{but only if it is permitted
    to use the funds in its technical digital pound wallet to
    do so.}
    This is primarily because a regulated FMI can handle personal data, become a 
    member of payment schemes, and provide common ecosystem services across 
    both the digital pound and commercial bank money.


    \item The CBDC system should support return and reversal
    payments to handle payment failures at UK payment schemes
    or at commercial banks, in order to support interoperability.

\end{itemizelessVspace}



\section{Use case for `Merchant initiated request to pay with interoperability 
 across the digital pound and commercial bank money' (U2)}
\label{sec:merch-requestpay}



In this section, we present the use case for a merchant initiated request to 
pay, with interoperability across the digital pound and commercial 
bank money. 
We then describe and evaluate design options for the specific capabilities for 
this use case.
\subsection{Use case details}
\label{subsec:merch-r2p-usecase}

\subsubsection{Use Case description}
\label{subsubsec:merch-r2p-usecaseDes}
A consumer (payer) orders a product from an ecommerce 
merchant (payee) and wishes to pay from his/her digital pound wallet
to the merchant's commercial bank account.
The ecommerce merchant initiates a request to pay which, after the 
authorisation by the consumer, results in the transfer of funds from 
the consumer's digital pound wallet to the merchant's commercial bank account.

Note that we focus on the standard case to 
simplify our preliminary evaluation.

\subsubsection{Use case variants}
\label{subsubsec:merch-r2p-variants}
We first identify a number of criteria that could potentially be used to 
develop variants for this use case. 
For each criterion, we highlight the standard case.
\begin{itemizelessVspace}
    \item Source and target account types:
    \begin{inneritemizelessVspace}
        \item From consumer's digital pound wallet to merchant's  
        commercial bank account (standard case).
        \item From consumer's commercial bank account to 
        merchant's digital pound wallet.
    \end{inneritemizelessVspace}
    \item Merchant type:
    \begin{inneritemizelessVspace}
        \item Ecommerce merchant (standard case).
        \item Brick-and-mortar store.
    \end{inneritemizelessVspace}
    \item Consumer's digital pound wallet type:
    \begin{inneritemizelessVspace}
        \item Wallet held in a smartphone app (standard case).
        \item Wallet held online at a PIP.
        \item Wallet held in a physical card.
    \end{inneritemizelessVspace}
    \item Authorisation decision by consumer on request to pay:
    \begin{inneritemizelessVspace}
        \item Authorise request to pay (standard case).
        \item Reject request to pay.
    \end{inneritemizelessVspace}

    \pagebreak

    \item Alternative methods for the consumer to provide payment account 
    details for the request to pay:
    \begin{inneritemizelessVspace}
        \item The consumer enters his/her digital pound wallet alias 
        on the merchant's online interface (standard case).
        \item The consumer enters his/her commercial bank account details
        on the merchant's online interface.
        \item The consumer uses his/her physical card at the terminal in the 
        merchant's brick-and-mortar store (e.g. using NFC).
        \item The merchant presents the request to pay
        on their payment interface (e.g. as a QR code) and the consumer scans
        the request using his/her smartphone app.
    \end{inneritemizelessVspace}
    
    \item Alternative options for communication between merchant's app and 
    PIP's digital pound wallet app on the consumer's smartphone 
    (only applicable for use case variant using an ecommerce merchant):
    \begin{inneritemizelessVspace}
        \item Using public/private network services (standard case).
        \item On-device app-to-app linking using a custom URL scheme.
    \end{inneritemizelessVspace}
    
    \item Alternative methods for consumer to authorise the
    request to pay transaction:
    \begin{inneritemizelessVspace}
        \item The consumer uses his/her PIP smartphone app to authorise 
        the transaction (standard case).
        \item The consumer uses the merchant's app/terminal to authorise the 
        transaction (e.g. PIN, NFC).
        \item The consumer is redirected to a web app hosted and controlled 
        by his/her PIP.
    \end{inneritemizelessVspace}
    
    \item Financial entity used by the consumer and the merchant:
    \begin{inneritemizelessVspace}
        \item The consumer's and merchant's wallet/account are managed by two 
        different financial entities (standard case).
        \item The consumer's and merchant's wallet/account are managed by the
        same financial entity.
    \end{inneritemizelessVspace}
    
    \item Funds availability in consumer's digital pound wallet 
    (only applicable for use case variant where consumer uses funds in 
    his/her digital pound wallet):
    \begin{inneritemizelessVspace}
        \item The consumer has sufficient funds in his/her digital pound wallet 
        (standard case).
        \item The consumer does not have sufficient funds in his/her 
        digital pound wallet, resulting in sweeping of funds from a linked commercial 
        bank account to his/her digital pound wallet (reverse waterfall 
        approach\footref{fn-ecb-wf-rwf}).
        \item The consumer does not have sufficient funds across his/her digital pound 
        wallet and commercial bank account, resulting in failure of the 
        transaction.
    \end{inneritemizelessVspace}
    \item Merchant's digital pound holding threshold (only applicable for 
    use case variant where the merchant receives funds 
    in its digital pound wallet):
    \begin{inneritemizelessVspace}
        \item The merchant does not exceed its defined digital pound  
        holding threshold after receiving the funds.
        \item The merchant would exceed its defined digital pound holding threshold, 
        resulting in a transfer of the funds that exceed the 
        threshold into a linked commercial bank account 
        (waterfall approach\footref{fn-ecb-wf-rwf}).
    \end{inneritemizelessVspace}
\end{itemizelessVspace}



\subsubsection{Actors}
\label{subsubsec:merch-r2p-actors}
\begin{itemizelessVspace}
    \item \emph{Consumer}: Person who orders a product from a merchant.
    
    \item \emph{Merchant}: Ecommerce merchant who sells products to consumers.
    
    \item \emph{Delivery partner}: Person or entity delivering ordered products 
    to the consumer.
    \item \emph{Consumer's digital wallet provider (PIP)}: Authorised and 
    regulated firm providing a user interface between a user (consumer) and 
     the digital pound ledger.
    
     \item \emph{Merchant's acquirer}: Financial institution that 
    provides a payment gateway, a payment processor and merchant accounts, 
    thereby allowing end-to-end processing and settlement of  
    transactions on behalf of merchants.
    
    \item \emph{Acquirer's partner PIP}: The merchant's acquirer may partner 
    with a PIP to access the digital pound core ledger APIs and perform permitted 
    operations on digital pound wallets (such as initiate a request to pay).
    
    \item \emph{Technical Service Provider (TSP)}: A TSP provides 
    technical services such as communication, technical onboarding, 
    and information processing/storage to support an authorised 
    financial services provider. 
    It does not have a direct relationship with, or provide services to, 
    end users.

    \item \emph{Financial Market Infrastructure (FMI)}: An FMI is a 
    multilateral system among participating institutions, including 
    the operator of the system, used for clearing, settling or 
    recording payments, securities, derivatives or other financial transactions.
    
    \item \emph{Faster Payment System (FPS)}: UK payment system facilitating 
    real-time payments.
    
    \item \emph{New Payment Architecture (NPA)}: The UK payments industry's 
    proposed new way of organising the clearing and settlement of interbank 
    payments made from one payment account to another.
    
    \item \emph{CBDC core ledger}: The CBDC core ledger would record issuance 
    and transfer of digital pounds. 
    
    \item \emph{RTGS}: The Real-Time Gross Settlement (RTGS) system is the 
    infrastructure operated by the Bank of England that holds accounts 
    for commercial banks, building societies and other institutions. 
    The sterling balances held in these accounts are used to settle the 
    sterling payment obligations which arise from that institution's 
    direct participation in one or more UK payment or securities 
    settlement systems.

\end{itemizelessVspace}



\subsubsection{Preconditions}
\label{subsubsec:merch-r2p-precond}
\begin{itemizelessVspace}
    \item Consumer holds a digital pound wallet at a PIP and 
    merchant holds an account with a commercial bank.
    \item Consumer has a digital pound wallet app on a 
    smartphone with network connectivity and 
    has successfully authenticated on the app.
\end{itemizelessVspace}



\subsubsection{Normal flow}
\label{subsubsec:merch-r2p-flows}
\begin{enumerate}[itemsep=1pt, parsep=2pt, topsep=1pt, partopsep=2pt]
    \item The use case begins with a consumer (payer) using an ecommerce 
    merchant (payee) app to order a product.
    
    \item At online checkout, the consumer selects to pay using his/her digital 
    pound wallet.
    
    \item The merchant app prompts the consumer to enter his/her digital 
    pound wallet alias on the payment form.
    
    \item The consumer enters his/her digital pound wallet alias (such as mobile phone number) 
    and places the order.
    
    \item The merchant's acquirer validates the consumer's digital pound 
    wallet alias and initiates a request to pay.
    
    \item The merchant app prompts the consumer to authorise 
    the payment request.
    
    \item The consumer accesses his/her digital pound wallet 
    on the PIP's smartphone app and reviews the payment request. 
    The payment request includes details, such as payee's name 
    (merchant's name), order amount and payment reference number.
    
    \item The consumer authorises the payment request using 
    a pre-configured authorisation method, such as biometric or OTP.
    
    \item Compliance checks, such as AML and fraud checks, are performed by 
    the consumer's PIP.
    
    \item On success of compliance checks, the order amount is transferred from 
    the consumer's digital pound wallet to the merchant's commercial bank account.
    
    \item The consumer is presented with the payment confirmation 
    and notified of the debit to his/her digital pound wallet.
    
    \item The merchant receives confirmation of successful funds transfer.
    
    \item The merchant updates the order status as confirmed, 
    and notifies the consumer that the funds have been received, and the 
    order has been successfully placed.
    
    \item The merchant ships the product to the consumer, the consumer 
    receives the product, and the order is marked as completed.
    
    \item The use case ends successfully.   
\end{enumerate}


\subsubsection{Subflows}
\label{subsubsec:merch-r2p-subflows}
The following subflows are referenced in this use case but are not 
elaborated further within this document:
\begin{itemizelessVspace}
    \item Integrate merchant payment gateway with digital 
    pound ecosystem.
    \item Add products to the merchant's shopping cart in the 
    merchant's app.
    \item Validate a consumer's digital pound alias.
    \item Perform AML and fraud check for 
    the outbound payment at the consumer's PIP.
    \changes{\item Perform order management, fulfilment, and delivery.}
\end{itemizelessVspace}


\subsubsection{Postconditions}
\label{subsubsec:merch-r2p-postcond}
\begin{itemizelessVspace}
    \item Successful completion:
    \begin{inneritemizelessVspace}
        \item Consumer's digital pound wallet is debited by the funds 
        transferred to the merchant.
        \item Merchant's commercial bank account is credited 
        with the funds from the consumer.
        \item Funds transfer is settled between the digital pound core ledger 
        and settlement ledger operated by the Bank of England.
        \changes{\item Consumer has the ordered product.}
    \end{inneritemizelessVspace}
    \item Failure:
    \begin{inneritemizelessVspace}
        \item No funds are moved, and no changes are made to the consumer's 
        digital pound wallet balance and the merchant's commercial bank account 
        balance.
        \item The product is not delivered to the consumer.
    \end{inneritemizelessVspace}
\end{itemizelessVspace}

\subsubsection{Additional requirements}
\label{subsubsec:merch-r2p-addReq}
\begin{itemizelessVspace}
    \item Personal data should not be available 
    to the Bank of England and appropriate privacy controls 
    should be applied to protect user privacy.
\end{itemizelessVspace}

\subsection{Design options for specific capabilities}
\label{subsec:merch-r2p-design-opts}

In this section, we list our design assumptions and present design options 
for two specific capabilities for this use case:
\begin{enumerate}[label=(\roman*), itemsep=1pt, parsep=2pt, topsep=1pt, partopsep=2pt]
    \item U2.S1. Request to pay for digital pound wallets, and 
    \item U2.S2. Clearing and settlement of funds transfer from digital pounds 
    to commercial bank money.
\end{enumerate}

\noindent
\\
Note that we consider only the standard case in our preliminary evaluation.


\subsubsection{Assumptions}
\label{subsubsec:merch-r2p-assumptions}
We make the following assumptions for the design options:
\begin{itemizelessVspace}
    \item The design options must align with the Bank of England's platform model.
    \item The consumer's PIP could be potentially identified using an alias 
    and PIP lookup service, which would be accessible to PIPs, ESIPs, 
    commercial banks and 
    other participants of the digital pound ecosystem.

    \item The merchant's acquirer may not be a PIP (as some acquirers may 
    not wish to become PIPs) and may use the
    services of a partner PIP for digital pound operations. 

    \item The UK FPS/NPA would be used to clear and settle funds transfers 
    from/to commercial bank accounts, as it provides immediate clearing 
    for low value payments.

    \item The Bank of England would integrate the new digital pound core 
    ledger with their existing reserve/settlement account ledger to 
    enable fund transfers between the two ledgers.
    
    \item Any credit to an intermediary organisation's\footref{fn-interm-dg-cr} 
    digital pound wallet must be notified in near real-time to the intermediary 
    organisation. This would enable it to perform further necessary  
    timely actions (such as credit their \changes{customers' digital pound wallets}).    

    \item Any credit to an intermediary organisation's digital pound wallet would not require 
    confirmation by the intermediary because it would not need to perform compliance checks. 

    \item Holding limits would not be imposed on digital pound 
    wallets of regulated financial institutions (such as PIPs, FMI and commercial banks).

\end{itemizelessVspace}


\subsubsection{Design options for `Request to pay for digital pound wallets' (U2.S1)}
\label{subsubsec:merch-r2p-design-r2p}
This use case requires the specific capability to initiate requests to pay 
from ecosystem participants (such as merchants' acquirers) to PIPs, and obtain 
authorisation from consumers to make the payment.

We present three design options for this specific capability. 

\myparagraph{Design option `Intermediated by the CBDC system' (U2.S1.D1)}
    In this design option, all digital pound messages are routed through 
    the CBDC system operated by the Bank of England. 
    The merchant's acquirer uses a partner PIP's services to send a request 
    to pay to the CBDC system. 
    The CBDC system then forwards the request to the consumer's PIP 
    which, in turn obtains the consumer's authorisation (via its user interface) 
    to initiate funds transfer.

\myparagraph{Design option `Direct peer-to-peer' (U2.S1.D2)}
    In this design option, direct communications between PIPs, and between PIPs 
    and other ecosystem participants (such as merchants' acquirers),
    are facilitated by a TSP. 
    PIPs expose standard, secure and reliable APIs over the 
    internet. 
    The discovery and registration of ecosystem 
    participants (such as PIPs and acquirers) and their APIs are 
    automated, with registrations processed and API 
    responses provided in near real-time, similar to the Open Banking model. 
    Dynamic client registration APIs provided by a TSP allow 
    registrations to be performed automatically using certificates and 
    software statement assertions 
    (SSAs\footnote{\label{ssa-def}
        A software statement assertion (SSA) is a JSON Web Token (JWT) 
        containing client metadata (such as organisation's identity) about 
        an instance of Third Party Provider (TPP) client software.
        The JWT is issued and signed by trusted entities 
        (adapted from \cite{open-banking-ssa}).
        })
    issued by the TSP or another trusted entity to authenticate other 
    ecosystem participants. 
    The merchant's acquirer onboards onto the TSP and uses the dynamic 
    client registration service to directly access the request to pay
    API provided by the consumer's PIP. 
    The consumer's PIP, in turn obtains the consumer's 
    authorisation (via its user interface) to initiate funds transfer.

\myparagraph{Design option `Using a \changes{common third-party network}' (U2.S1.D3)}
    In this design option, a third party (such as a TSP or an FMI) operates 
    a \changes{common} network, similar to SWIFT and card networks. 
    All PIPs and other ecosystem participants onboard and connect to 
    the \changes{common} network operated by the third party. 
    This network may use a hub-and-spoke, point-to-point or hybrid topology.
    The merchant's acquirer and consumer's PIP onboard
    onto the network. 
    The merchant's acquirer sends the request to pay to the 
    consumer's PIP via the network.
    The consumer's PIP, in turn obtains the consumer's 
    authorisation (via its user interface) to initiate funds transfer.

\noindent
\newline
Appendix \ref{app:uc2-seq-dig-comm-topology} presents the sequence diagrams 
for each design option above.

\subsubsection{Design options for `Clearing and settlement of funds transfer 
from digital pounds to commercial bank money' (U2.S2)}
\label{subsubsec:merch-r2p-design-Settlement}
This use case requires the specific capability of clearing and settlement 
of funds transfer from digital pounds to commercial bank money so that
funds can be transferred from the consumer's digital 
pound wallet to the merchant's commercial bank account 
after the consumer has authorised the request to pay. 

We present five design options for this specific capability.
Note we do not explore the design option where clearing and settlement
of funds transfer is provided by PIPs which are either DCNSPs or indirect FPS 
scheme participants because this design 
option is unsuitable (see Table \ref{table:table-p2p-designOpts-settlement}).

    \myparagraph{Design option `Provided by the CBDC system' (U2.S2.D1)}
    In this design option, the CBDC system operated by the Bank of England 
    provides the capability to clear and settle funds transfers from 
    digital pounds to commercial bank money.
    The Bank of England would have an FPS-reachable sort code and settlement account 
    in the reserve/settlement ledger (operated by the Bank of England).

    In this use case, after obtaining the consumer's authorisation 
    for the request to pay, the consumer's PIP 
    submits a payment instruction to the CBDC system 
    to transfer funds from the consumer's digital pound wallet 
    to the merchant's commercial bank account. 
    The merchant's commercial bank account details are included 
    in the payment instruction from the consumer's PIP.
    The CBDC system debits funds from the consumer's digital pound 
    wallet and initiates an FPS/NPA instant payment to transfer the funds
    to the merchant's commercial bank account. 

    The Bank of England would need to maintain sufficient funds in its 
    FPS-reachable account in the reserve/settlement ledger. 

    \myparagraph{Design option `Provided by merchant acquirer's 
    partner financial institution which is a PIP' (U2.S2.D2)}
    In this design option, the clearing and settlement of funds transfers 
    from digital pounds to commercial bank money would be 
    provided by the merchant acquirer's partner financial institution 
    which is both a PIP and a direct participant of UK payment systems.
    Note, the merchant acquirer's partner financial institution
    would hold both a digital pound wallet and 
    a settlement account at the Bank of England.
    
    In this use case, after obtaining the consumer's authorisation for the
    request to pay, the consumer's PIP submits a payment request to 
    the CBDC system to transfer funds from the consumer's digital pound 
    wallet to the partner financial institution's digital pound wallet. 
    This instruction does not include the consumer's personal data
    and the merchant's details.
    The partner financial institution retrieves the consumer's personal data
    and the merchant's details from the consumer's PIP.
    The partner financial institution then initiates an FPS 
    instant payment to transfer funds to the merchant's commercial 
    bank account. 
    
    The partner financial institution would need to maintain 
    sufficient funds in its FPS settlement account for instant funds 
    transfer settlement.

    Note, if the partner financial institution is not a direct 
    participant of the UK payment system (such as FPS), it would need to hold 
    an account with another bank which is a direct participant of UK payment 
    system, increasing complexity for no additional benefits. 
    We have, therefore, not included this design option in our analysis.

    \myparagraph{Design option `Provided by payer's PIP which 
    is also a commercial bank' (U2.S2.D3)}
    In this design option, the clearing and settlement of funds transfers 
    from digital pounds to commercial bank money would be 
    provided by the consumer's (payer) PIP which is a commercial bank and
    a direct participant of UK payment systems.

    In this use case, after obtaining the consumer's authorisation for the
    request to pay, the consumer's PIP submits a payment request to 
    the CBDC system to transfer funds from the consumer's digital pound 
    wallet to its own digital pound wallet. 
    The consumer's PIP then initiates an FPS instant payment to 
    transfer funds to the merchant's commercial 
    bank account. 

    The consumer's PIP would need to maintain 
    sufficient funds in its FPS settlement account for instant funds 
    transfer settlement.

    Note, if the consumer's PIP is not a direct participant of the UK payment 
    systems (such as FPS), it would need to hold an account with another 
    commercial bank which is a direct participant of UK payment systems, 
    which would increase complexity. 
    We have not included this design option in our analysis.

    \myparagraph{Design option `Provided by an FMI' (U2.S2.D4)}
    In this design option, the clearing and settlement of funds transfers 
    from digital pounds to commercial bank money would be provided 
    by an FMI. 
    The FMI would be a participant of both the UK payment systems and the CBDC 
    system in order to clear and settle payments between digital pounds and 
    commercial bank money.
    The FMI would hold (and manage liquidity across) technical settlement 
    accounts on both the Bank of England's reserve/settlement ledger and 
    the digital pound core ledger. 
    PIPs would onboard onto the FMI, potentially including an exchange 
    of cryptographic keys.

    In this use case, after obtaining the consumer's authorisation for the
    request to pay, the consumer's PIP submits a payment instruction
    (including the encrypted consumer's personal data and merchant's details) to 
    the CBDC system 
    to transfer funds from the consumer's digital pound wallet to the 
    FMI's technical digital pound wallet. 
    The payment notification from the CBDC system to the FMI 
    includes the encrypted personal data. 
    In another approach, the FMI could directly retrieve the encrypted data 
    (including the consumer's personal data and the merchant's details) 
    from the consumer's PIP.
    The FMI then initiates an FPS instant payment to transfer funds 
    to the merchant's commercial bank account. 

    The FMI would need to maintain 
    sufficient funds in its FPS settlement account for instant funds 
    transfer settlement.
    
    \myparagraph{Design option `Provided by an enhanced payment system' (U2.S2.D5)}
    In this design option, the clearing and settlement of funds transfers 
    from digital pounds to commercial bank money would be 
    provided by an enhanced payment system (such as NPA) potentially 
    operated by a third party. 
    PIPs and commercial banks would onboard onto the enhanced payment system, 
    including an exchange of cryptographic keys.

    In this use case, after obtaining the consumer's authorisation for the
    request to pay, the consumer's PIP submits a payment request to 
    the enhanced payment system to transfer funds from the consumer's digital 
    pound wallet to the merchant's commercial bank account. 
    The enhanced payment system sends an instruction (which does 
    not include either the consumer's personal data or the merchant's details) 
    to the CBDC system to debit the consumer's digital pound wallet. 
    The enhanced payment system then settles the funds transfer between 
    the Bank of England's dedicated 
    digital pound settlement account and the merchant's commercial bank's 
    settlement account in the reserve/settlement ledger. 
    Lastly, the enhanced payment system instructs the merchant's commercial bank to 
    credit funds to the merchant's commercial bank account.

    Note, the settlement of funds between the Bank of England's dedicated digital
    pound settlement account and commercial banks' settlement accounts may 
    be deferred and use
    batching and netting because the RTGS system may not support retail payment 
    volumes and 24x7 operations. 
    This deferred settlement would require the Bank of England and commercial 
    banks to manage the resulting settlement risk.
    \\
    \noindent
    \newline
    Appendix \ref{app:uc2-seq-dig-settlement} presents the sequence diagrams 
    for each design option above.



\subsection{Evaluation of design options}
\label{subsec:merch-r2p-eval-designOpts}
In this section, we present our preliminary evaluation of each 
design option for the two specific capabilities of this use case.
Our preliminary evaluation comprises pros, cons and a suitability rating 
for each design option.

\subsubsection{Preliminary evaluation of design options for 
`Request to pay for digital pound wallets' (U2.S1)}
\label{subsubsec:merch-r2p-eval-designOpts-r2p}
Table \ref{table:table-merch-r2p-designOpts-r2p} presents our preliminary 
evaluation of the design options to support 
request to pay for digital pound wallets.



  \begin{center}
    \begingroup
    \renewcommand{\arraystretch}{1.75}     
    \captionsetup{width=15cm}
      
      \begin{longtable}{ >{\raggedright}p{0.14\textwidth} >{\raggedright}p{0.11\textwidth} >{\raggedright}p{0.32\textwidth} >{\raggedright\arraybackslash}p{0.34\textwidth} }
       
       \toprule
       \textbf{Design option} & \textbf{Suitability rating} & \textbf{Pros} & \textbf{Cons} \\  [0.5ex] 
       \midrule
       \endfirsthead
  
       \multicolumn{4}{r}{\emph{Continued from previous page}} \\
       \toprule
       \textbf{Design option} & \textbf{Suitability rating} & \textbf{Pros} & \textbf{Cons} \\  [0.5ex] 
       \midrule
       \endhead
  
       \bottomrule
       \multicolumn{4}{r}{\emph{Continued on next page}} \\
       \endfoot
  
       \bottomrule
       \caption{Preliminary evaluation of design options to 
       support request to pay for digital pound wallets (U2.S1).}
       \label{table:table-merch-r2p-designOpts-r2p}
       \endlastfoot
  
        Intermediated by the CBDC system (U2.S1.D1)& 

        Partially suitable & 
        
        \begin{compactimize}
            \item Simple design for PIPs because they would rely on 
             the CBDC system and not require multiple point-to-point 
             integrations.
        \end{compactimize}  &

        \begin{compactimize}
            \item Introduces additional complexity at the CBDC system 
            (e.g. PIP key management and retrieval) 
            to ensure that personal data in the request 
            is not available to the Bank of England.
            \item Increases traffic and dependency on CBDC system 
            for ancillary operations (e.g. request to pay).
            \item Non-PIP ecosystem participants 
            would depend on PIPs
            to initiate requests to pay.
        \end{compactimize} \\ 
  
        Direct peer-to-peer (U2.S1.D2)& 
  
        Suitable & 

        \begin{compactimize}
            \item Users' personal data is not available to the Bank of England.
            \item Reduces traffic on the CBDC system for
            ancillary operations 
            (e.g. request to pay).
            \item Allows ecosystem participants (e.g. merchants' acquirers) 
            to directly communicate with PIPs.
            \item Similar to the established Open Banking model.
        \end{compactimize} &
  
        \begin{compactimize}
          \item All APIs exposed by PIPs may not be consistently secure, 
          reliable and as per the standards.
          \item Stringent security measures
          are required to mitigate cyber risks caused by
          use of the public internet.
          \item Overhead and complexity of operating a TSP
          service.
        \item Ecosystem participants would need to maintain multiple
          peer-to-peer integrations with every PIP.
        \end{compactimize} \\   

        Using a \changes{common third-party network} (U2.S1.D3)& 
  
        Suitable & 

        \begin{compactimize}
            \item Secure, reliable, and consistent communication channel.
            \item Participant authentication is managed by the network.
            \item Third-party network operator could be an FMI that also provides 
            funds clearing and settlement, which would simplify 
            the end-to-end payment processing.
            \item Provides a layer of security to PIPs' systems against cyber 
            threats.
        \end{compactimize} &
  
        \begin{compactimize}
          \item Higher cost of establishing and operating a \changes{common} third-party network.
          \item Options for customisation could be limited by the third 
          party.
          \item Could be a central point of failure if a hub and spoke 
          model is implemented.
          \item Users' personal data would be available to the TSP, 
          creating a risk of exposure to other third parties.

        \end{compactimize} \\ 
  
      \end{longtable}
  
    \endgroup
        
  \end{center}




\subsubsection{Preliminary evaluation of design options for
`Clearing and 
settlement of funds transfer from digital pounds to
commercial bank money' (U2.S2)}
Table \ref{table:table-merch-r2p-designOpts-settlement} presents our preliminary 
evaluation of the design options to provide clearing and settlement of funds transfer 
from digital pounds to commercial bank money.


  \begin{center}
    \begingroup
    \renewcommand{\arraystretch}{1.75}     
    \captionsetup{width=15cm}
      
      \begin{longtable}{ >{\raggedright}p{0.14\textwidth} >{\raggedright}p{0.11\textwidth} >{\raggedright}p{0.32\textwidth} >{\raggedright\arraybackslash}p{0.34\textwidth} }
       
       \toprule
       \textbf{Design option} & \textbf{Suitability rating} & \textbf{Pros} & \textbf{Cons} \\  [0.5ex] 
       \midrule
       \endfirsthead
  
       \multicolumn{4}{r}{\emph{Continued from previous page}} \\
       \toprule
       \textbf{Design option} & \textbf{Suitability rating} & \textbf{Pros} & \textbf{Cons} \\  [0.5ex] 
       \midrule
       \endhead
  
       \bottomrule
       \multicolumn{4}{r}{\emph{Continued on next page}} \\
       \endfoot
  
       \bottomrule
       \caption{Preliminary evaluation of design options to provide 
       clearing and settlement of funds transfer from digital pounds 
       to commercial bank money (U2.S2)}
       \label{table:table-merch-r2p-designOpts-settlement}
       \endlastfoot
  
        Provided by the CBDC system (U2.S2.D1) & 

        Unsuitable & 
        
        \begin{compactimize}
            \item Simple design for the PIPs because they can rely on the 
            CBDC system to settle payments across digital pounds 
            and commercial bank money.
        \end{compactimize}  &

        \begin{compactimize}
            \item Users' personal data (such as payer and 
            payee details) contained in the message formats of most payment 
            systems would be available to the Bank of England\footref{ft-pet-cbdc-privacy}. 
            \item Introduces higher complexity in the CBDC system to build 
            and maintain integrations with UK payment systems.
        \end{compactimize} 
        \\ 
  
        Provided by merchant acquirer's 
        partner financial institution which is a 
        PIP (U2.S2.D2) & 
  
        Partially suitable & 

        \begin{compactimize}
            \item Users' personal data is not available to the
            Bank of England.
            \item Aligns with existing merchant and acquirer payment clearing and 
            settlement processes.
            \item If the merchant's acquirer is both a PIP and a commercial bank then
             it can clear and settle payments via a shorter end-to-end process.
        \end{compactimize} &
  
        \begin{compactimize}
          \item The partner financial institution would need to maintain sufficient 
          liquidity in its reserve/settlement account for settlement of the
          funds transfer.
          \item \changess{The CP states that financial firms' access to
          the digital pound could be restricted, which may prevent
          the merchant acquirer's partner financial institution
          from using funds in its own digital pound wallet to
          support interoperability.}
          \item Users' personal data is available to the partner
          financial institution.
          \item \changes{Acquirer's partner institution would be a
          potential point of delays and failure.}
          \item \changes{The payer would receive
          payment confirmation before the funds are transferred
          to the merchant's account}
          \changess{(similar to some existing payment methods
          such as card payments).}
        \end{compactimize} \\

        Provided by payer's PIP which is also a commercial bank (U2.S2.D3)& 
  
        Partially suitable & 

        \begin{compactimize}
            \item PIPs that are existing participants 
            of UK payment systems could provide interoperability on their own 
            via a shorter end-to-end process.
            \item Users' personal data is not available to the
            Bank of England.
        \end{compactimize} &
  
        \begin{compactimize}
          \item PIPs that are not existing participants of UK payment 
          would be dependent on their partner commercial bank and 
          may also incur higher costs.
          \item PIPs would need to maintain sufficient 
          liquidity in their reserve/settlement account for settlement of 
          funds transfer.
          \item \changess{The CP states that financial firms'
          access to the digital pound could be restricted, which
          may prevent PIPs from using funds in their own digital
          pound wallets to support interoperability.}
        \end{compactimize} \\   

        Provided by an FMI (U2.S2.D4)& 
  
        \changess{Partially suitable} & 

        \begin{compactimize}
            \item PIPs and commercial banks 
            can rely on the FMI to settle payments across 
            digital pounds and commercial bank money.
            \item Provides a consistent interoperability service that could be 
            used by other ecosystem participants.
            \item Users' personal data is not available to 
            the Bank of England.
        \end{compactimize} &
  
        \begin{compactimize}
          \item Cost and complexity of establishing and operating the FMI service.
          \item The FMI would need to ensure sufficient liquidity in its 
          reserve/settlement account to settle funds transfer.
          \item \changess{The CP states that financial firms' access to
          the digital pound could be restricted, which may prevent
          an FMI from using funds in its own 
          digital
          pound wallet to support interoperability.}
        \end{compactimize} \\ 

        Provided by an enhanced payment system (U2.S2.D5) & 
  
        Partially suitable & 

        \begin{compactimize}
            \item PIPs and commercial banks  
            can rely on the enhanced payment system to settle payments across 
            digital pounds and commercial bank money.
            \item Users' personal data is not available to
            the Bank of England.
        \end{compactimize} &
  
        \begin{compactimize}
            \item Cost and complexity of enhancing the payment system.
        \end{compactimize} \\ 
      \end{longtable}
  
    \endgroup
        
  \end{center}



\vspace{-4mm}

\subsection{Initial insights}
\label{subsec:merch-r2p-key-thoughts}

Our initial insights from the preliminary evaluation include:
\begin{itemizelessVspace}
    \item Design options U2.S1.D2 (Direct peer-to-peer) and U2.S1.D3 
    (Using a \changes{common third-party network}) could be more suitable for providing 
    request to pay and other such ancillary communications (which include 
    personal data) between PIPs and 
    other ecosystem participants. 
    Further analysis could explore these design options 
    for providing other ancillary communications required for other use cases.

    \item \changess{Several design options that provide
    clearing and settlement of funds transfers from
    digital pounds to commercial bank money depend on
    financial intermediaries (such as PIPs, commercial banks
    or an FMI) holding digital pounds (either for themselves
    or temporarily on behalf of users) in order to support
    interoperability.
    Note, however, the CP states that financial firms'
    access to the digital pound could be restricted,
    which may prevent them from using these funds to
    support interoperability.}

    \item Design option U2.S2.D4 (Provided by an FMI) 
    could be the most suitable option for providing interoperability,
    \changess{but only if it is permitted to use the funds in its
    technical digital pound wallet to do so.}
    This is primarily because a regulated FMI can handle personal data, become a 
    member of payment schemes, and provide common ecosystem services across 
    both the digital pound and commercial bank money.
    
    \item An FMI operating a \changes{common third-party network} 
    (as described in design option U2.S1.D3 `Using a \changes{common third-party network}'),
    and providing interoperability across digital pounds and commercial bank money 
    (as described in design option U2.S2.D4 `Provided by an FMI') would simplify
    the experience for ecosystem participants including consumers, PIPs and merchants.

    \item Policy and technical arrangements should be analysed
    to support payment reversals on the digital pound for scenarios
    such as payment rejections, timeouts and other failure scenarios, particularly 
    when such payment reversals could potentially result in negative 
    balance in a digital pound wallet. 
\end{itemizelessVspace}



\section{Use case for `Lock digital pounds and pay on physical delivery from 
digital pounds to commercial bank money' (U3)}
\label{sec:lock-pay}


In this section, we present the use case for locking digital pounds at 
the time of placing an order with an ecommerce merchant, and transferring 
funds to the merchant's commercial bank account at the time of 
physical delivery of goods.
We then describe and evaluate design options for the specific 
capabilities for 
this use case. 
\subsection{Use case details}
\label{subsec:lock-pay-details}

\subsubsection{Use case description}
\label{subsubsec:lock-pay-description}
A consumer (payer) orders a product from an ecommerce merchant (payee) 
and wishes to pay from his/her digital pound wallet
to the merchant's commercial bank account on physical delivery of the 
ordered product. 
The ecommerce merchant initiates a request to lock 
funds in the consumer's digital pound wallet. 
Upon authorisation by the consumer, the required funds are 
locked\footnote{Locking of funds results in reducing the available balance 
in the consumer's digital pound wallet while keeping the ledger balance 
the same. 
An alternative to locking funds is escrowing funds, where the funds 
are transferred from the consumer's digital pound wallet to the digital 
pound wallet of a trusted third party.
Similar to releasing locked funds, the trusted third party releases the 
escrowed funds to the beneficiary at the time of physical delivery.
\changes{This use case elaborates only locking functionality and not
escrow functionality.}} 
and the order is confirmed. 
At the time of physical delivery, the delivery agent 
requests the consumer to release the locked funds and pay. 
Upon authorisation from the consumer, the locked funds are released 
and transferred from the
consumer's digital pound wallet to the merchant's 
commercial bank account. 
The delivery agent receives confirmation of payment and hands over 
the product to the consumer.

Note that we focus on the standard case to  
simplify our preliminary evaluation.

\subsubsection{Use case variants}
\label{subsubsec:lock-pay-usecaseVar}
We first identify a number of criteria that could potentially be used to 
develop variants for this use case. 
For each criterion, we highlight the standard case.
\begin{itemizelessVspace}
    \item Source and target account types:
    \begin{inneritemizelessVspace}
        \item From consumer's digital pound wallet to merchant's 
        commercial bank account (standard case).
        \item From consumer's commercial bank account to merchant's 
        commercial bank account.
        \item From consumer's commercial bank account to merchant's 
        digital pound wallet.
        \item From consumer's digital pound wallet to merchant's 
        digital pound wallet.
    \end{inneritemizelessVspace}

    \item Merchant type:
    \begin{inneritemizelessVspace}
        \item Ecommerce merchant (standard case).
        \item Brick-and-mortar store.
    \end{inneritemizelessVspace}

    \pagebreak

    \item Order delivery status:
    \begin{inneritemizelessVspace}
        \item Successful delivery to consumer (standard case).
        \item Failed delivery to consumer.
    \end{inneritemizelessVspace}

    \item Number of products in an order:
    \begin{inneritemizelessVspace}
        \item Only one product to be delivered to the consumer (standard case).
        \item Multiple products to be delivered to the consumer separately.
    \end{inneritemizelessVspace}

    \item Consumer's digital pound wallet type:
    \begin{inneritemizelessVspace}
        \item Wallet held in a smartphone app (standard case).
        \item Wallet held online at a PIP.
        \item Wallet held in a physical card.
    \end{inneritemizelessVspace}

    \item Authorisation decision by consumer on request to lock:
    \begin{inneritemizelessVspace}
        \item Authorise request to lock (standard case).
        \item Reject request to lock.
    \end{inneritemizelessVspace}
    

    \item Alternative methods for the consumer to provide payment account 
    details for the request to lock:
    \begin{inneritemizelessVspace}
        \item The consumer enters his/her digital pound wallet alias 
        on the merchant's online interface (standard case).
        \item The consumer enters his/her commercial bank account 
        details on the merchant's online interface.
        \item The consumer uses his/her physical card at the terminal in the 
        merchant's brick-and-mortar store (e.g. using NFC).
        \item The merchant presents the request to lock
        on their payment interface (e.g. as a QR code) and the consumer scans 
        the request using his/her smartphone app.
    \end{inneritemizelessVspace}

    \item Financial entity used by the consumer and the merchant:
    \begin{inneritemizelessVspace}
        \item The consumer's and merchant's wallet/account are managed by two 
        different financial entities (standard case).
        \item The consumer's and merchant's wallet/account are managed by the
        same financial entity.
    \end{inneritemizelessVspace}

    \item Funds availability in consumer's digital pound wallet 
    (only applicable for use case variant where consumer uses funds in 
    his/her digital pound wallet):
    \begin{inneritemizelessVspace}
        \item The consumer has sufficient funds in his/her digital 
        pound wallet (standard case).
        \item The consumer does not have sufficient funds in his/her 
        digital pound wallet, resulting in sweeping of funds from a linked 
        commercial bank account to his/her digital pound wallet 
        (reverse waterfall approach\footref{fn-ecb-wf-rwf}) before locking 
        the funds.
        \item The consumer does not have sufficient funds across his/her 
        digital pound wallet and commercial bank account, 
        resulting in failure to lock funds.
    \end{inneritemizelessVspace}
             
    \item Merchant's digital pound holding threshold (only applicable 
    for use case variant where the merchant receives funds in its 
    digital pound wallet):
    \begin{inneritemizelessVspace}
        \item The merchant does not exceed its defined digital pound 
        holding threshold after receiving the funds.
        \item The merchant would exceed its defined digital pound holding threshold,  
        resulting in a transfer of the funds that exceed the threshold into a 
        linked commercial bank account 
        (waterfall approach\footref{fn-ecb-wf-rwf}).
    \end{inneritemizelessVspace}

    \item Alternative methods for consumer to authorise the request to lock 
    transaction:
    \begin{inneritemizelessVspace}
        \item The consumer uses his/her PIP smartphone app 
        to authorise the funds lock (standard case).
        \item The consumer uses the merchant's app/terminal to authorise 
        the funds lock (e.g. PIN, OTP).
        \item The consumer is redirected to a web 
        app hosted and controlled by his/her PIP to authorise the funds lock.
    \end{inneritemizelessVspace}

    \item Alternative methods to confirm the release of locked funds 
    on physical delivery:
    \begin{inneritemizelessVspace}
        \item Consumer triggers release of locked funds via his/her PIP's 
        app and the delivery agent is notified of successful transfer 
        of funds (standard case).
        \item Delivery agent initiates a request to release locked 
        funds, which is routed to the consumer's PIP app where he/she 
        authorises the request. The delivery agent is 
        notified of successful transfer of funds.
        \item Merchant provides consumer with a delivery confirmation 
        code which he/she provides to the delivery agent. The delivery 
        agent verifies the code, triggering the release of locked 
        funds. The delivery agent is notified of successful transfer of funds.
    \end{inneritemizelessVspace} 

\end{itemizelessVspace}

\subsubsection{Actors}
\label{subsubsec:lock-pay-actors}
\begin{itemizelessVspace}
    \item \emph{Consumer}: Person who orders a product from a merchant.

    \item \emph{Merchant}: Ecommerce merchant who sells 
    products to consumers.

    \item \emph{Delivery partner}: Person or entity delivering ordered 
    products to the consumer.

    \item \emph{Consumer's digital wallet provider (PIP)}: Authorised 
    and regulated firm providing a user interface between 
    a user (consumer) and the digital pound ledger.

    \item \emph{Merchant's acquirer}: Financial institution that 
    provides a payment gateway, a payment processor and merchant 
    accounts, thereby allowing end-to-end processing and 
    settlement of transactions on behalf of merchants.

    \item \emph{Acquirer's partner PIP}: The merchant's acquirer may 
    partner with a PIP to access the digital pound core ledger 
    APIs and perform permitted operations on digital pound 
    wallets (such as initiate a request to lock).

    \item \emph{Technical Service Provider (TSP)}: A TSP provides 
    technical services such as communication, technical onboarding, 
    and information processing/storage to support an authorised 
    financial services provider.
    It does not have a direct relationship with, or provide services 
    to, end users.

    \item \emph{Financial Market Infrastructure (FMI)}: An FMI is a 
    multilateral system among participating institutions, including 
    the operator of the system, used for clearing, settling or 
    recording payments, securities, derivatives or other financial 
    transactions.

    \item \emph{CBDC core ledger}: The CBDC core ledger would record 
    issuance and transfer of digital pounds.

\end{itemizelessVspace}

\subsubsection{Preconditions}
\label{subsubsec:lock-pay-precond}

\begin{itemizelessVspace}
    \item Consumer holds a digital pound wallet at a PIP 
    and merchant holds an account with a commercial bank.

    \item Consumer has a digital pound wallet app on a 
    smartphone with network connectivity and 
    has successfully authenticated on the app.

\end{itemizelessVspace}

\subsubsection{Normal flow}
\label{subsubsec:lock-pay-flows}
\begin{enumerate}[itemsep=1pt, parsep=2pt, topsep=1pt, partopsep=2pt]
    \item The use case begins with a consumer (payer) using an 
    ecommerce merchant (payee) app to order a product.

    \item At online checkout, the consumer selects to pay on physical 
    delivery of the product using his/her digital pound wallet.

    \item The merchant app prompts the consumer to enter his/her 
    digital pound wallet alias on the payment form.

    \item The consumer enters his/her digital pound wallet alias 
    (such as mobile phone number) and places the order.
    
    \item The merchant's acquirer validates the consumer's digital 
    pound alias and initiates a request to lock funds in
    the consumer's digital pound wallet.

    \item The merchant app prompts the consumer to 
    authorise the request to lock funds.

    \item The consumer receives a notification on his/her PIP's 
    smartphone app to authorise the request to lock funds.

    \item The consumer accesses his/her digital pound 
    wallet on the PIP's smartphone app and reviews the lock request. 
    The lock request includes details such as the merchant's name,
    lock amount, lock expiry date and reference number.

    \item The consumer authorises the lock request using a 
    pre-configured authorisation method, such as biometric or OTP.

    \item Compliance checks, such as AML and fraud checks, are performed 
    by the consumer's PIP.

    \item  On success of compliance checks, the order amount is locked 
    in the consumer's digital pound wallet. 
    The merchant's commercial bank account is recorded 
    as the beneficiary of the locked funds.

    \item The consumer is presented with the lock confirmation 
    and notified of funds locked in his/her digital pound wallet.

    \item The merchant receives confirmation of successful 
    funds lock.

    \item The merchant updates the order status as confirmed, and
    notifies the consumer that the funds have been locked and that 
    the order has been successfully placed.

    \item The merchant dispatches the product and the
    delivery agent transports the product to the consumer's address.

    \item The delivery agent requests (e.g. by presenting a QR code) the 
    consumer to release the locked funds and pay.

    \item The consumer checks the product is as per the order and uses 
    the PIP's smartphone app to initiate the release of the locked funds 
    (e.g. by scanning the QR code presented by the delivery agent).

    \item This triggers the release of the locked funds and the funds 
    transfer to the merchant's commercial bank account.

    \item The ledger balance in the consumer's digital pound wallet 
    is updated and reflects the debit of the order amount (the 
    available balance does not change).

    \item The merchant receives confirmation of successful receipt of 
    funds in their commercial bank account.

    \item The merchant notifies the delivery agent of successful 
    receipt of funds.

    \item The agent hands over the product to the consumer and marks 
    the delivery as complete.
    
    \item The merchant marks the order as completed.
    
    \item The use case ends successfully.
\end{enumerate}

\subsubsection{Subflows}
\label{subsubsec:lock-pay-subflows}
The following subflows are referenced in this use case but are not 
elaborated further within this document:
\begin{itemizelessVspace}
    \item Integrate merchant payment gateway with digital pound 
    ecosystem.

    \item Add products to the merchant's shopping cart in the 
    merchant's app.

    \item Validate a consumer's digital pound alias. 
        
    \item Perform AML and fraud check for 
    the outbound payment at the consumer's PIP.

    \item Perform order management, fulfilment, and delivery.

\end{itemizelessVspace}


\pagebreak

\subsubsection{Postconditions}
\label{subsubsec:lock-pay-postcond}
\begin{itemizelessVspace}
    \item Successful completion:
    \begin{inneritemizelessVspace}
        \item Consumer's digital pound wallet funds are transferred to 
        the merchant's commercial bank account.

        \item Funds transfer is settled between the digital pound core ledger and 
        settlement ledger operated by the Bank of England.

        \item Consumer has the ordered product.
    \end{inneritemizelessVspace}
    \item Failure:
    \begin{inneritemizelessVspace}
        \item No funds are moved, and no changes are made to the consumer's 
        digital pound wallet balance and the merchant's commercial bank account 
        balance.

        \item The funds locked for this order are unlocked and now 
        available to the consumer.

        \item The product is not delivered to the consumer (and is 
        returned if it was dispatched).
    \end{inneritemizelessVspace}
\end{itemizelessVspace}

\subsubsection{Additional requirements}
\label{subsubsec:lock-pay-addReq}
\begin{itemizelessVspace}
    \item Personal data should not be available 
    to the Bank of England and appropriate privacy controls 
    should be applied to protect user privacy.
\end{itemizelessVspace}

\subsection{Design options for specific capabilities}
\label{subsec:lock-pay-design-opts}

In this section, we list our design assumptions and present design options 
for three specific capabilities for this use case:
\begin{enumerate}[label=(\roman*), itemsep=1pt, parsep=2pt, topsep=1pt, partopsep=2pt]
\item U3.S1. Request to lock funds in digital pound wallets,
\item U3.S2. Lock funds and confirm lock status to the merchant, and 
\item U3.S3. Release locked funds from digital pound wallets and initiate funds transfers.
\end{enumerate}

\noindent
The specific capability to clear and settle the funds 
transfer from the consumer's digital pound wallet to the merchant's 
commercial bank account has already been covered in Section 
\ref{subsubsec:merch-r2p-design-Settlement} and
so will not be described here.

\changes{Note that we consider only the standard case in our preliminary evaluation.}


\subsubsection{Assumptions}
\label{subsubsec:lock-pay-assumptions}
We make the following assumptions for the design options:
\begin{itemizelessVspace}
    \item The design options must align with the Bank of England's platform model.
    
    \item The consumer's PIP could be potentially identified using an alias 
    and PIP lookup service, which would be accessible to PIPs, ESIPs, 
    commercial banks and other participants of the digital pound ecosystem.

    \item The merchant's acquirer may not be a PIP (as some acquirers may 
    not wish to become PIPs) and may use the
    services of a partner PIP for digital pound operations. 

    \item The future policies and regulations introduced for the digital pound
    will permit locking/escrowing consumers' digital pounds until 
    delivery.

\end{itemizelessVspace}

\subsubsection{Design options for `Request to lock funds in digital 
pound wallets' (U3.S1)}
\label{subsubsec:lock-pay-design-req2lock}
This use case requires the specific capability to initiate
requests to lock from ecosystem participants (such as the merchant's
acquirer) to PIPs in order to obtain authorisation from consumers to 
place the lock.

We present three design options for this 
specific capability. 

\myparagraph{Design option `Intermediated by the CBDC system' (U3.S1.D1) }
    In this design option, all digital pound 
    messages are routed through the CBDC system operated by 
    the Bank of England. 
    The merchant's acquirer uses a partner 
    PIP's services to send a request to lock to the CBDC 
    system. 
    The CBDC system then forwards the request to the consumer's 
    PIP which, in turn obtains the consumer's 
    authorisation (via its user interface) to lock funds.
\myparagraph{Design option `Direct peer-to-peer' (U3.S1.D2)} 
    In this design option, direct communications between PIPs, and between PIPs 
    and other ecosystem participants (such as merchant's acquirers), 
    are facilitated by a TSP. 
    PIPs expose standard, secure and 
    reliable APIs over the internet. 
    The discovery and registration of 
    ecosystem participants (such as PIPs and acquirers) and their 
    APIs are automated, with registrations processed and API 
    responses provided in near real-time, similar to the Open Banking model. 
    Dynamic client registration APIs provided by a TSP allow 
    registrations to be performed automatically using certificates 
    and SSAs\footref{ssa-def}  
     issued by the TSP or another trusted entity to authenticate other 
    ecosystem participants. 
    The merchant's acquirer onboards onto the TSP and use the 
    dynamic client registration service to directly access the
    request to lock API provided 
    by the consumer's PIP. The consumer's PIP, in turn obtains the 
    consumer's authorisation (via its user interface) to lock funds.

\myparagraph{Design option `Using a \changes{common third-party network}' 
    (U3.S1.D3)}
    In this design option, a third party (such as a TSP or an FMI) operates
    a \changes{common} network, similar to SWIFT and card networks. 
    All PIPs and other ecosystem participants onboard and connect 
    to the \changes{common} network operated by the third party. 
    This network may use a hub-and-spoke model, 
    point-to-point or hybrid topology. 
    The merchant's acquirer and consumer's PIP onboard
    onto the network. The merchant's acquirer sends the 
    request to lock to the consumer's PIP via the network. The 
    consumer's PIP, in turn obtains the consumer's authorisation 
    (via its user interface) to lock funds.

\noindent
\\
Appendix \ref{app:uc3-seq-dig-r2l-topology} presents the sequence diagrams 
for each design option above.

\subsubsection{Design options for `Locking funds in digital pound
wallets and confirming lock status to the merchant' (U3.S2)}
\label{subsubsec:lock-pay-design-conf-lock}
This use case requires the specific capability to lock funds 
after a consumer provides authorisation and notify the merchant 
of the successful funds lock. 
Design assumptions specific to these design options are as follows:
\begin{itemizelessVspace}
    \item For the design options where the PIP records locks 
    (design options U3.S2.D2, U3.S2.D3 and U3.S2.D4), 
    the payment APIs provided by the CBDC system include 
    a `minimum available balance' field and ensure that 
    after the payment is completed, the consumer's digital pound 
    wallet balance does not fall below the amount specified in this field.
    The PIPs can specify the total locked amount on a given wallet 
    in the `minimum available balance' field on all payment API calls 
    to ensure that the locked funds are always available in 
    the consumer's digital pound wallet. 
    \item For the design option where the FMI records locks 
    (design options U3.S2.D5), the CBDC system operated by the 
    Bank of England would provide technical accounts, 
    which do not impose holding limits, to FMIs.
\end{itemizelessVspace}
\noindent
We present the following five design options for this specific capability.

\myparagraph{Design option `Locking and confirming via the CBDC system' (U3.S2.D1)} 
    The CBDC system operated by the Bank of England records a funds lock 
    and routes all communications between PIPs. 
    The consumer's PIP sends a lock funds 
    instruction to the CBDC system. The CBDC system locks the 
    funds and sends the lock confirmation to the acquirer's 
    partner PIP, which in turn sends it to the acquirer.

    In an alternate flow, the lock could be maintained at the CBDC 
    system and the lock's beneficiary details could be 
    maintained with an external trusted entity (such as consumer's 
    PIP). However, in this approach, there is risk of updates 
    to the lock's beneficiary details independent of the lock, 
    which could result in diverting locked funds to a different 
    beneficiary.

\myparagraph{Design option `Locking at payer's PIP and confirming via the CBDC system' 
    (U3.S2.D2)} 
    The consumer's PIP records a funds lock in its system, while the CBDC 
    system operated by the Bank of England routes all communications 
    between PIPs. 
    The consumer's PIP records the funds lock in its 
    system and sends a funds lock confirmation to the CBDC 
    system which, in turn routes it to the acquirer's partner 
    PIP, which in turn sends it to the acquirer.

\myparagraph{Design option `Locking at payer's PIP and confirming using a direct 
    peer-to-peer network' (U3.S2.D3)} 
    The consumer's PIP records a funds lock in its system, 
    while the confirmation is communicated directly peer-to-peer and 
    is facilitated by a Technical Service Provider (TSP). 
    PIPs and other ecosystem participants expose 
    standard, secure and reliable APIs over the internet. 
    The discovery and registration of 
    ecosystem participants (such as PIPs and acquirers) and their 
    APIs are automated, with registrations processed and API 
    responses provided in near real-time, similar to the Open Banking model. 
    Dynamic client registration APIs provided by a TSP allow 
    registrations to be performed automatically using certificates 
    and SSAs\footref{ssa-def} issued by the TSP or another trusted 
    entity to authenticate other ecosystem participants.
    The consumer's PIP records 
    the funds lock in its systems and sends a funds lock 
    confirmation directly to the acquirer, using the APIs 
    exposed by the acquirer.

\myparagraph{Design option `Locking at payer's PIP and confirming
via a \changes{common third-party network}' (U3.S2.D4)}
    The consumer's PIP records a funds lock in its system, while the 
    communication of confirmation is routed to the merchant via 
    a \changes{common} network operated by a third party (such as a TSP/FMI). 
    All PIPs and ecosystem participants onboard and
    connect to the \changes{common} network operated by the third party. 
    This network may operate in a hub-and-spoke model, 
    point-to-point or hybrid topology.
    The consumer's PIP records the funds lock in its systems and sends 
    a funds lock confirmation to the acquirer via the \changes{common} 
    network.

\myparagraph{Design option `Locking and confirming using an FMI' (U3.S2.D5)} 
    An FMI escrows funds, records a funds lock and 
    routes the lock confirmation to the merchant. 
    All PIPs and ecosystem 
    participants onboard onto the FMI. 
    After obtaining authorisation from the consumer on the request 
    to lock, the consumer's PIP sends a lock funds instruction 
    to the FMI. 
    The FMI instructs the core ledger to transfer the required 
    funds from 
    the consumer's digital pound wallet to the FMI's 
    technical digital pound wallet.
    On successful transfer, the FMI sends the lock confirmation 
    to the merchant's acquirer.

\noindent
\newline
Appendix \ref{app:uc3-seq-dig-lock-conf-topology} presents the sequence diagrams 
for each design option above.


\subsubsection{Design options for `Releasing locked funds from 
digital pound wallets and initiating funds transfers' (U3.S3)}
\label{subsubsec:lock-pay-design-rel-lock}
This use case requires the specific capability to
release locked funds and initiate funds transfer to the merchant's 
commercial bank account at the time of physical delivery. 
The consumer instructs his/her PIP to release and transfer the 
locked funds based on a request from the delivery agent 
(e.g. by scanning a QR code presented by the delivery agent). 
If the consumer's PIP controls the funds lock, it releases the 
locked funds and initiates a funds transfer to the 
merchant's commercial bank account. 
If another entity, such as the CBDC system or an FMI 
(as described in Section \ref{subsubsec:lock-pay-design-conf-lock}), controls 
the funds lock then the consumer's PIP requests that entity to 
release the locked funds and initiate a funds transfer to the 
merchant's commercial bank account.  
Note the funds transfer is cleared and settled using one of 
the design options covered in section 
\ref{subsubsec:merch-r2p-design-Settlement}.

Design assumptions specific to these design options are as follows:
\begin{itemizelessVspace}
    \item For the design option where the PIP releases locks 
    (design options U3.S3.D2), 
    the payment APIs provided by the CBDC system include 
    a `minimum available balance' field and ensure that 
    after the payment is completed, the consumer's digital pound 
    wallet balance does not fall below the amount specified in this field.
    The PIPs can specify the total locked amount on a given wallet 
    in the `minimum available balance' field on all payment API calls 
    to ensure that the locked funds are always available in 
    the consumer's digital pound wallet. 
    \item For the design option where the FMI releases locks 
    (design options U3.S3.D3), the CBDC system operated 
    by the Bank of England would 
    provide technical accounts, which do not impose holding 
    limits, to FMIs.
\end{itemizelessVspace}

\noindent 
We present the following three design options for this specific capability. 
\myparagraph{Design option `Release locked funds and initiate funds 
transfer using the CBDC system' (U3.S3.D1)} 
    The CBDC system operated by the Bank 
    of England records the funds lock and provides an API to release 
    the locked funds and initiate the payment to the merchant. 
    After the successful physical delivery of the product to 
    the consumer, the consumer's PIP sends a release and pay 
    instruction to the CBDC system. 
    The CBDC system releases the locked funds
    and initiates the payment to the merchant's commercial bank 
    account. 

\myparagraph{Design option `Release locked funds and initiate 
funds transfer using the payer's PIP' (U3.S3.D2)} 
    The consumer's PIP records the funds 
    lock. After the successful physical delivery of the product
    to the consumer, the consumer's PIP releases the locked funds, 
    sends an instruction (which includes the reduced `minimum available 
    balance') to the CBDC system to debit the consumer's 
    digital pound wallet, and 
    initiates the payment to the merchant's commercial bank account.

\myparagraph{Design option `Release locked funds and initiate 
funds transfer using an FMI' (U3.S3.D3)} 
    An FMI escrows funds in a technical digital pound wallet and 
    provides an API to release 
    the locked funds and initiate the payment to the merchant. 
    All PIPs and ecosystem participants onboard onto the FMI. 
    After the successful physical delivery of the product to the 
    consumer, the consumer's PIP sends a release funds and pay 
    instruction to the FMI. The FMI uses the escrowed funds to 
    initiate the payment to the 
    merchant's commercial bank account.

\noindent 
\newline   
Appendix \ref{app:uc3-seq-dig-rel-lock-topology} presents the sequence diagrams 
for each design option above.


\subsection{Evaluation of design options}
\label{subsec:lock-pay-eval-designOpts}
In this section, we present our preliminary evaluation of each 
design option for the three specific capabilities of this use case.
Our preliminary evaluation comprises pros, cons and a suitability rating 
for each design option.

\subsubsection{Preliminary evaluation of design options for 
`Request to lock funds in digital pound wallets' (U3.S1)}
\label{subsubsec:lock-pay-eval-designOpts-req2lock}
Table \ref{table:table-lock-pay-eval-designOpts-r2l} presents
a preliminary evaluation of the design options to support request 
to lock funds in digital pound wallets. 

\vspace{-5mm}

\begin{center}
    \begingroup
    \renewcommand{\arraystretch}{1.75}     
    \captionsetup{width=15cm}
      
      \begin{longtable}{ >{\raggedright}p{0.14\textwidth} >{\raggedright}p{0.11\textwidth} >{\raggedright}p{0.32\textwidth} >{\raggedright\arraybackslash}p{0.34\textwidth} }
       
       \toprule
       \textbf{Design option} & \textbf{Suitability rating} & \textbf{Pros} & \textbf{Cons} \\  [0.5ex] 
       \midrule
       \endfirsthead
  
       \multicolumn{4}{r}{\emph{Continued from previous page}} \\
       \toprule
       \textbf{Design option} & \textbf{Suitability rating} & \textbf{Pros} & \textbf{Cons} \\  [0.5ex] 
       \midrule
       \endhead
  
       \bottomrule
       \multicolumn{4}{r}{\emph{Continued on next page}} \\
       \endfoot
  
       \bottomrule
       \caption{Preliminary evaluation of design options to support request 
       to lock funds in digital pound wallets (U3.S1).}
       \label{table:table-lock-pay-eval-designOpts-r2l}
       \endlastfoot
  
        Intermediated by the CBDC system (U3.S1.D1)& 

        Partially suitable & 
        
        \begin{compactimize}
            \item Simple design for PIPs because they would rely on 
            the CBDC system and not require multiple 
            point-to-point integrations.
    
            \item Participant authentication managed by the CBDC system.
        \end{compactimize}  &

        \begin{compactimize}
            \item Introduces additional complexity at the
            CBDC system 
            to ensure that personal data 
            in the request to lock 
            is not available to the Bank of England.
            \item Increases traffic and dependency on CBDC system 
            for ancillary operations (e.g. request to lock).
            \item Non-PIP ecosystem participants 
            would depend on PIPs
            to initiate requests to lock.
        \end{compactimize} \\

        Direct peer-to-peer (U3.S1.D2) & 
  
        Suitable & 

        \begin{compactimize}
            \item Users' personal data is not available to the Bank of England.
            \item Reduces traffic on the CBDC system for
            ancillary operations (e.g. request to lock).
            \item Allows ecosystem participants (e.g. merchant's acquirer) 
            to directly communicate with PIPs.
            \item Similar to the direct peer-to-peer communication 
            model used in Open Banking.
        \end{compactimize} &
  
        \begin{compactimize}
          \item All APIs exposed by PIPs may not be consistently secure, 
          reliable and as per standards.
          \item Stringent security measures
          are required to mitigate cyber risks caused by
          use of the public internet.
          \item Overhead and complexity of operating a TSP
          service.
        \item Ecosystem participants would need to maintain multiple
          peer-to-peer integrations with every PIP.
        \end{compactimize} \\   

        Using a \changes{common third-party network} (U3.S1.D3) & 
  
        Suitable & 

        \begin{compactimize}
            \item Secure, reliable, and consistent communication channel.
            \item Participant authentication is managed by the network.
            \item Third-party network operator could be an FMI that also provides 
            funds locking and release, which would simplify 
            the end-to-end payment processing.
            \item Provides a layer of security to PIPs' systems against cyber 
            threats.
            \changes{\item Reduces traffic on the CBDC system for
            ancillary operations (e.g. request to lock).}
        \end{compactimize} &
  
        \begin{compactimize}
          \item Higher cost of establishing and operating a \changes{common}
          third-party network.
          \item Options for customisation could be limited by 
          the third party.
          \item Could be a central point of failure if a hub and spoke 
          model is implemented.
          \item Users' personal data would be available to the TSP, 
          creating a risk of exposure to other third parties.
        \end{compactimize} \\ 
  
      \end{longtable}
  
    \endgroup
        
  \end{center}

\subsubsection{Preliminary evaluation of design options for `Locking funds in 
digital pound wallets and confirming lock status to the merchant' (U3.S2)}
\label{subsubsec:lock-pay-eval-designOpts-conf-lock}

Table \ref{table:table-lock-pay-eval-designOpts-conf-lock} presents our
preliminary evaluation of the design options to \changes{support} funds locking 
in the consumer's digital pound wallet and 
confirmation of lock status to the merchant. 

\begin{center}
    \begingroup
    \renewcommand{\arraystretch}{1.75}     
    \captionsetup{width=15cm}
      
      \begin{longtable}{ >{\raggedright}p{0.14\textwidth} >{\raggedright}p{0.11\textwidth} >{\raggedright}p{0.32\textwidth} >{\raggedright\arraybackslash}p{0.34\textwidth} }
       
       \toprule
       \textbf{Design option} & \textbf{Suitability rating} & \textbf{Pros} & \textbf{Cons} \\  [0.5ex] 
       \midrule
       \endfirsthead
  
       \multicolumn{4}{r}{\emph{Continued from previous page}} \\
       \toprule
       \textbf{Design option} & \textbf{Suitability rating} & \textbf{Pros} & \textbf{Cons} \\  [0.5ex] 
       \midrule
       \endhead
  
       \bottomrule
       \multicolumn{4}{r}{\emph{Continued on next page}} \\
       \endfoot
  
       \bottomrule
       \caption{Preliminary evaluation of design options to \changes{support 
       funds locking in the consumer's digital pound wallet and confirmation 
       of lock status to the merchant} (U3.S2).}
       \label{table:table-lock-pay-eval-designOpts-conf-lock}
       \endlastfoot
  
        Locking and confirming via the CBDC
        system (U3.S2.D1) & 

        Partially suitable & 
        
        \begin{compactimize}
            \item Maintaining locks in the CBDC system is
            simpler as the core 
            ledger records balances and transactions for
            digital pound wallets.
            \item Maintaining locks at the CBDC system
            provides a 
            consistent service to PIPs. \vspace{-6mm}
        \end{compactimize}  &

        \begin{compactimize}
            \item May introduce additional complexity at the
            CBDC 
            system (e.g. PIP key management and retrieval) to
            ensure 
            that personal 
            data included in the request to lock is not
            available to 
            the Bank of England.
        \end{compactimize} \\ 

        Locking at payer's PIP and confirming via the
        CBDC system (U3.S2.D2) & 
  
        Partially suitable & 

        \begin{compactimize}
            \item Users' personal data is not available to
            the Bank of England.
            \item Standardised communication channel to
            confirm funds lock.
        \end{compactimize} &
  
        \begin{compactimize}
          \item Maintaining the lock at payer's PIPs increases complexity and creates 
          a risk of inconsistent operations across PIPs.
          \item All transactions would need to be validated by the 
          payer's PIP.
          \item Core ledger pay/lock APIs would be more complex
          to  
          support `minimum available balance'.
          \item Non-PIPs would be dependent on their partner PIP to 
          receive lock confirmation.
          \item \changes{May introduce additional complexity at the
          CBDC system (e.g. PIP key management and retrieval)
          to ensure that personal 
          data included in the lock confirmation is not
          available to the Bank of England.}  
        \end{compactimize} \\   

        Locking at payer's PIP and confirming using a direct
        peer-to-peer network (U3.S2.D3) &
  
        Partially suitable & 

        \begin{compactimize}
            \item Users' personal data is not available to the Bank of England.
            \item Similar to the direct peer-to-peer communication 
            model used in Open Banking.
            \item Non-PIP ecosystem participants can receive lock 
            confirmations directly without depending on a partner PIP.
        \end{compactimize} &
  
        \begin{compactimize}
          \item Maintaining the lock at payer's PIPs increases complexity and creates 
          a risk of inconsistent operations across PIPs.
          \item All transactions would need to be validated by the 
          payer's PIP.
          \item Core ledger pay/lock APIs would be more complex
          to  
          support `minimum available balance'.
          \item All APIs exposed by PIPs may not be consistently secure, 
          reliable and as per standards.
          \item Stringent security measures are required
          to mitigate cyber risks.
          \item Overhead and complexity of operating a TSP service.
        \end{compactimize} \\ 

        Locking at payer's PIP and confirming via a 
        \changes{common third-party network} (U3.S2.D4) &
  
        Partially suitable & 

        \begin{compactimize}
            \item Users' personal data is not available to the Bank of England.
            \item Secure, reliable, and consistent communication channel.
            \item Participant authentication managed by the network.
            \item Provides a layer of security to PIPs' systems 
            against cyber threats.
            \item Non-PIP ecosystem participants can receive lock 
            confirmations without depending on a partner PIP.
        \end{compactimize} &
  
        \begin{compactimize}
          \item Maintaining the lock at payer's PIPs increases complexity and creates 
          a risk of inconsistent operations across PIPs.
          \item All transactions would need to be validated by the 
          payer's PIP.
          \item Core ledger pay/lock APIs would be more complex
          to  
          support `minimum available balance'.
          \item Higher cost of establishing and operating a \changes{common}
          third-party network.
          \item Options for customisation could be limited by 
          the third party.
          \item Could be a central point of failure if a hub-and-spoke 
          model is implemented.
          \item User's personal data would be available to the TSP, 
          creating a risk of exposure to other third parties.
        \end{compactimize} \\
        Locking and confirming using an FMI (U3.S2.D5) &
  
        Suitable & 

        \begin{compactimize}
            \item Users' personal data is not available to the Bank of England.
            \item Simple design for PIPs and other ecosystem participants as they 
            can rely on the FMI to record locks and confirm lock status.
            \item Secure, reliable, and consistent communication channel.
            \item Non-PIP ecosystem participants can receive lock 
            confirmations directly from the FMI without depending on a
            partner PIP.            
            \item The FMI could also support locking of commercial bank money.
        \end{compactimize} &
  
        \begin{compactimize}
          \item Cost and complexity of establishing and operating 
          the FMI service.
        \end{compactimize} \\
      \end{longtable}
    \endgroup
  \end{center}

\vspace{-6mm}

\subsubsection{Preliminary evaluation of design options for `Releasing 
locked funds from digital pound wallets
and initiating funds transfers' 
(U3.S3)}
\label{subsubsec:lock-pay-eval-design-rel-lock}

Table 7 presents
our preliminary evaluation of the design options to release locked funds 
and initiate funds transfer at physical delivery.


\begin{center}
    \begingroup
    \renewcommand{\arraystretch}{1.75}     
    \captionsetup{width=15cm}
        
      \begin{longtable}{ >{\raggedright}p{0.14\textwidth} >{\raggedright}p{0.11\textwidth} >{\raggedright}p{0.32\textwidth} >{\raggedright\arraybackslash}p{0.34\textwidth} }
       
       \toprule
       \textbf{Design option} & \textbf{Suitability rating} & \textbf{Pros} & \textbf{Cons} \\  [0.5ex] 
       \midrule
       \endfirsthead
    
       \multicolumn{4}{r}{\emph{Continued from previous page}} \\
       \toprule
       \textbf{Design option} & \textbf{Suitability rating} & \textbf{Pros} & \textbf{Cons} \\  [0.5ex] 
       \midrule
       \endhead
    
       \bottomrule
       \multicolumn{4}{r}{\emph{Continued on next page}} \\
       \endfoot
    
       \bottomrule
       \caption{Preliminary evaluation of design options to 
       release locked funds from digital pound wallets and 
       initiate funds transfers (U3.S3).}
       \label{table:table-release-lock-eval-designOpts}
       \endlastfoot
    
        Release locked funds and initiate funds transfer
        using the CBDC system (U3.S3.D1) & 

        Unsuitable & 
            
        \begin{compactimize}
            \item Maintaining locks in the CBDC system is simpler as the core 
            ledger records balances and transactions for digital pound wallets.
            \item Provides a consistent shared service to PIPs to 
            release locked funds and initiate funds transfer
            to the merchant's commercial bank account. \vspace{-6mm}
        \end{compactimize}  &

        \begin{compactimize}
            \item Personal data included in the funds lock
            \changes{could} be available to the Bank of England
            at the time of initiating the funds transfer.
        \end{compactimize} \\ 

        Release locked funds and initiate funds transfer
        using the payer's PIP (U3.S3.D2) &

        Partially suitable & 

        \begin{compactimize}
            \item Users' personal data is not available to
            the Bank of England.
        \end{compactimize} &

        \begin{compactimize}
            \item Releasing the lock at the payer's PIP increases 
            complexity and creates  
            a risk of inconsistent operations across PIPs.
            \item All transactions would need to be validated by the 
            payer's PIP.
            \item Core ledger pay/lock APIs would be more complex
            to support `minimum available balance'.  
        \end{compactimize} \\   

        Release locked funds and initiate funds transfer
        using an FMI (U3.S3.D3) & 

        Suitable & 

        \begin{compactimize}
            \item Users' personal data is not available to the Bank of England.
            \item Simple design for PIPs and other ecosystem participants as they 
            can rely on the FMI to release the locked funds and initiate the 
            payment to the merchant's commercial bank account.
        \end{compactimize} &
  
        \begin{compactimize}
            \item Cost and complexity of establishing and operating 
            the FMI service. 
        \end{compactimize} \\   

      \end{longtable}
    
    \endgroup

  \end{center}




\subsection{Initial insights}
\label{subsec:lock-pay-key-thoughts}
Our initial insights from the preliminary evaluation include:
\begin{itemizelessVspace}
    \item Design options U3.S1.D2 (Direct peer-to-peer) 
    and U3.S1.D3 (Using a \changes{common third-party network}) could 
    be more suitable for providing request to lock and other 
    such ancillary communications (which include 
    personal data) between PIPs and other ecosystem participants. 
    Further analysis could explore these design options 
    for providing other ancillary communications 
    required for other use cases.

    \item Maintaining locks at the PIP would be complex and difficult to 
    implement consistently. 
    It would also require all payments from, and funds locks on, 
    user digital pound wallets to be initiated/authorised by the PIPs 
    with support provided in the CBDC system APIs to enforce  
    a `minimum available balance'. 
    Maintaining locks at the CBDC system operated 
    by the Bank of England may require a 
    complex solution for protecting users' personal data. 
    Maintaining locks at the FMI could provide a consistent funds locking 
    capability across both the digital pound and commercial bank money.

    \item An FMI operating the \changes{common third-party network} (design option 
    U3.S1.D3) and providing funds locking, release and transfer 
    capabilities (design options U3.S2.D5 and U3.S3.D3) would enable 
    ecosystem participants to develop innovative services that require 
    funds transfers across different forms of money to be  
    synchronised with real-world events. 
\end{itemizelessVspace}


\endgroup



\section{Summary and conclusions}
\label{sec:summconclusions}

In this paper, we continued to analyse the design options for
supporting functional consistency by focusing on three
key capabilities:
(i) communication between PIPs and other ecosystem participants
(ii) funds locking, and 
(iii) interoperability between the digital pound and commercial bank money. 
We explored these key capabilities via three payments
use cases: 
  (i) Person-to-person push payment with interoperability across 
  the digital pound and commercial bank money,
  (ii) Merchant initiated request to pay with interoperability 
  across the digital pound and commercial bank money, and
  (iii) Lock digital pounds and pay on physical delivery from 
  digital pounds to commercial bank money.

We presented design options that provide the specific capabilities
for each use case, performed a preliminary evaluation of these design
options, and drew initial insights.
Figure 1 below summarises our preliminary
evaluation of the design options.
In the figure, 
the specific capabilities for each use case are
grouped under key capabilities, and
the design options are categorised based on whether the
specific capability is provided by the central bank, 
by PIPs (optionally in partnership with other entities),
or by a third party (such as a TSP or an FMI).

We draw the following initial conclusions from our
preliminary evaluation:

\begin{itemizelessVspace}

  \item \changess{None of the design options presented are fully
  suitable for providing interoperability between digital
  pounds and commercial bank money.
  This is primarily because several design options depend
  on financial intermediaries (such as PIPs, commercial
  banks or an FMI) holding digital pounds either for themselves
  or temporarily on behalf of users in order to support
  interoperability.
  If financial intermediaries are restricted from doing so,
  a suitable option to provide interoperability could be
  an enhanced payment system (such as NPA) that integrates
  with and provides settlement between commercial bank money
  and digital pounds.
  However, there would be increased cost and complexity
  for this design option.}

  \item A regulated third-party such as a TSP or an FMI
  could be a suitable option for supporting most of the specific
  capabilities for \changes{the three} use cases.
  An FMI facilitating communication
  between PIPs and other ecosystem participants, and
  providing funds locking, release and clearing and
  settlement of fund transfers across the digital pound and commercial
  bank money 
  could simplify the experience of ecosystem participants
  including consumers and merchants,
  simplify operating platforms for PIPs and the central bank,
  and facilitate the creation of innovative services.

  \item PIPs could integrate directly with other ecosystem
  participants using a peer-to-peer model 
  facilitated by a direct client registration service
  to provide a suitable means of communication.
  Clearing and settling funds transfers from commercial bank
  money to digital pounds could be greatly simplified
  if all commercial banks could access the CBDC system without
  having to become PIPs
  (e.g. by extending the platform model to
  add a `PIP Lite' participant).
    
  \item Although the CBDC system operated by the Bank of England
  would provide foundational capabilities (such as payments
  between digital pound wallets), it could be unsuitable or
  partially suitable for supporting the specific
  capabilities for the use cases.
  This is primarily because providing these specific capabilities
  could result in users' personal data becoming available to the
  Bank of England.

\end{itemizelessVspace}


\begin{figure}[ht!]
  \label{fig:evaluation-summary}
  \begin{center}
      \includegraphics[width=0.85\paperwidth]{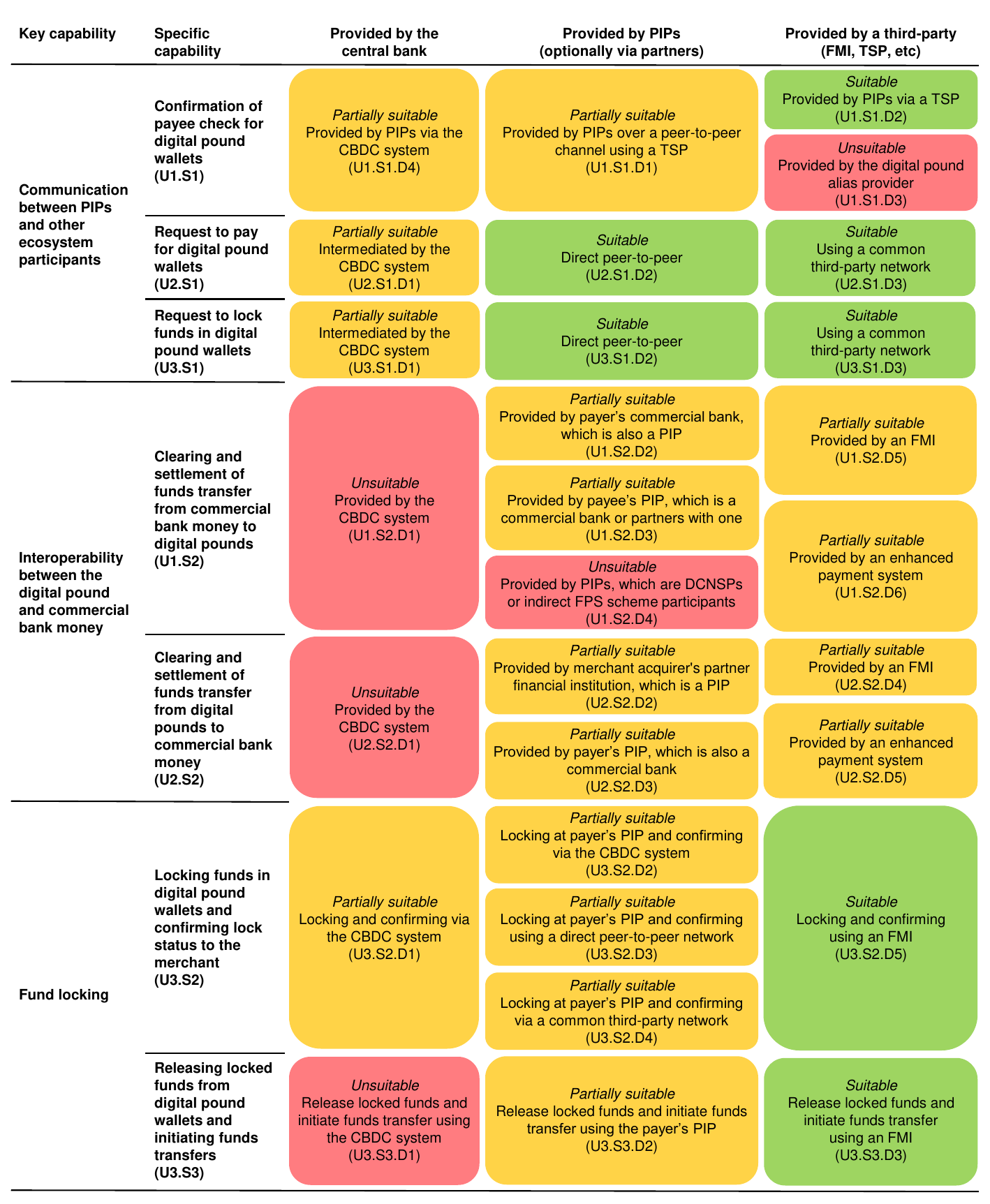}
  \end{center}
  \vspace{-4mm}
  \caption{\footnotesize{Summary of our preliminary evaluation
  of the suitability of the
  design options to support each specific capability.}}
\end{figure}

\noindent
Our previous paper \cite{barc-cbdc-func-cons} concluded that
no single design option could provide functional consistency
across digital pounds and commercial bank money and, instead, a
complete solution would need to combine the
suitable design option(s) for each key capability and include
common ecosystem services provided by an FMI and TSPs.
Our key contribution to the design space for the digital
pound from this subsequent paper is the insight that
an FMI could be a suitable option to:
(i) facilitate communication between PIPs and other
ecosystem participants,
(ii) provide funds locking and release across both the digital
pound and commercial bank money, and 
(iii) clear and settle funds transfers across the digital
pound and commercial bank money.
Such FMI services would 
simplify the experience of ecosystem participants including
consumers and merchants,
simplify the operating platforms for both PIPs and the central bank,
and facilitate the creation of innovative services.


\section{Further work}
\label{subsec:summary-and-further-work}

Ongoing industry design and experimentation is needed to
select design options that 
align with the Bank of England's platform model,
meet the desired goals for privacy and innovation, and
support functional consistency across digital
pounds and commercial bank money.

Next steps could include further technical, legal, commercial and
operational analyses and experimentation to explore:

\begin{itemizelessVspace}

  \item \changess{whether financial intermediaries could
  hold funds in digital pound wallets (either for themselves
  or temporarily on behalf of users) in order to provide
  interoperability between digital pounds and commercial
  bank money,}

  \item a common ecosystem services layer that supports
  functional consistency,
  including providing access to funds locking on both
  digital pounds and commercial bank money,
  

  \item potential enhancements to the RTGS service to
  facilitate funding and liquidity management across commercial bank money and
  digital pounds in order to support interoperability,

  \item extending the existing Confirmation of Payee service
  for commercial bank accounts to support digital pound wallets.

  \item building and operating a new or enhanced
  payment rail that natively operates across the digital pound
  core ledger and commercial bank ledgers to
  provide interoperability, and

  \item extending the platform model to include a
  `PIP lite' participant model for some
  commercial banks to ease interoperability,

\end{itemizelessVspace}

\noindent
We hope the use cases, design options and analysis presented in
this paper
will aid the design and experimentation for the digital pound,
and look forward to ongoing industry engagement.

\vspace{5mm}

\noindent \textbf{Acknowledgements:} 
We thank Paul Lucas (IBM), David MacKeith (Deloitte), 
Richard Brown (R3), Edwin Aoki (PayPal),
Tim Moncrieff (Visa), 
Gary King (Lloyds), 
Julia Demidova (FIS), Alain Martin (Thales), 
Adrian Field (OneID), Akshay Kant (Chainalysis), 
Bejoy Das Gupta (eCurrency) and Mitch Cohen (Consultant) 
for helpful input and feedback on this paper.

\pagebreak


\pretolerance=-1
\tolerance=-1
\emergencystretch=0pt

\bibliography{uk-cbdc-usecase-paper}
\bibliographystyle{plain}


\pagebreak

\appendix
\begin{landscape}

  

\section{Appendix - Component model for sequence diagrams}
\label{app:comp-model}

\subsection{Actors}
\label{app:actors}


  \begin{center}
    \begingroup
    \renewcommand{\arraystretch}{1.75}     
    \captionsetup{width=20cm}
      
      \begin{longtable}{ >{\raggedright}p{0.15\textwidth} >{\raggedright}p{0.2\textwidth} >{\arraybackslash}p{0.6\textwidth} }
       
       \toprule
       \textbf{Actor reference} & \textbf{Actor} & \textbf{Description} \\  [0.5ex] 
       \midrule
       \endfirsthead
  
       \multicolumn{3}{r}{\textit{Continued from previous page}} \\
       \toprule
       \textbf{Actor reference} & \textbf{Actor} & \textbf{Description} \\  [0.5ex] 
       \midrule
       \endhead
  
       \bottomrule
       \multicolumn{3}{r}{\textit{Continued on next page}} \\
       \endfoot
  
       \bottomrule
       \caption{Actors.}
       \label{table:table-append1-actors}
       \endlastfoot
  
  
        A001&
        Payer&
        Person initiating funds transfer.\\
       
        A002&	
        Payee&	
        A payee is a beneficiary of funds transfer.\\

        A006&
        Delivery Partner&
       Person or entity delivering ordered products to the consumer.
      \end{longtable}
  
    \endgroup
        
  \end{center}

\subsection{Components}
\label{app:components}


  \begin{center}
    \begingroup
    \renewcommand{\arraystretch}{1.75}     
    \captionsetup{width=20cm}
      
      \begin{longtable}{ >{\raggedright}p{0.15\textwidth} >{\raggedright}p{0.2\textwidth} >{\arraybackslash}p{0.6\textwidth} }
       
       \toprule
       \textbf{Component reference} & \textbf{Component} & \textbf{Description} \\  [0.5ex] 
       \midrule
       \endfirsthead
  
       \multicolumn{3}{r}{\textit{Continued from previous page}} \\
       \toprule
       \textbf{Component reference} & \textbf{Component} & \textbf{Description} \\  [0.5ex] 
       \midrule
       \endhead
  
       \bottomrule
       \multicolumn{3}{r}{\textit{Continued on next page}} \\
       \endfoot
  
       \bottomrule
       \caption{Components.}
       \label{table:table-append1-comp}
       \endlastfoot
  
  
        C001&
        Core Ledger&
        The CBDC core ledger would record issuance and transfer
        of digital pounds.\\
		
        C002&	
        Payee PIP System&	  
        Infrastructure and operations of a PIP that facilitates the payee to access and use   
        his/her digital pounds. \\

        C003&
        Payer PIP System&
        Infrastructure and operations of a PIP that facilitates the payer to access and use
        his/her digital pounds. \\

        C004&
        Identity Verifier&
        A service verifying user's identity. \\

        C005&
        Payer App&
        The smartphone application provided to the payer by his/her PIP to 
        facilitate the payer to access and use digital pounds. \\

        C006&
        Payee App&
        The smartphone application provided to the payee by his/her PIP to 
        facilitate the payee to access and use digital pounds. \\

        C007&
        Alias and PIP Lookup Service&
        A service operated by CBDC core ledger operator providing \linebreak digital pounds alias 
        validation and returning details, such as digital pound wallet core ledger identifier, 
        PIP name and PIP endpoints for a given digital pound wallet alias. \\
 
        C008&
        Central Router&
        Central routing service that routes calls from PIPs to 
        \linebreak appropriate counterparties. \\

        C010&
        Central Router and Choreographer&
        Centralised routing service provided by TSP/FMI that route calls 
        from PIPs to appropriate counterparties.  \\

        C011&
        Alias and PIP Lookup Service&
        A service operated by a third party (such as a TSP or an FMI) \linebreak 
        providing digital pounds alias 
        validation and returning details, such as digital pound core ledger identifier, 
        PIP name and PIP endpoints for a given digital pound wallet alias. \\

        C012&
        Dynamic Client Registration Service&
        A Technical Service Provider facilitates the communication \linebreak
        between PIPs by providing DCR (Dynamic Client \linebreak Registration) capabilities for supporting 
        automated \linebreak onboarding and authentication between PIPs. \\

        C013&
        Payee PIP's Partner Bank PIP System&
        PIP system used by the payee PIP's partner commercial bank (which is also a PIP)
        to perform operations on digital pound wallets, 
        such as transfer of funds between digital pound wallets (counterpart of EC012).\\

        C014&
        Payer's Bank PIP System&
        PIP System used by the payer's commercial bank (which is also a PIP) 
        to perform operations on digital pound wallets, 
        such as transfer of funds between digital pound wallets (counterpart of EC013). \\

        C015&
        Acquirer's Partner PIP&
        
        Acquirer may wish to partner with a PIP to access the \linebreak digital pound core 
        ledger APIs to initiate operations on the \linebreak digital pound wallets (such as 
        to initiate a request to pay to the payer's PIP). \\

        C016&
        Acquirer's Partner Bank PIP System&
        PIP system used by the acquirer's partner bank (i.e. acquirer may wish to 
        partner with a bank which is also a PIP to \linebreak transfer 
        and settle funds from a digital pound wallet to a \linebreak commercial bank account)
        to perform operations on digital pound wallets (counterpart of EC014). \\
      
        \changes{C017}&
        \changes{Payer Bank's Partner PIP} &
        \changes{Payer's commercial bank may wish to partner with a PIP to \linebreak 
        access the digital pound core 
        ledger APIs in order to initiate \linebreak operations on digital pound wallets 
        (such as confirmation of payee checks).} 
        \\
      \end{longtable}
  
    \endgroup
        
  \end{center}


  \pagebreak

  \subsection{External components}
  \label{app:extComp}


  \begin{center}
    \begingroup
    \renewcommand{\arraystretch}{1.75}     
    \captionsetup{width=20cm}
      
      \begin{longtable}{ >{\raggedright}p{0.15\textwidth} >{\raggedright}p{0.2\textwidth} >{\arraybackslash}p{0.6\textwidth} }
       
       \toprule
       \textbf{Component reference} & \textbf{Component} & \textbf{Description} \\  [0.5ex] 
       \midrule
       \endfirsthead
  
       \multicolumn{3}{r}{\textit{Continued from previous page}} \\
       \toprule
       \textbf{Component reference} & \textbf{Component} & \textbf{Description} \\  [0.5ex] 
       \midrule
       \endhead
  
       \bottomrule
       \multicolumn{3}{r}{\textit{Continued on next page}} \\
       \endfoot
  
       \bottomrule
       \caption{External components.}
       \label{table:table-append1-extcomp}
       \endlastfoot
  

        EC001&
        Faster Payments Scheme&
        The existing FPS scheme used to move commercial bank money between 
        commercial bank accounts.  \\
        
        EC002 &
        User Retail Bank System&
        The commercial bank used by a payer to transfer funds into/out of the CBDC 
        ecosystem. \\
        
        EC003&
        Payer PIP Partner Bank System&
        Payer PIP's partner commercial bank providing interoperability across 
        digital pounds and commercial bank money.\\

        EC004&
        FMI Clearing and Settlement System&
        Financial Market Infrastructure providing clearing and \linebreak settlement service
        across all forms of money, including digital pounds and commercial bank 
        money. It could potentially be a new market infrastructure.\\

        EC005&
        FMI Escrow System&
        Financial Market Infrastructure providing funds escrow \linebreak services across all 
        forms of money, including digital pounds and commercial bank money. 
        It could potentially be a new market infrastructure. \\

        EC006&
        Enhanced Payment System&
        The enhanced payment system providing interoperability across 
        commercial bank money and digital pounds. \\

        EC007&
        RTGS System&
        The RTGS system operated by the Bank of England. \\

        EC008&
        Merchant App&
        Ecommerce merchant's smartphone app used by the consumer to order products online. \\

        EC009&
        Merchant's acquirer&
        Financial institution that 
        provides a payment gateway, a \linebreak payment processor and merchant accounts, 
        thereby allowing end-to-end processing and settlement of  
        transactions on behalf of merchants. \\

        EC010&
        Merchant's Bank System&
        The backend system of a commercial bank holding the \linebreak merchant's account. \\

        EC011&
        Payee PIP's Partner Bank System&
        The backend system of payee PIP's partner commercial bank to 
        perform operations on commercial bank money, such as
        transfer of funds between commercial bank accounts. \\

        EC012&
        Payee PIP's Partner Bank (also a PIP) System&
        The backend system of payee PIP's partner commercial bank (which is also a PIP) for 
        performing operations on commercial bank money, such as
        transfer of funds between commercial bank accounts (counterpart of C013). \\

        EC013&
        Payer's Bank (also a PIP) System&
        The backend system of the payer's commercial bank (which is also a PIP)  
        to perform operations on commercial bank money, such as
        transfer of funds between commercial bank accounts (counterpart of C014). \\

        EC014&
        Acquirer's Partner Bank (also a PIP) System &
        The backend system of an acquirer's partner bank (which is also a PIP) 
        to perform operations on commercial bank money \linebreak (counterpart of C016).

    \end{longtable}
  
    \endgroup
        
  \end{center}



\section{Appendix - Sequence diagrams for use case `Person-to-person push payment with interoperability across 
the digital pound and commercial bank money' (U1)}
\label{app:uc1-sequence-diagrams}
\subsection{Design options for `Confirmation of payee check for digital pound wallets' (U1.S1)}
\label{app:uc1-sequence-diagram-a}
\begin{figure}[h!]
    \label{fig:OptionU1.S1.D1}
    \begin{center}
        \includegraphics[width=0.69\paperwidth]{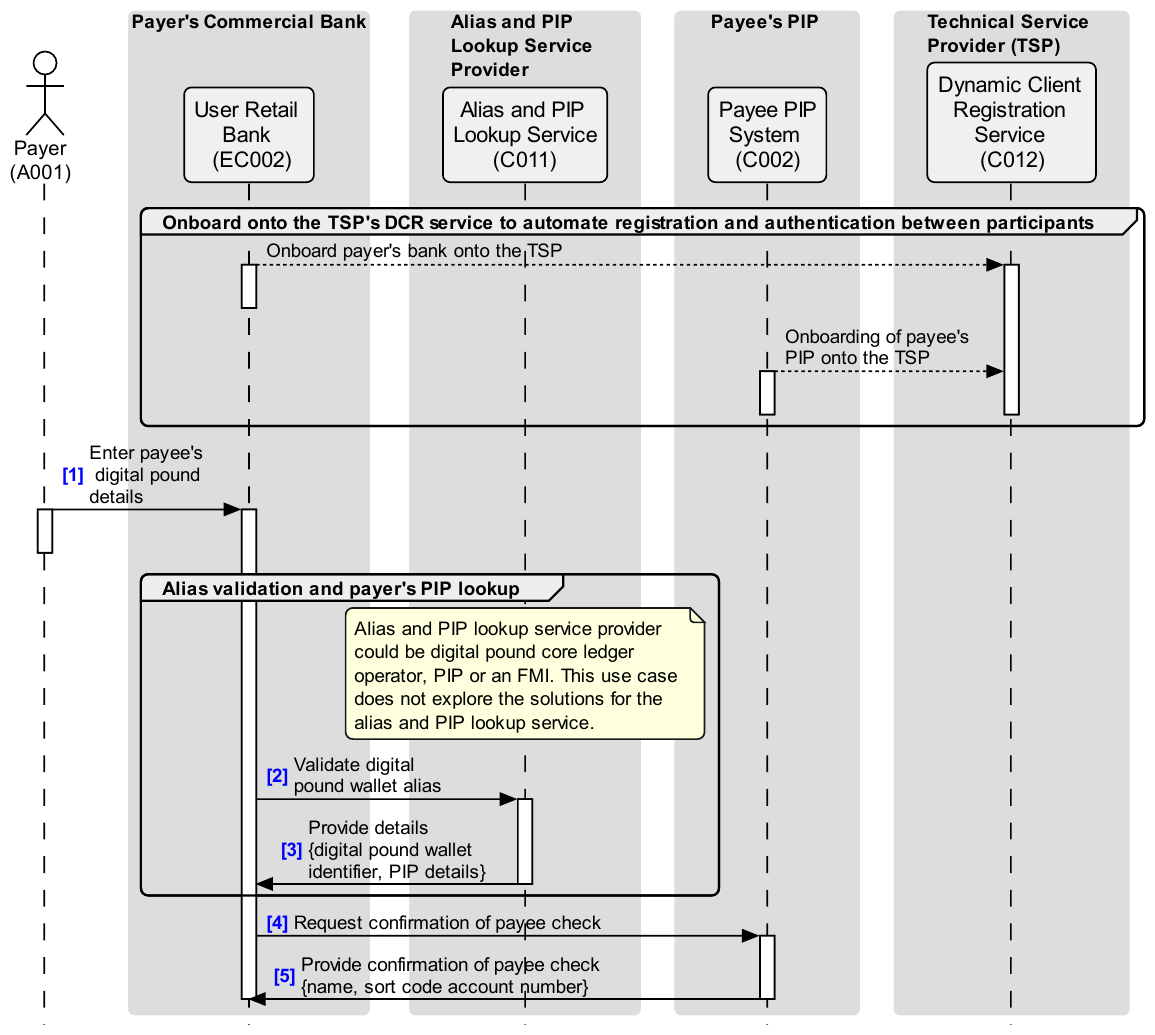}
    \end{center}
    \vspace{-4mm}
    \caption{\footnotesize{Sequence diagram for design option `Provided by PIPs over peer-to-peer channel using a TSP' (U1.S1.D1).}}
\end{figure}
\pagebreak

\begin{figure}[h!]
    \label{fig:OptionU1.S1.D2}
    \begin{center}
        \includegraphics[width=0.75\paperwidth]{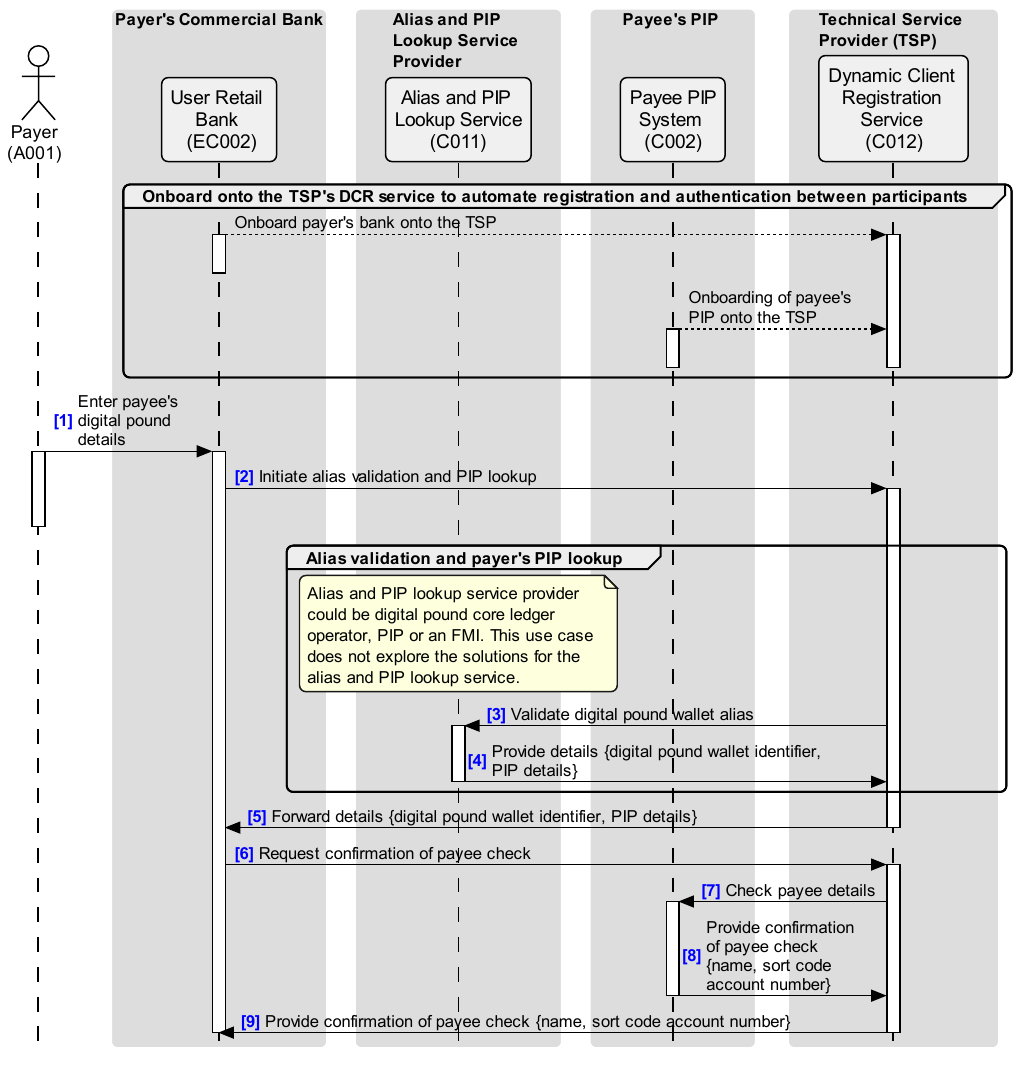}
    \end{center}
    \vspace{-4mm}
    \caption{\footnotesize{Sequence diagram for design option `Provided by PIPs via a TSP' (U1.S1.D2).}}
\end{figure}
\pagebreak

\begin{figure}[h!]
    \label{fig:OptionU1.S1.D3}
    \begin{center}
        \includegraphics[width=0.70\paperwidth]{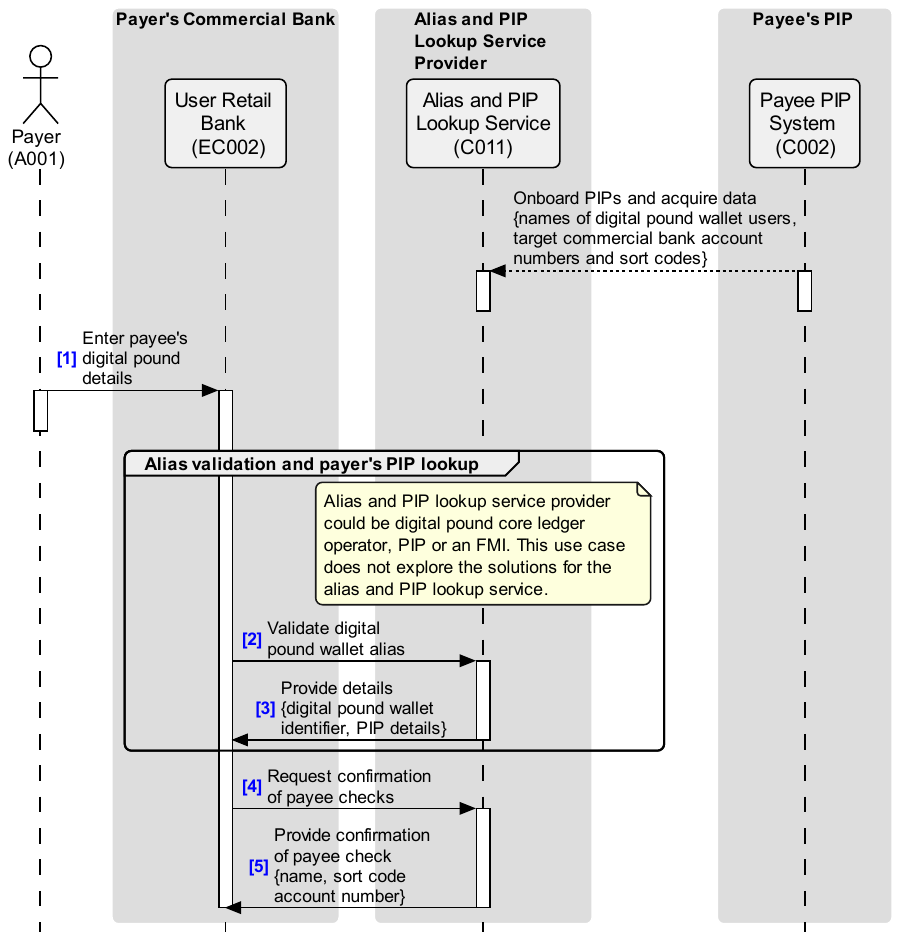}
    \end{center}
    \vspace{-4mm}
    \caption{\footnotesize{Sequence diagram for design option `Provided by the digital pound alias service provider' (U1.S1.D3).}}
\end{figure}
\pagebreak

\begin{figure}[h!]
    \label{fig:OptionU1.S1.D4}
    \begin{center}
        \includegraphics[width=1.00\paperwidth]{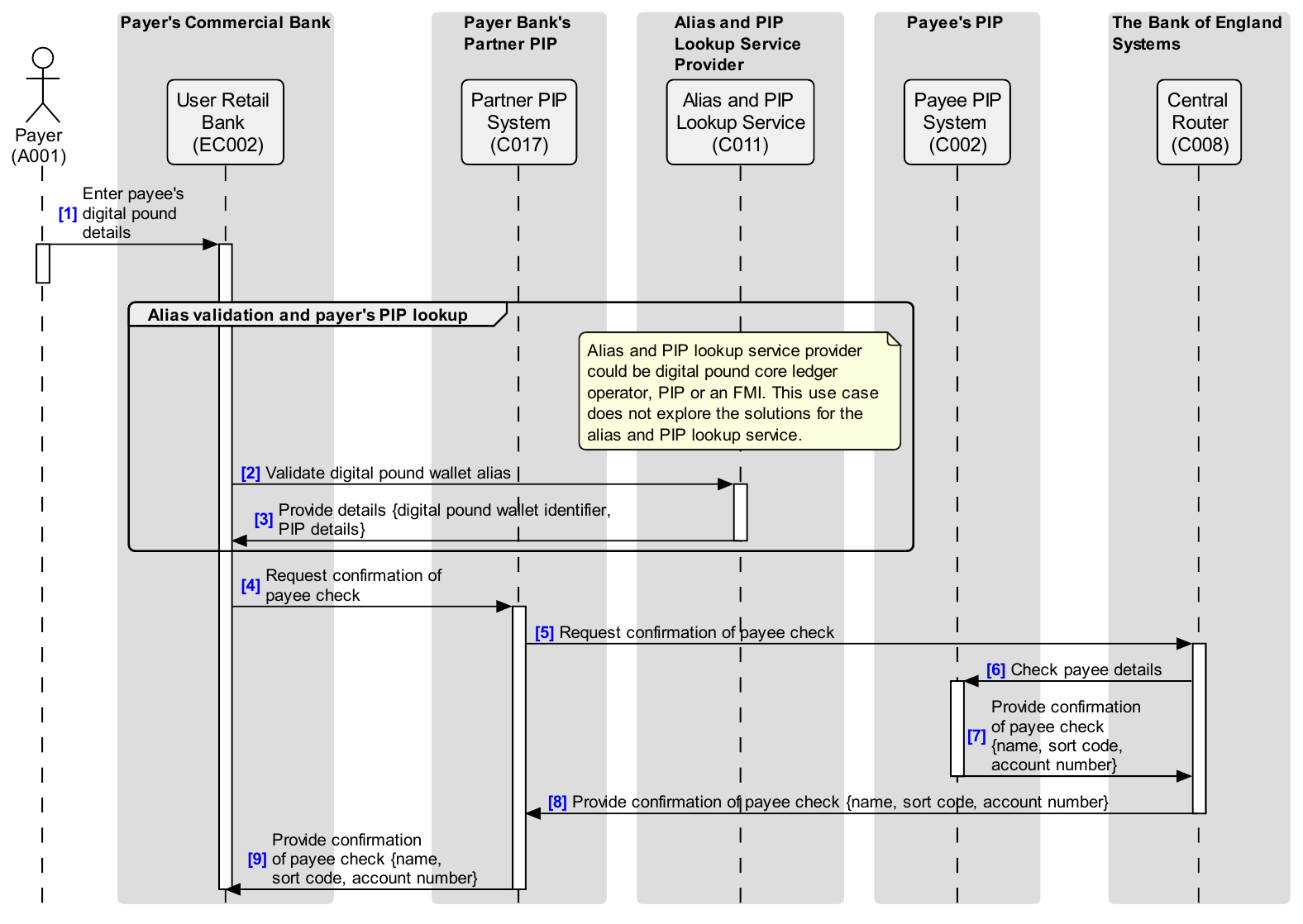}
    \end{center}
    \vspace{-4mm}
    \caption{\footnotesize{Sequence diagram for design option `Provided by PIPs via the CBDC system' (U1.S1.D4).}}
\end{figure}
\pagebreak

\subsection{Design options for `Clearing and settlement of funds transfer from 
commercial bank money to digital pounds' (U1.S2) }
\label{app:uc1-sequence-diagram-b}
\begin{figure}[h!]
    \label{fig:OptionU1.S2.D1}
    \begin{center}
        \includegraphics[width=1.1\paperwidth]{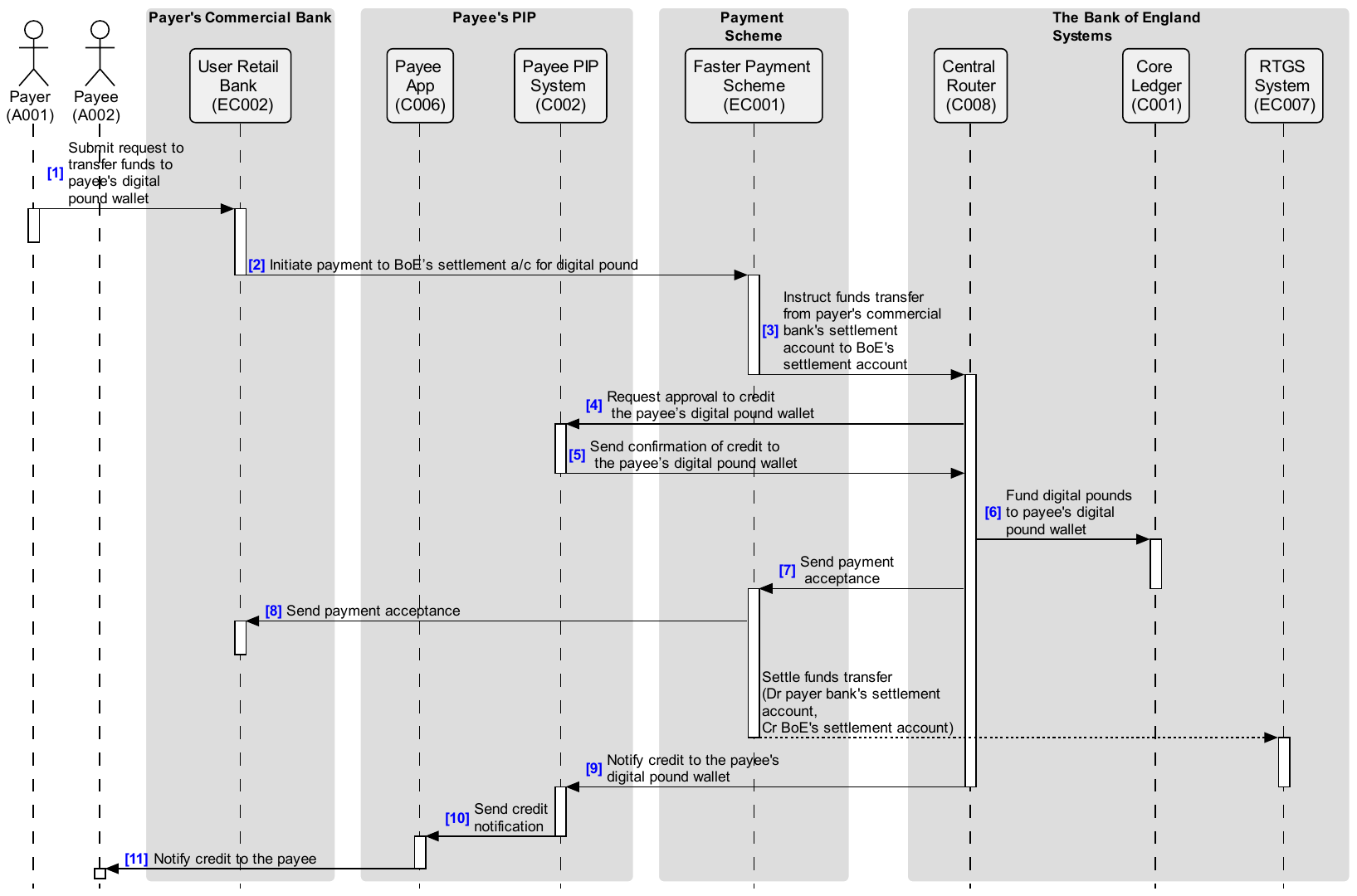}
    \end{center}
    \vspace{-4mm}
    \caption{\footnotesize{Sequence diagram for design option `Provided by the CBDC system' (U1.S2.D1).}}
\end{figure}
\pagebreak

\begin{figure}[h!]
    \label{fig:OptionU1.S2.D2}
    \begin{center}
        \includegraphics[width=0.99\paperwidth]{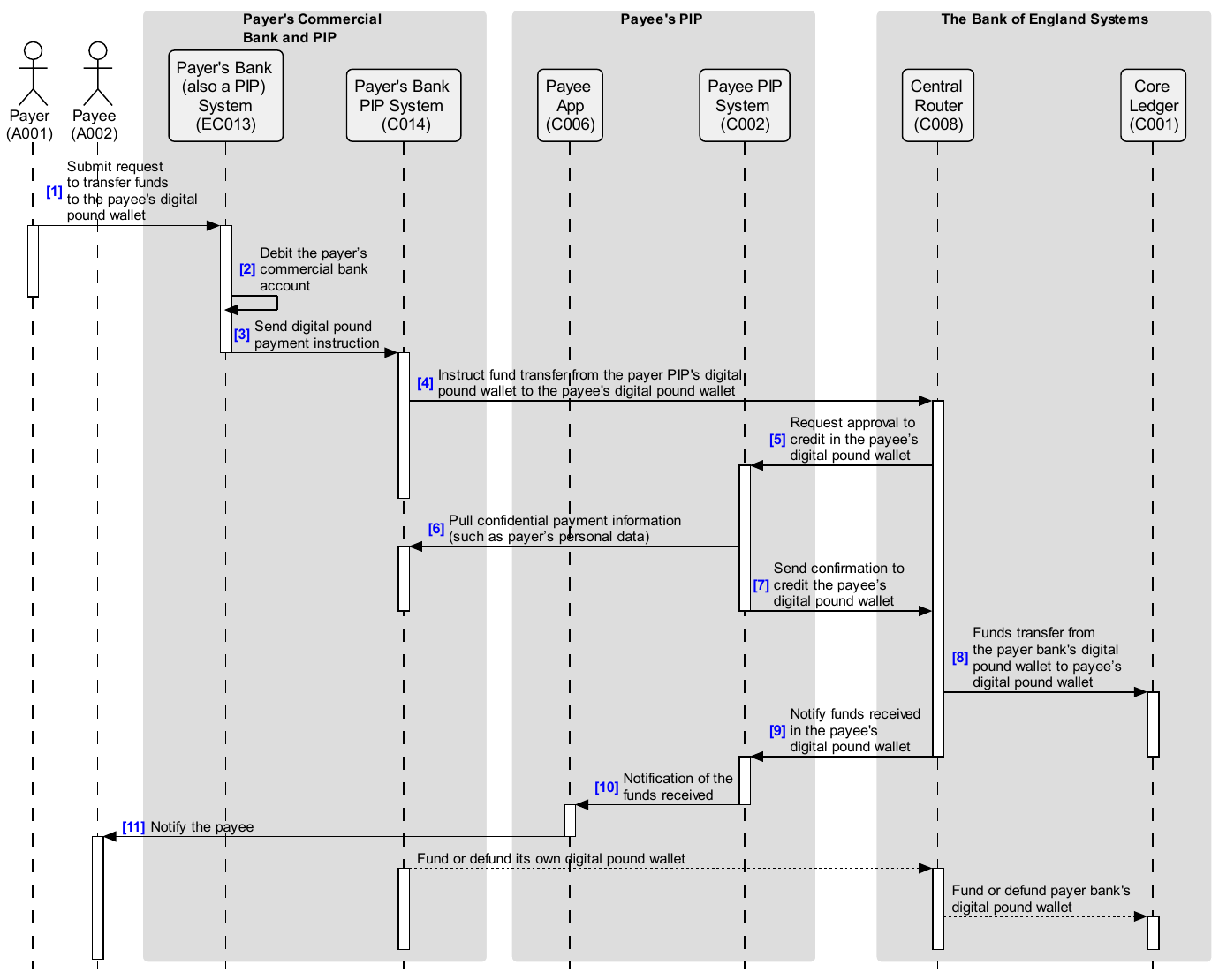}
    \end{center}
    \vspace{-4mm}
    \caption{\footnotesize{Sequence diagram for design option `Provided by payer's commercial bank which is 
    also a PIP' (U1.S2.D2).}}
\end{figure}
\pagebreak

\begin{figure}[h!]
    \label{fig:OptionU1.S2.D3}
    \begin{center}
        \includegraphics[width=0.99\paperwidth]{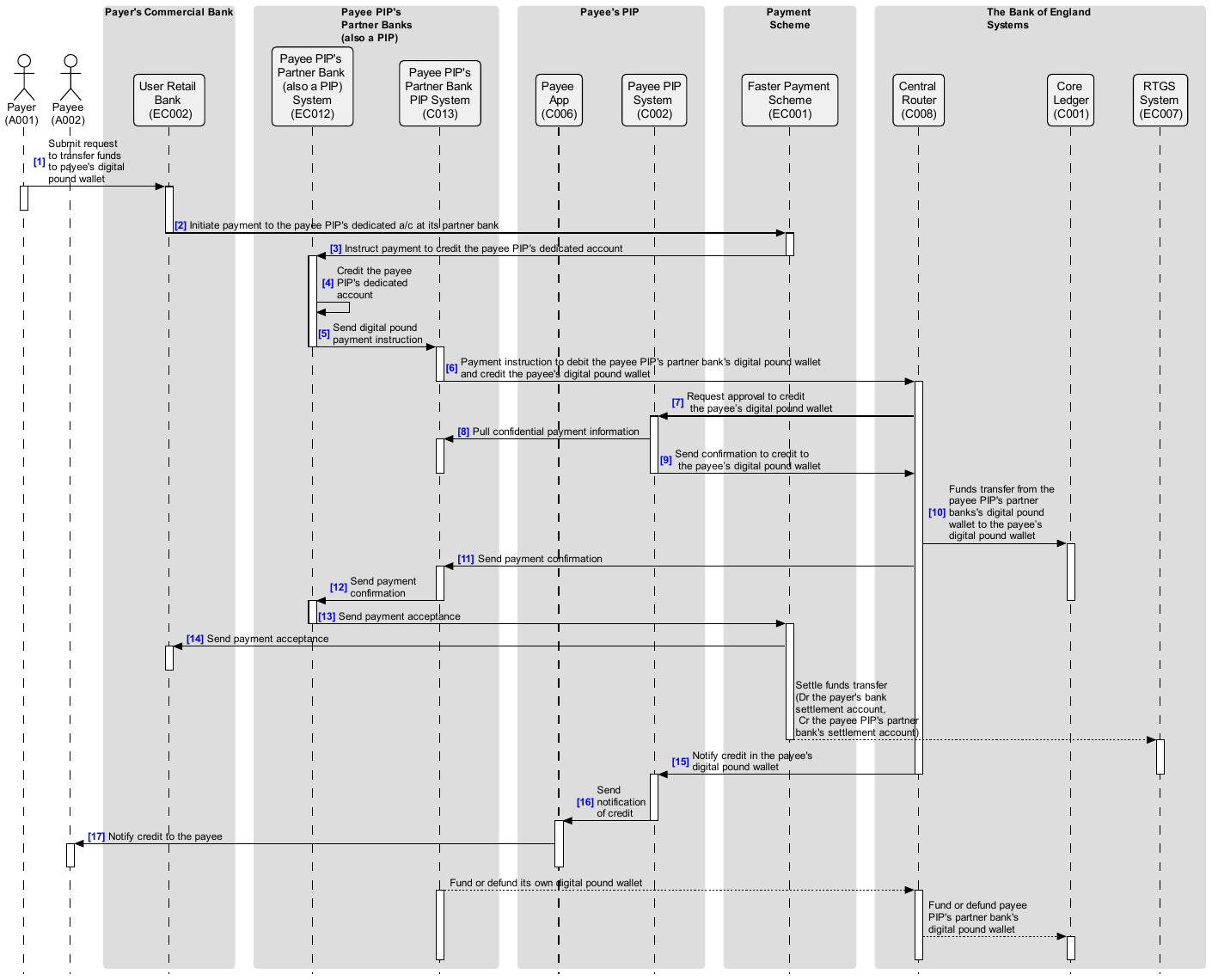}
    \end{center}
    \vspace{-4mm}
    \caption{\footnotesize{Sequence diagram for design option `Provided by payee's PIP, which is either a 
    commercial bank or partners with a commercial bank which is a PIP' (U1.S2.D3).}}
\end{figure}
\pagebreak

\begin{figure}[h!]
    \label{fig:OptionU1.S2.D4}
    \begin{center}
        \includegraphics[width=1.2\paperwidth]{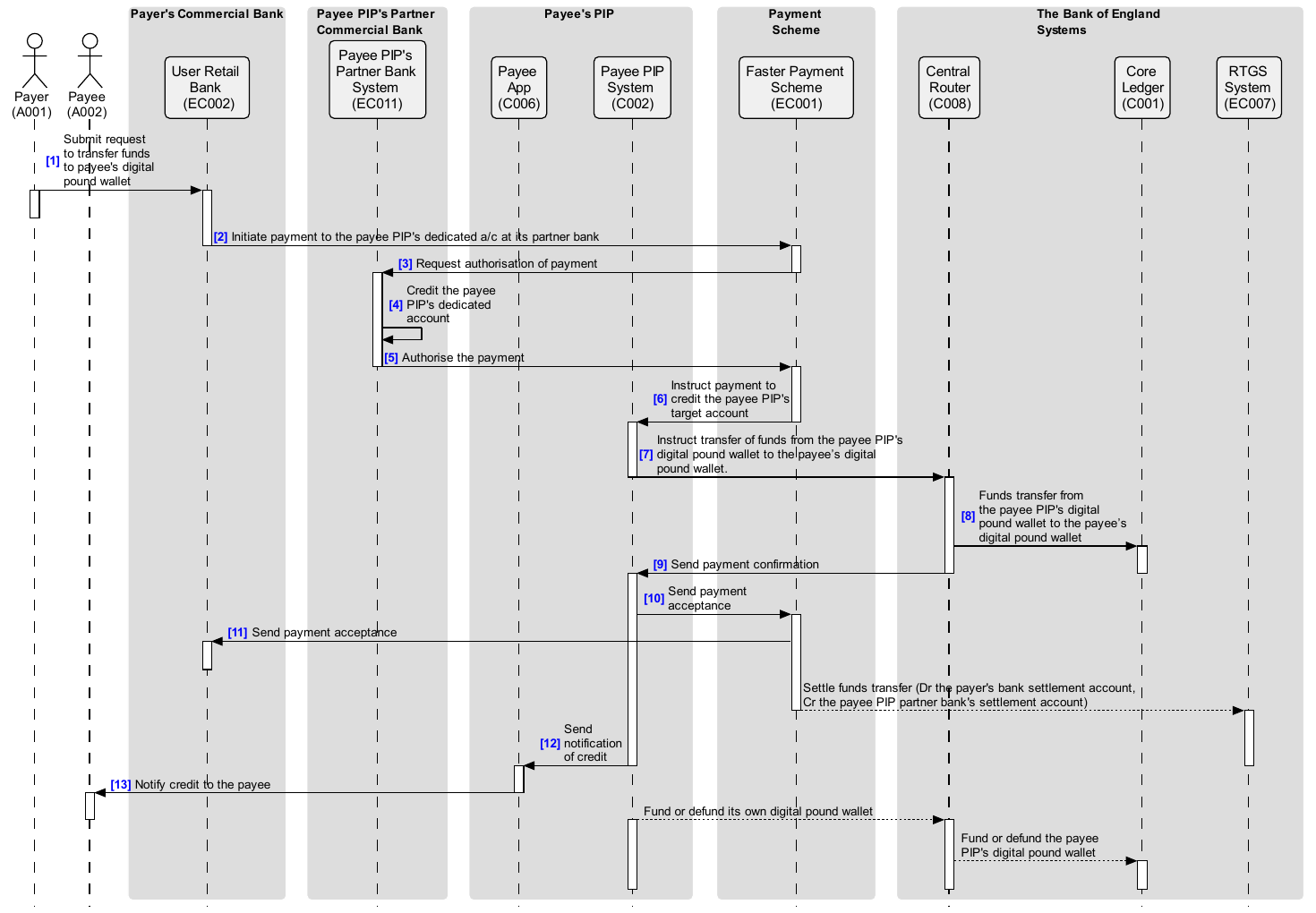}
    \end{center}
    \vspace{-4mm}
    \caption{\footnotesize{Sequence diagram for \changes{DCNSP model of} design option `Provided by PIPs which are either directly connected 
    non-settling participants (DCNSP) or indirect FPS scheme participants' (U1.S2.D4).}}
\end{figure}
\pagebreak

\begin{figure}[h!]
    \label{fig:OptionU1.S2.D5}
    \begin{center}
        \includegraphics[width=1.0\paperwidth]{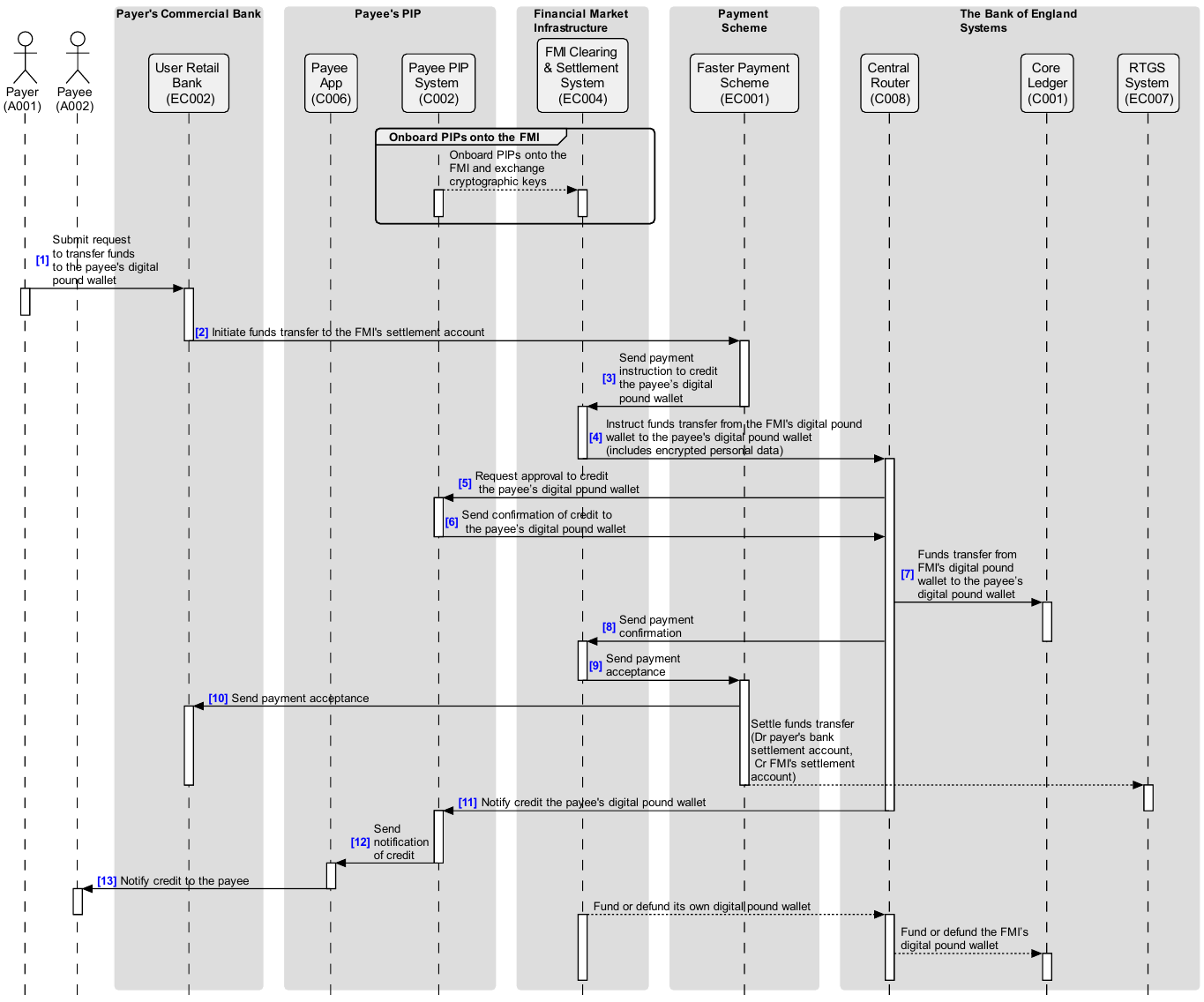}
    \end{center}
    \vspace{-4mm}
    \caption{\footnotesize{Sequence diagram for design option `Provided by an FMI' (U1.S2.D5).}}
\end{figure}
\pagebreak

\begin{figure}[h!]
    \label{fig:OptionU1.S2.D6}
    \begin{center}
        \includegraphics[width=1.1\paperwidth]{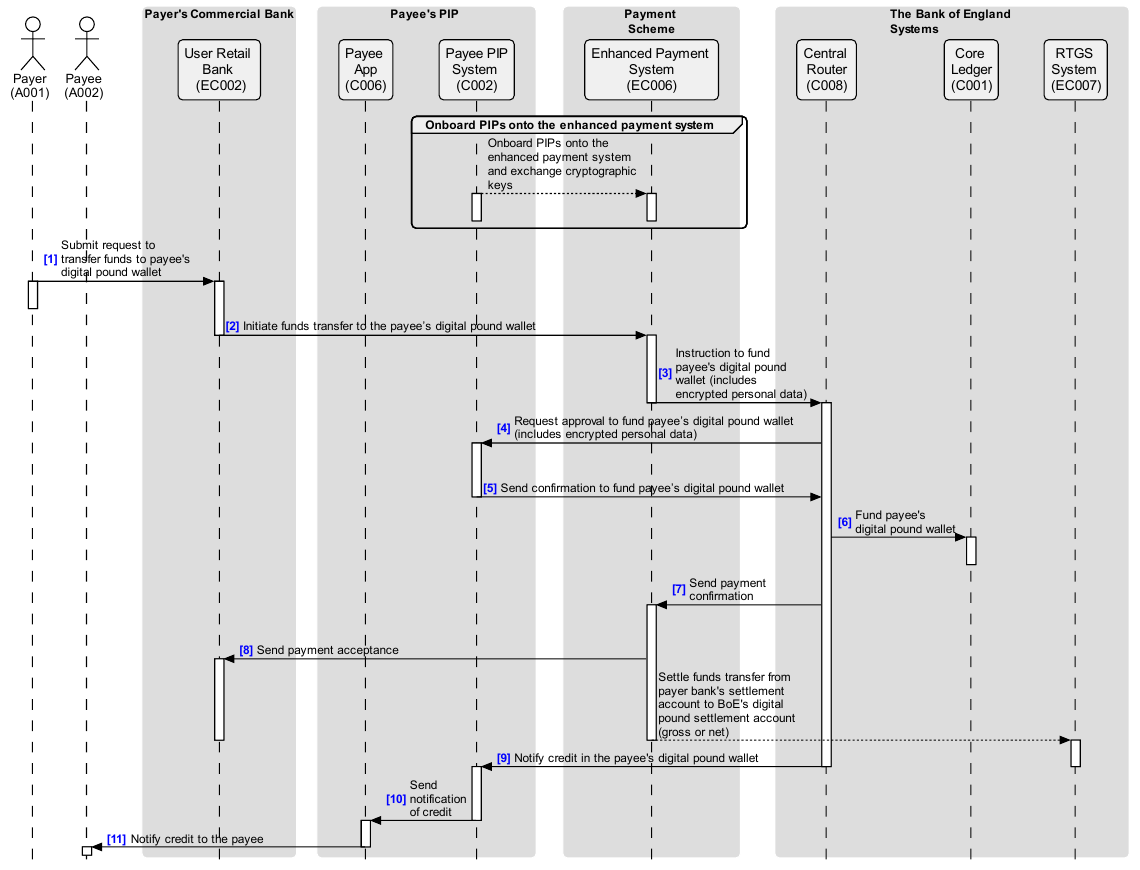}
    \end{center}
    \vspace{-4mm}
    \caption{\footnotesize{Sequence diagram for design option `Provided by an enhanced payment system' (U1.S2.D6).}}
\end{figure}
\pagebreak

\section{Appendix - Sequence diagrams for use case `Merchant initiated request to pay with interoperability
across the digital pound and commercial bank money' (U2)}
\label{app:uc2-sequence-diagrams}

\subsection{Design options for ‘Request to pay for digital pound wallets’ (U2.S1)}
\label{app:uc2-seq-dig-comm-topology}
\begin{figure}[h!]
    \label{fig:OptionU2S1D1}
    \begin{center}
        \includegraphics[width=0.88\paperwidth]{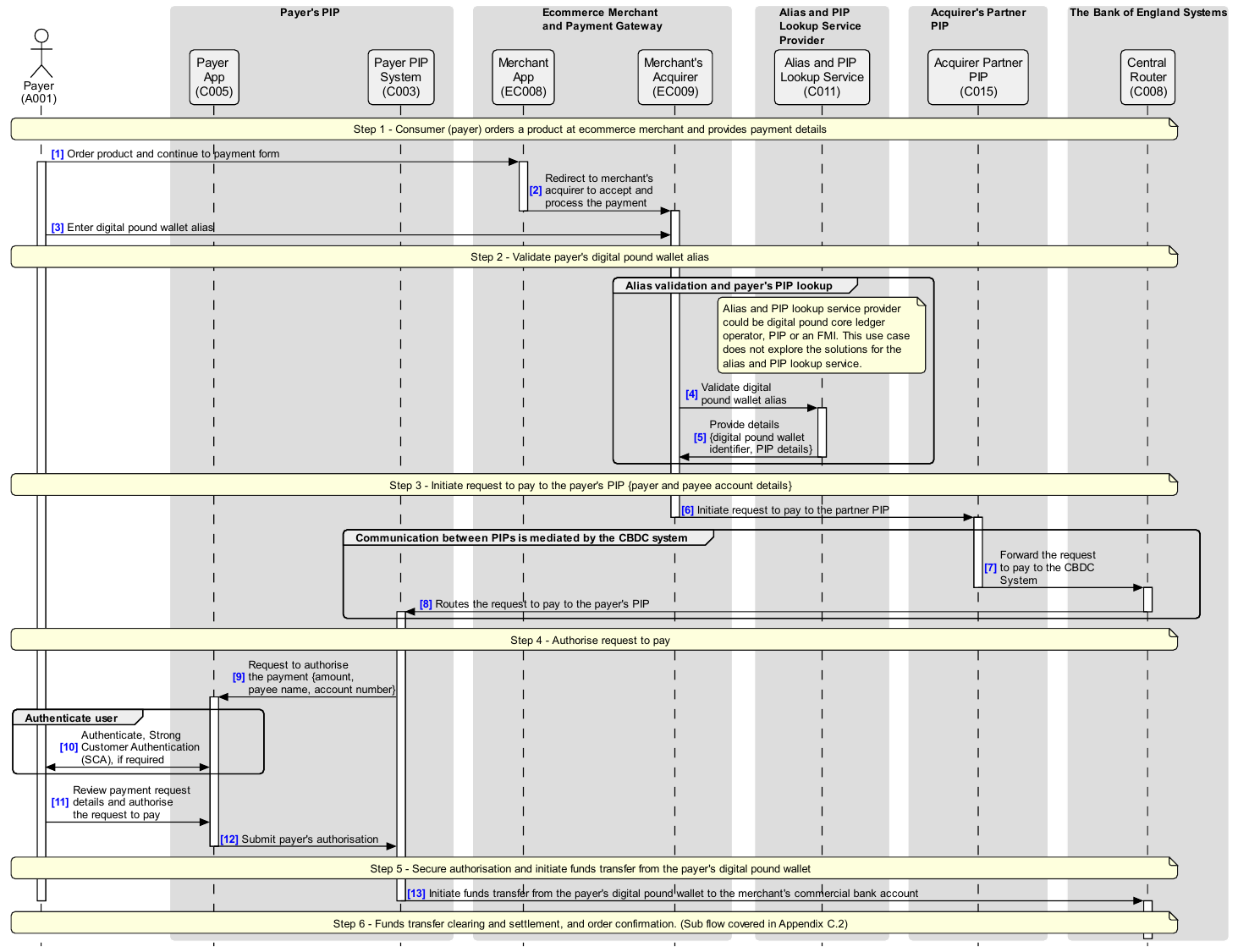}
    \end{center}
    \vspace{-4mm}
    \caption{\footnotesize{Sequence diagram for design option 
    `Intermediated by the CBDC system' (U2.S1.D1).}}
\end{figure}
\pagebreak

\begin{figure}[h!]
    \label{fig:OptionU2S1D2}
    \begin{center}
        \includegraphics[width=0.9\paperwidth]{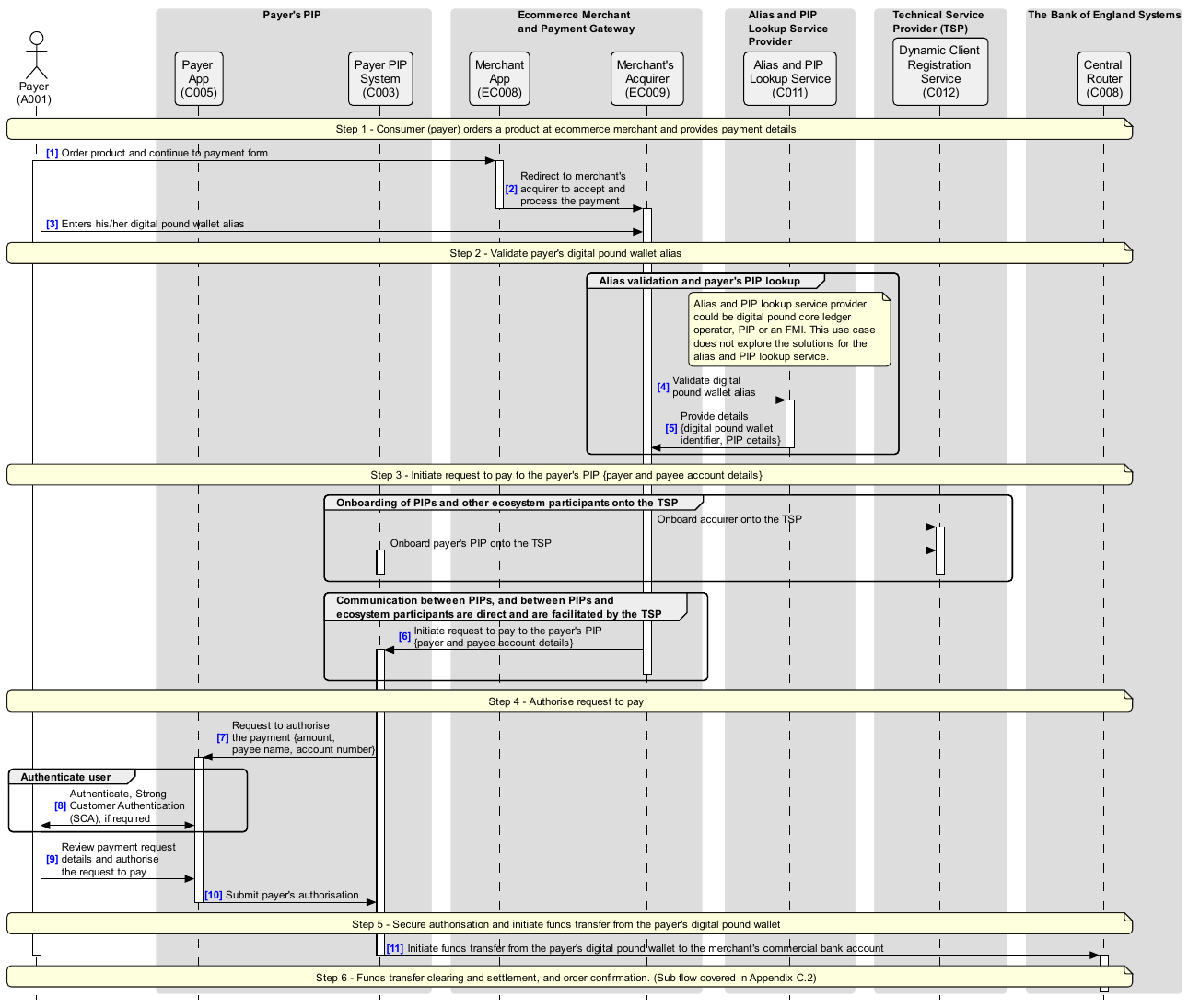}
    \end{center}
    \vspace{-4mm}
    \caption{\footnotesize{Sequence diagram for design option 
    `Direct peer-to-peer' (U2.S1.D2).}}
\end{figure}
\pagebreak

\begin{figure}[h!]
    \label{fig:OptionU2S1D3}
    \begin{center}
        \includegraphics[width=0.88\paperwidth]{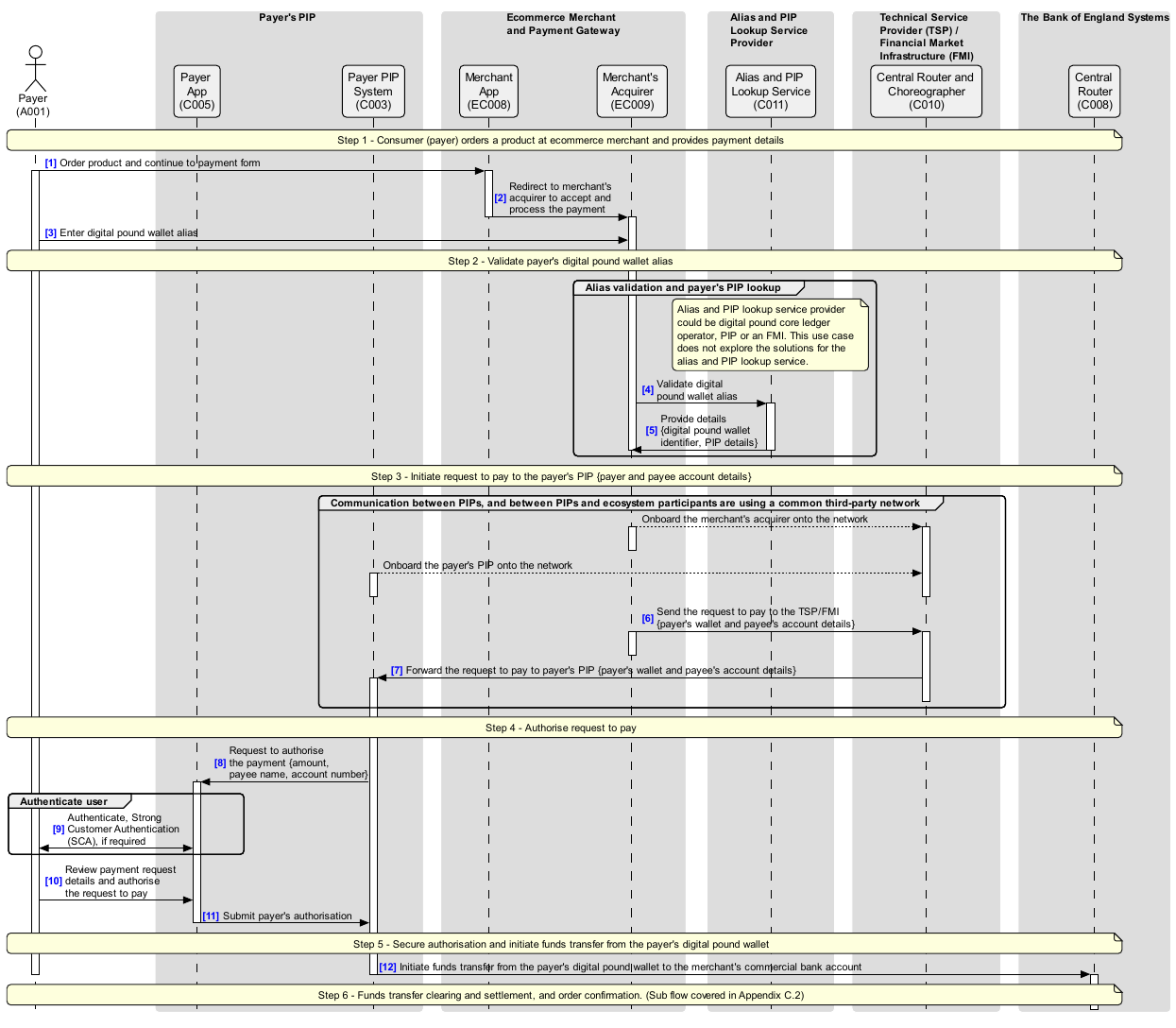}
    \end{center}
    \vspace{-4mm}
    \caption{\footnotesize{Sequence diagram for design option 
    `Using a \changes{common third-party network}' (U2.S1.D3).}}
\end{figure}
\pagebreak

\subsection{Design Options for 'Clearing and settlement of funds transfer from digital pounds 
to commercial bank money' (U2.S2)}
\label{app:uc2-seq-dig-settlement}
\begin{figure}[h!]
    \label{fig:OptionU2S2D1}
    \begin{center}
        \includegraphics[width=1.0\paperwidth]{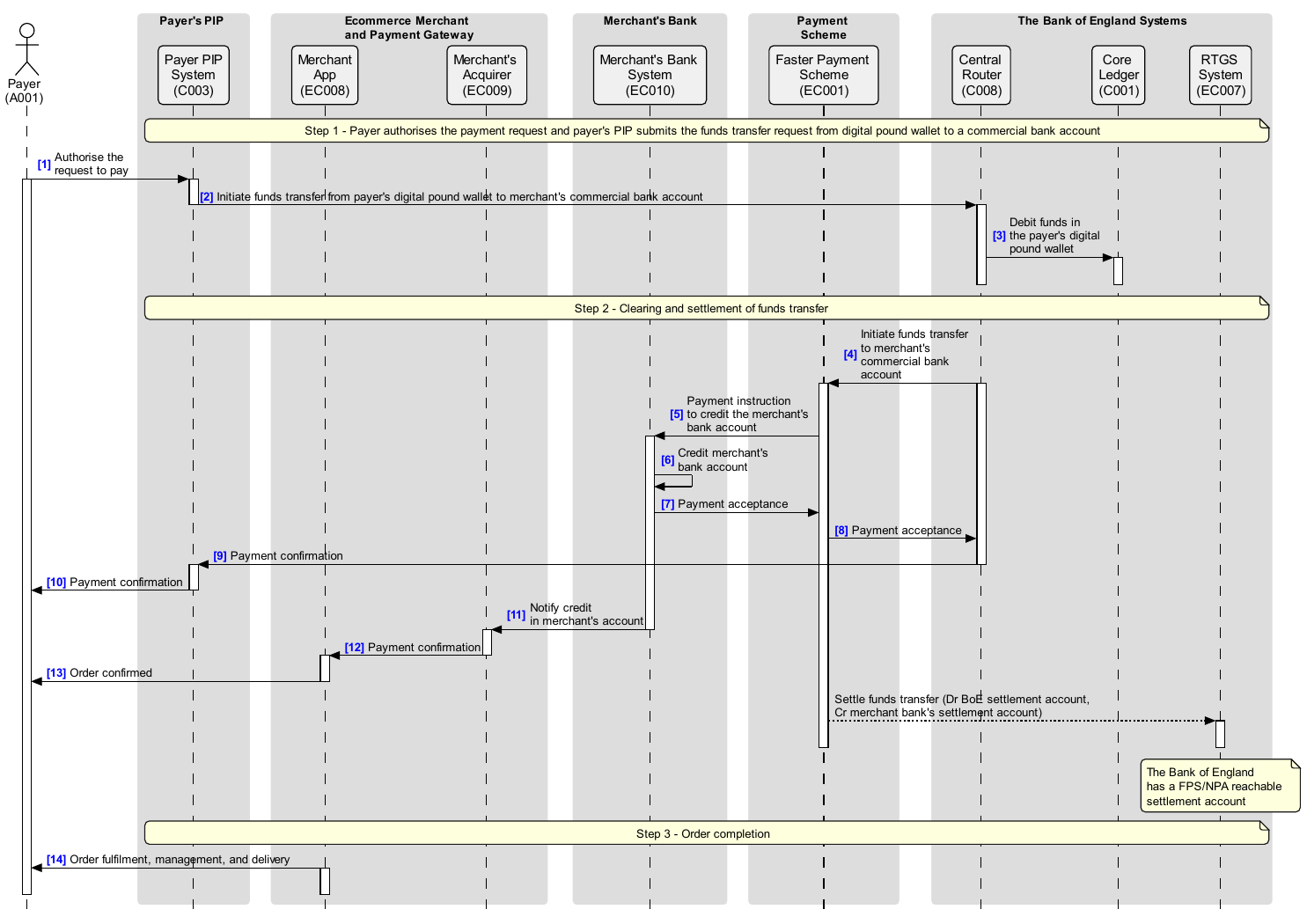}
    \end{center}
    \vspace{-4mm}
    \caption{\footnotesize{Sequence diagram for design option 
    `Provided by the CBDC system' (U2.S2.D1).}}
\end{figure}
\pagebreak

\begin{figure}[h!]
    \label{fig:OptionU2S2D2}
    \begin{center}
        \includegraphics[width=1.1\paperwidth]{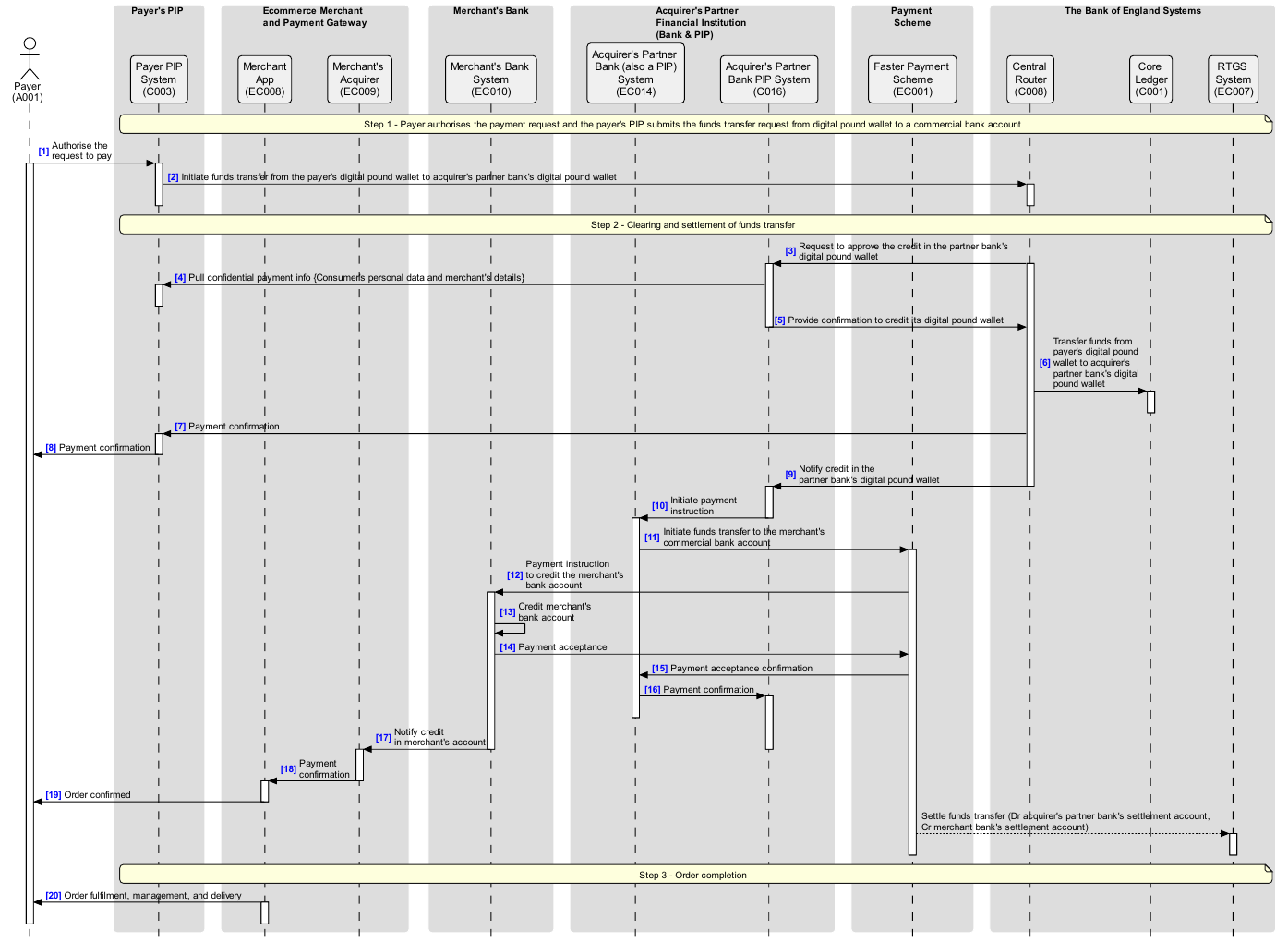}
    \end{center}
    \vspace{-4mm}
    \caption{\footnotesize{Sequence diagram for design option 
    `Provided by merchant acquirer's parter financial institution 
    which is a PIP' (U2.S2.D2).}}
\end{figure}
\pagebreak

\begin{figure}[h!]
    \label{fig:OptionU2S2D3}
    \begin{center}
        \includegraphics[width=1.2\paperwidth]{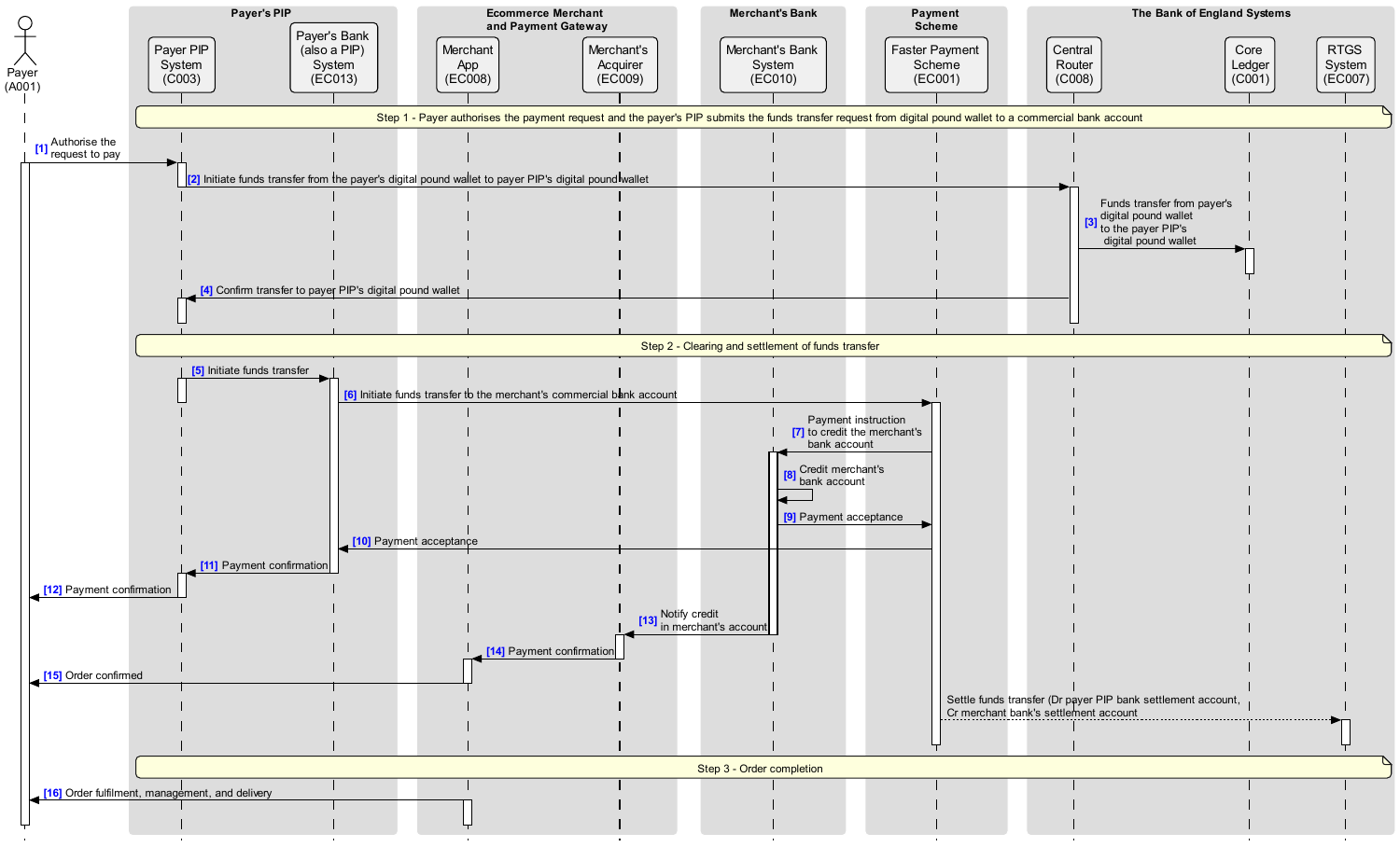}
    \end{center}
    \vspace{-4mm}
    \caption{\footnotesize{Sequence diagram for design option 
    `Provided by payer's PIP which is also a commercial bank' (U2.S2.D3).}}
\end{figure}
\pagebreak

\begin{figure}[h!]
    \label{fig:OptionU2S2D4}
    \begin{center}
        \includegraphics[width=1.2\paperwidth]{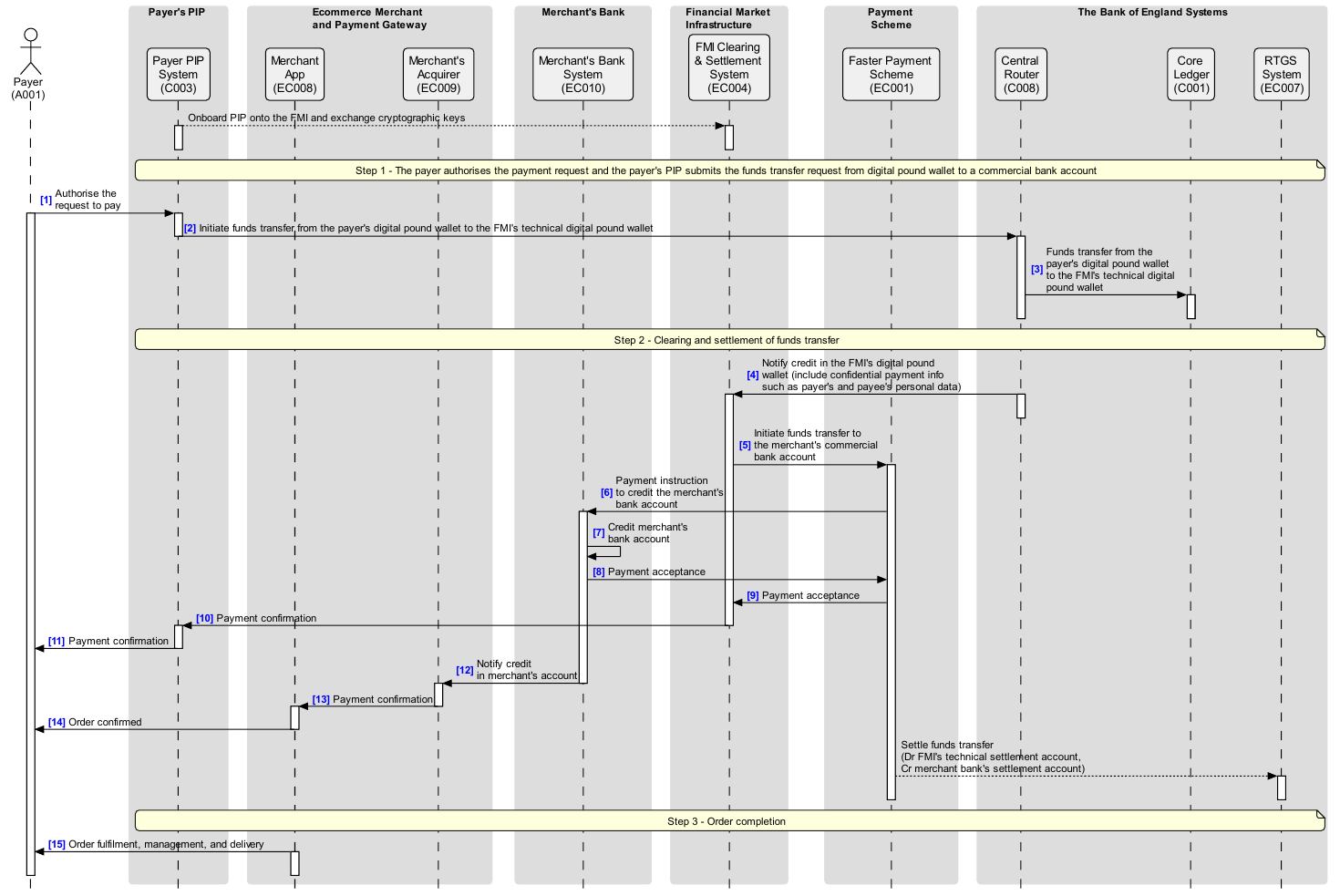}
    \end{center}
    \vspace{-4mm}
    \caption{\footnotesize{Sequence diagram for design option 
    `Provided by an FMI' (U2.S2.D4).}}
\end{figure}
\pagebreak

\begin{figure}[h!]
    \label{fig:OptionU2S2D5}
    \begin{center}
        \includegraphics[width=1.2\paperwidth]{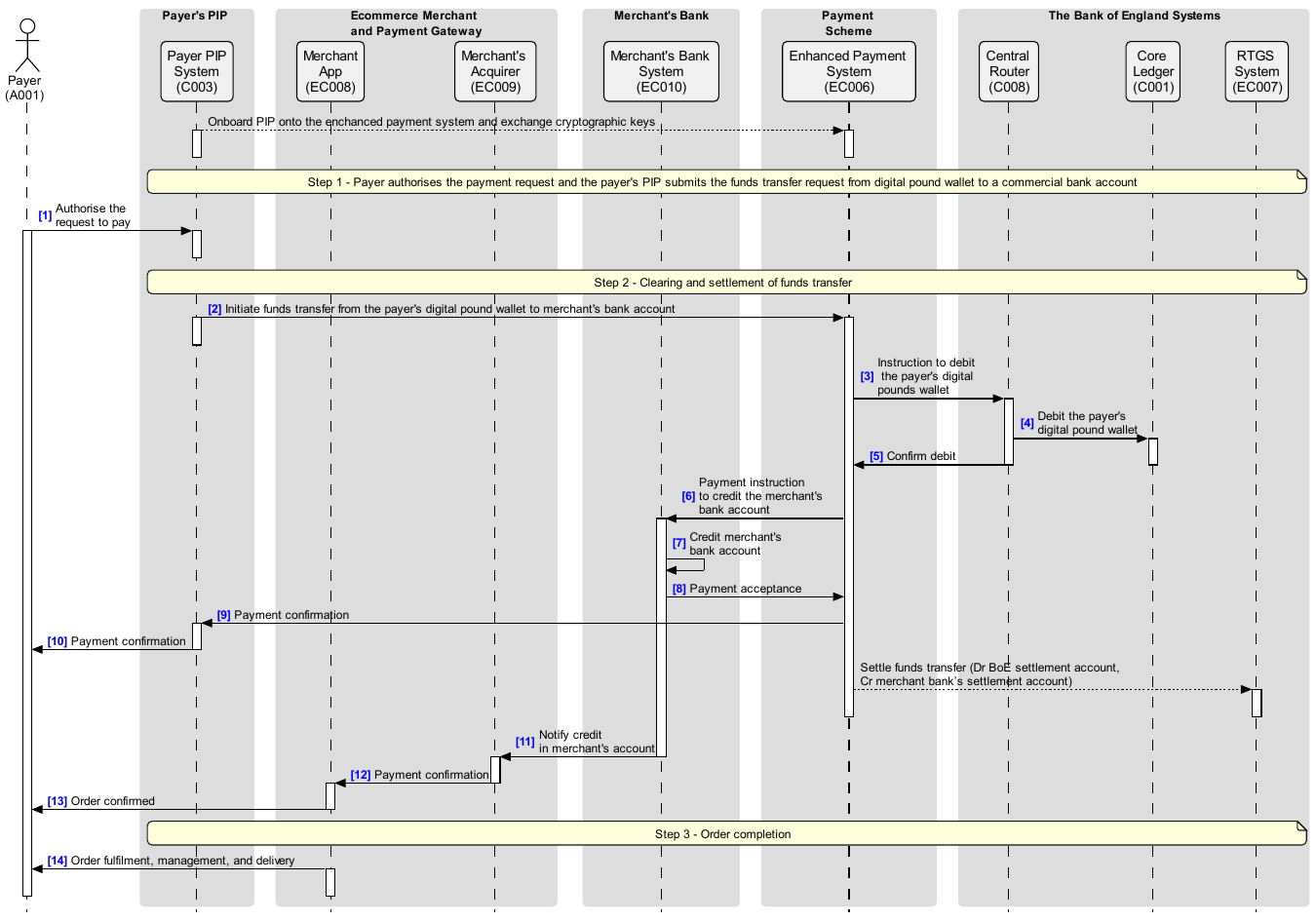}
    \end{center}
    \vspace{-4mm}
    \caption{\footnotesize{Sequence diagram for design option 
    `Provided by an enhanced payment system' (U2.S2.D5).}}
\end{figure}
\pagebreak

  

\section{Appendix - Sequence diagrams for use case `Lock digital pounds and pay on physical delivery from
digital pounds to commercial bank money' (U3)}
\label{app:uc3-sequence-diagrams}

\subsection{Design options for `Request to lock funds in digital 
pound wallets' (U3.S1)}
\label{app:uc3-seq-dig-r2l-topology}
\begin{figure}[h!]
\label{fig:DesignOptionU3.S1.D1}
\begin{center}
    \includegraphics[width=0.85\paperwidth]{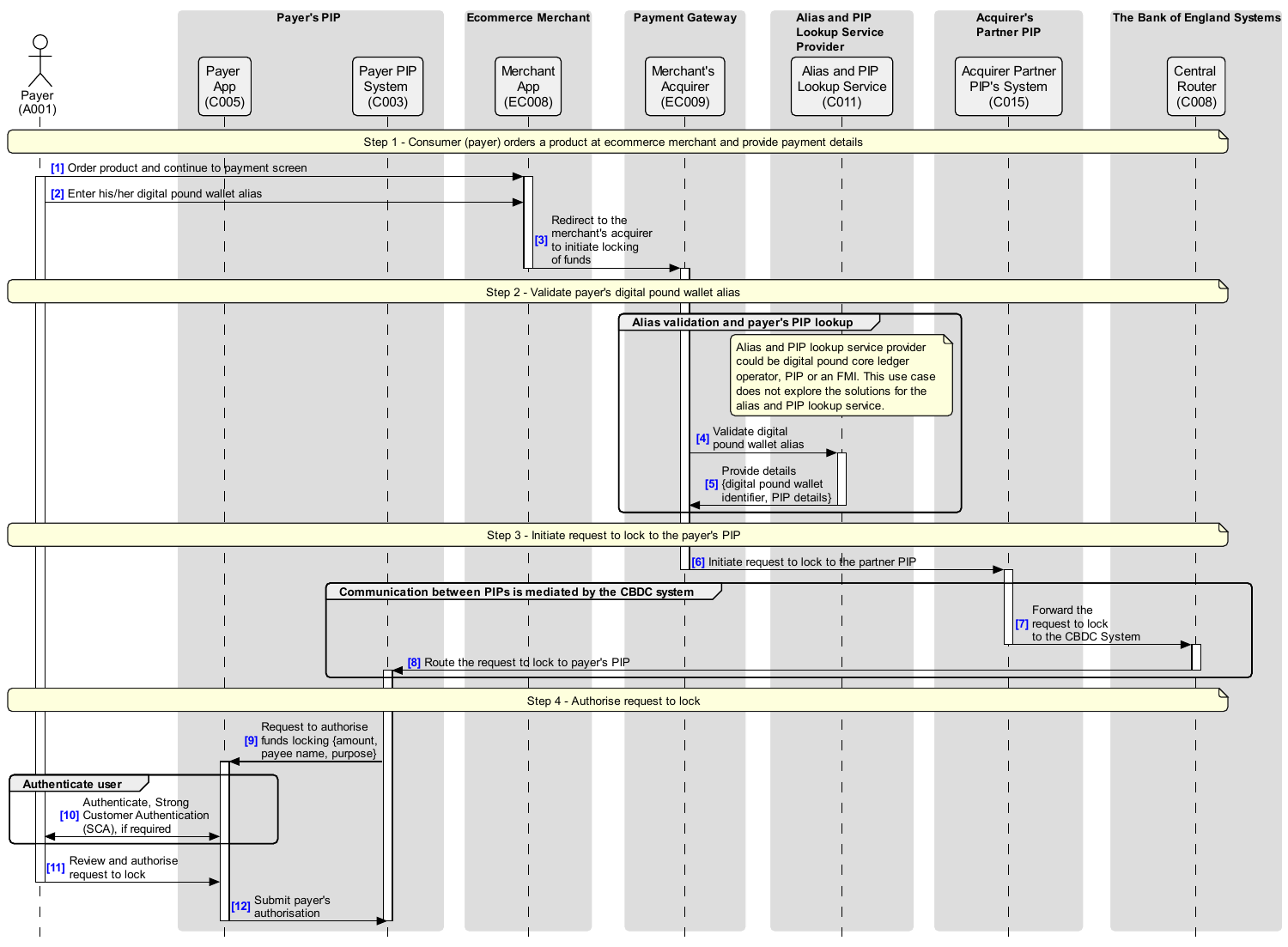}
\end{center}
\vspace{-4mm}
\caption{\footnotesize{Sequence diagram for design option 
`Intermediated by the CBDC system' (U3.S1.D1).}}
\end{figure}
\pagebreak

\begin{figure}[h!]
\label{fig:DesignOptionU3.S1.D2}
\begin{center}
    \includegraphics[width=0.80\paperwidth]{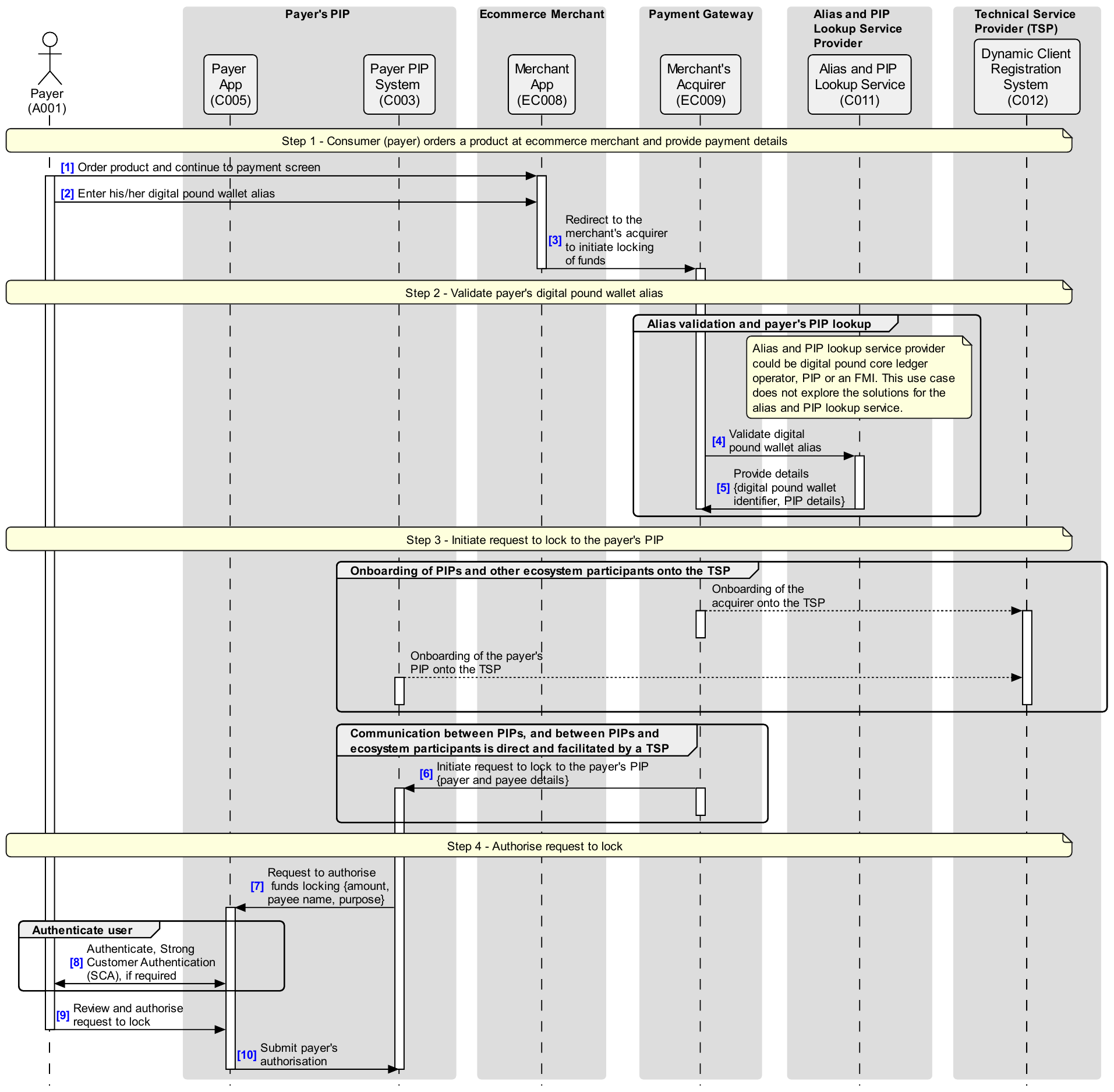}
\end{center}
\vspace{-4mm}
\caption{\footnotesize{Sequence diagram for design option `Direct peer-to-peer' (U3.S1.D2).}}
\end{figure}
\pagebreak

\begin{figure}[h!]
\label{fig:DesignOptionU3.S1.D3}
\begin{center}
    \includegraphics[width=0.79\paperwidth]{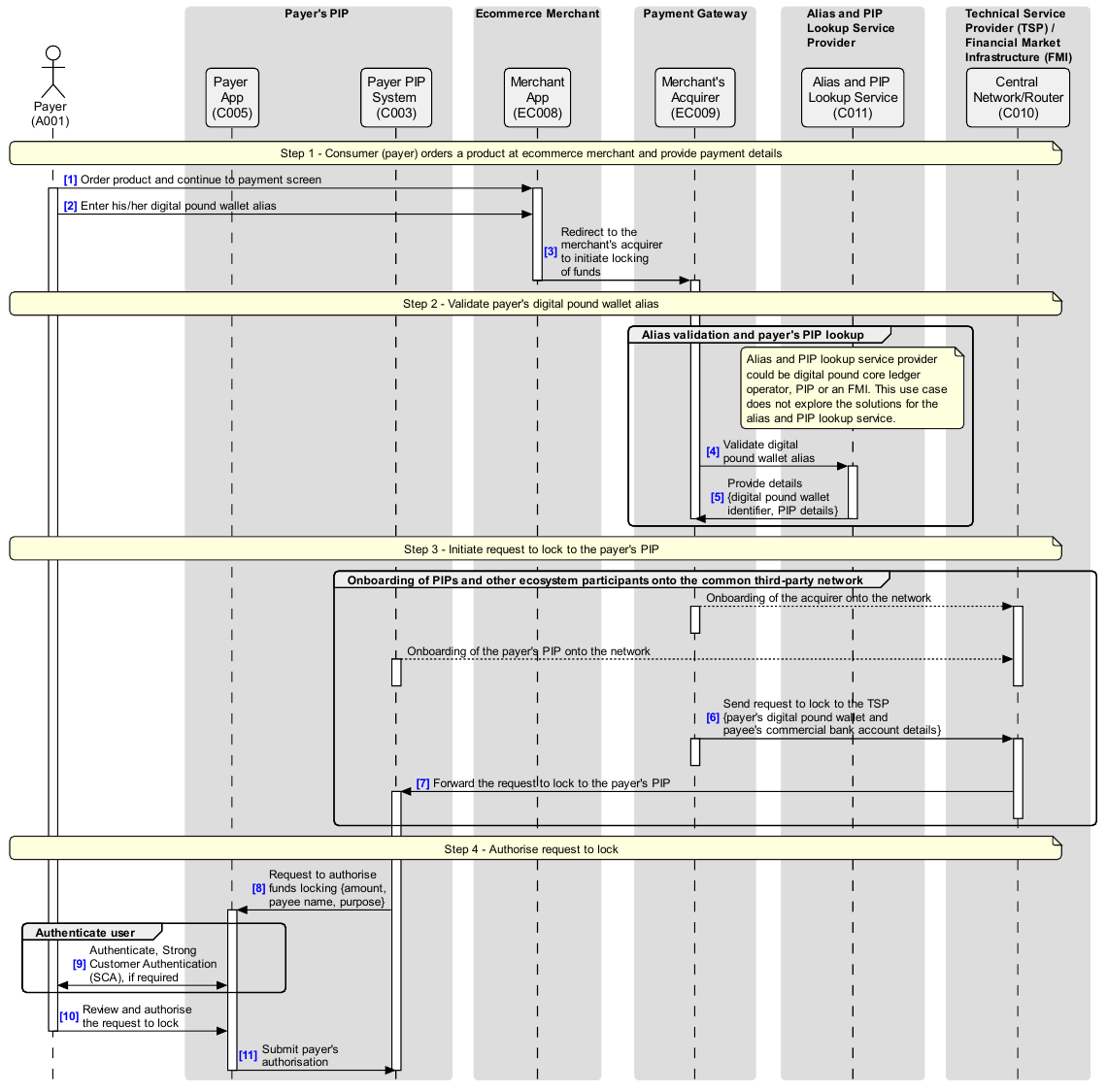}
\end{center}
\vspace{-4mm}
\caption{\footnotesize{Sequence diagram for design option 
`Using a \changes{common third-party network}' (U3.S1.D3).}}
\end{figure}
\pagebreak

\subsection{Design options for `Locking funds in digital pound
wallets and confirming lock status to the merchant' (U3.S2)}
\label{app:uc3-seq-dig-lock-conf-topology}
\begin{figure}[h!]
\label{fig:DesignOptionU3.S2.D1}
\begin{center}
    \includegraphics[width=1.2\paperwidth]{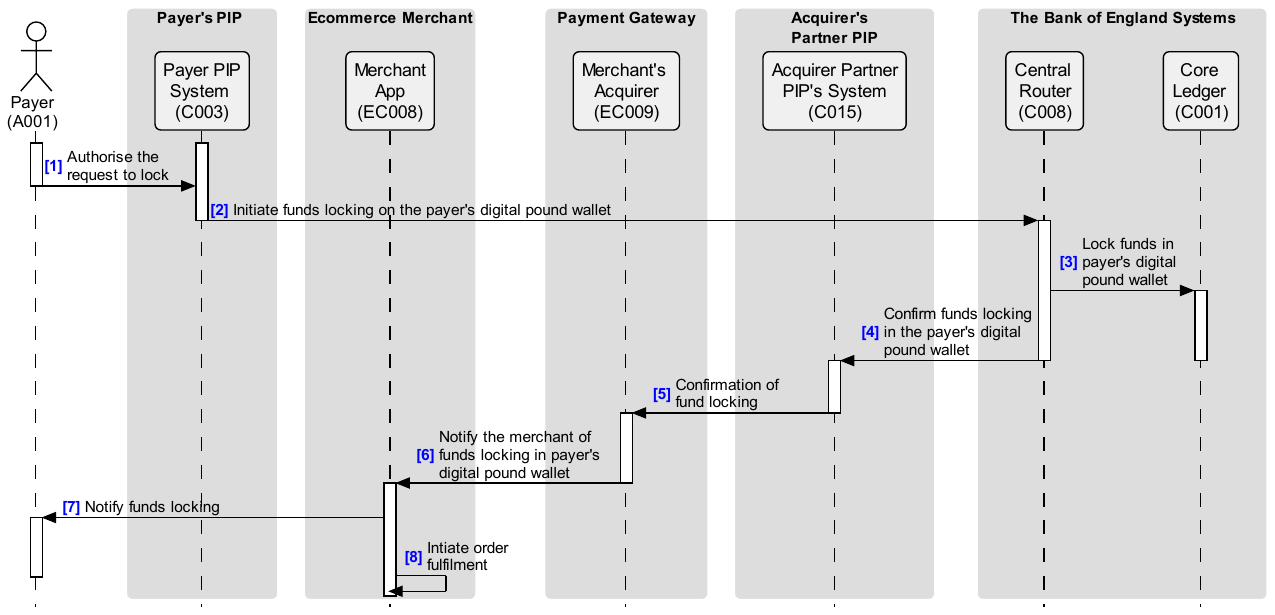}
\end{center}
\vspace{-4mm}
\caption{\footnotesize{Sequence diagram for design option 
`Locking and confirming via the CBDC system' (U3.S2.D1).}}
\end{figure}
\pagebreak

\begin{figure}[h!]
\label{fig:DesignOptionU3.S2.D2}
\begin{center}
    \includegraphics[width=1.2\paperwidth]{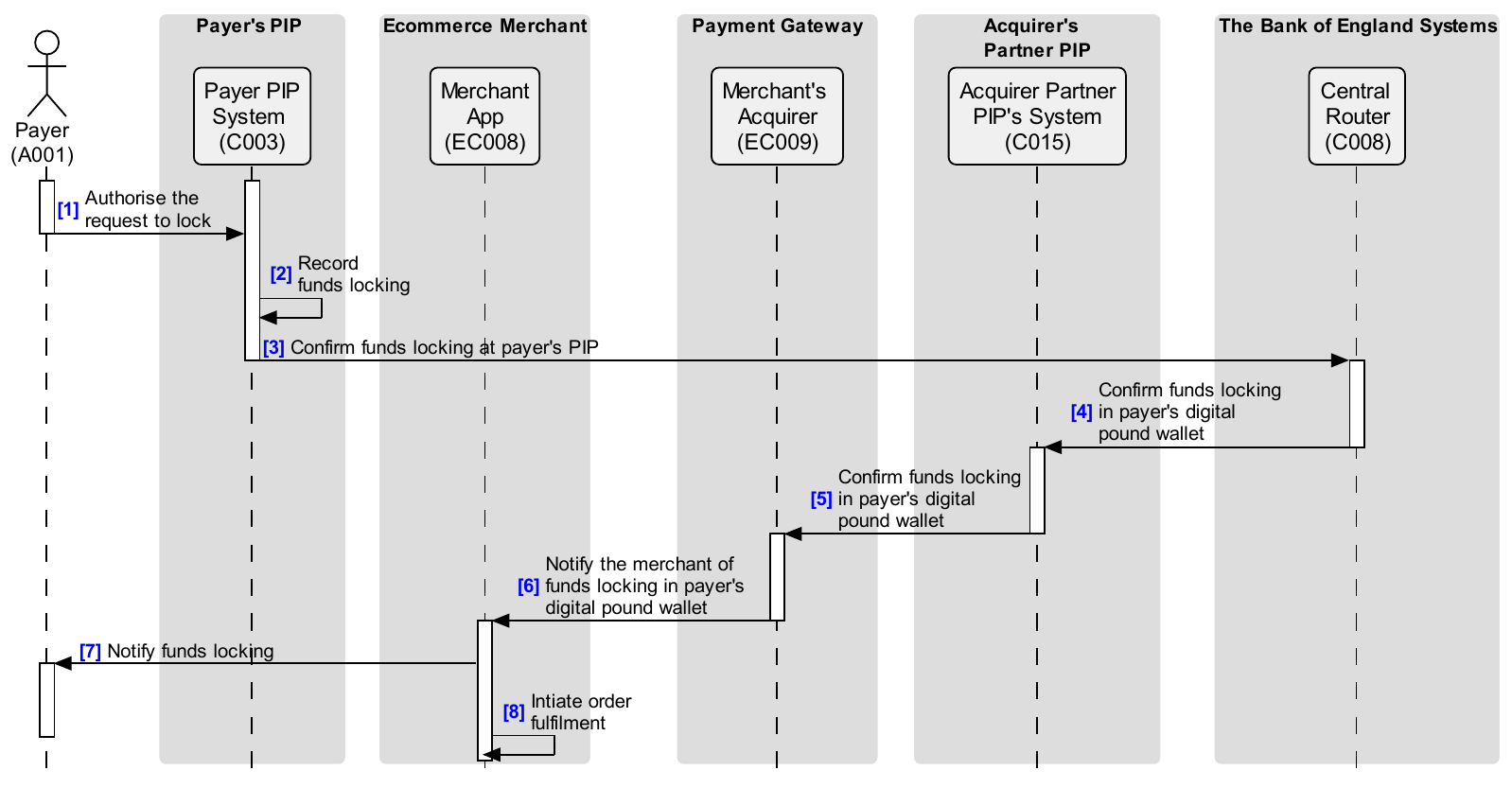}
\end{center}
\vspace{-4mm}
\caption{\footnotesize{Sequence diagram for design option 
`Locking at payer's PIP and confirming via the CBDC system' (U3.S2.D2).}}
\end{figure}
\pagebreak

\begin{figure}[h!]
\label{fig:DesignOptionU3.S2.D3}
\begin{center}
    \includegraphics[width=0.87\paperwidth]{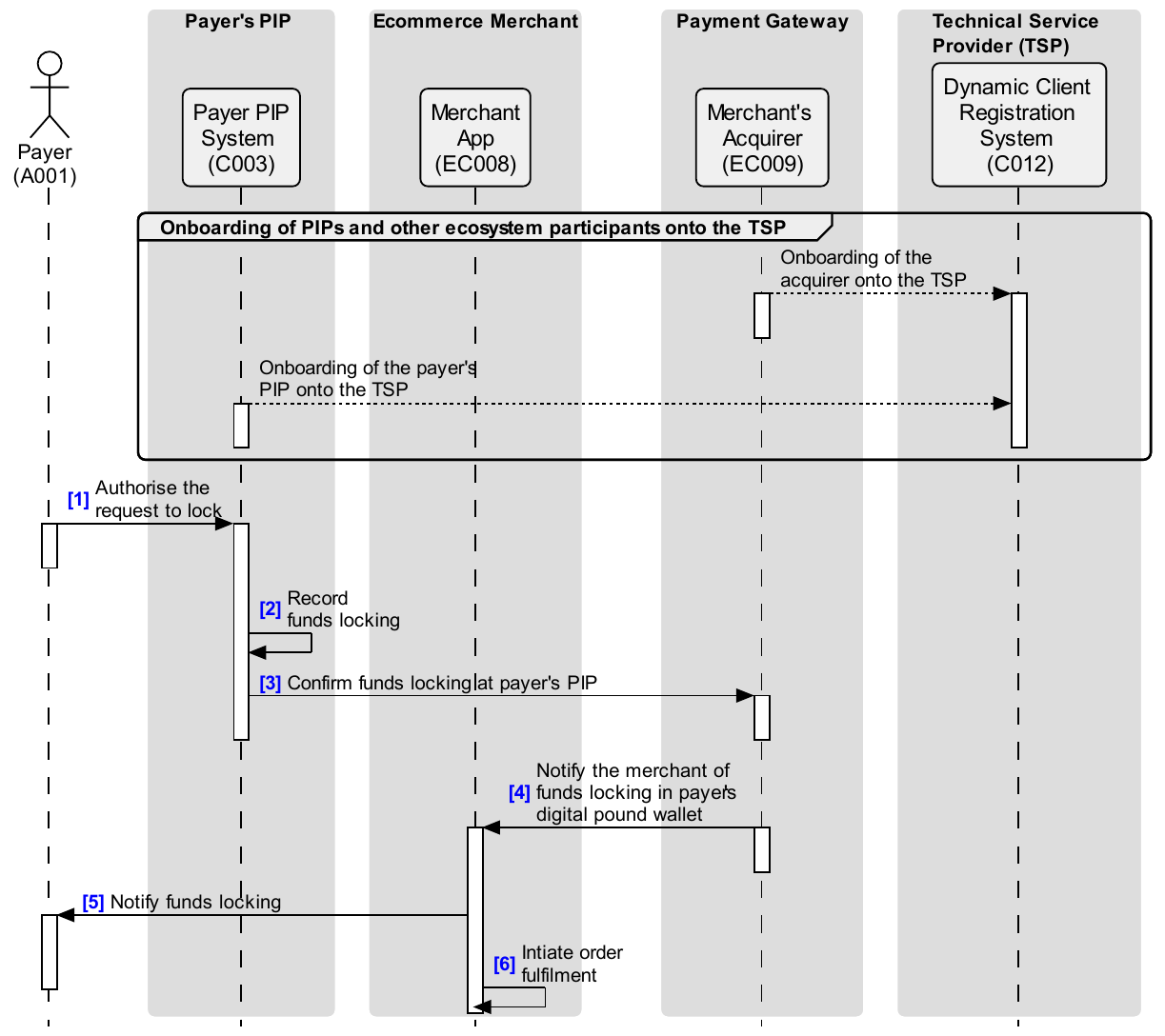}
\end{center}
\vspace{-4mm}
\caption{\footnotesize{Sequence diagram for design option 
`Locking at payer's PIP and confirming using a direct peer-to-peer network' (U3.S2.D3).}}
\end{figure}
\pagebreak

\begin{figure}[h!]
\label{fig:DesignOptionU3.S2.D4}
\begin{center}
    \includegraphics[width=0.85\paperwidth]{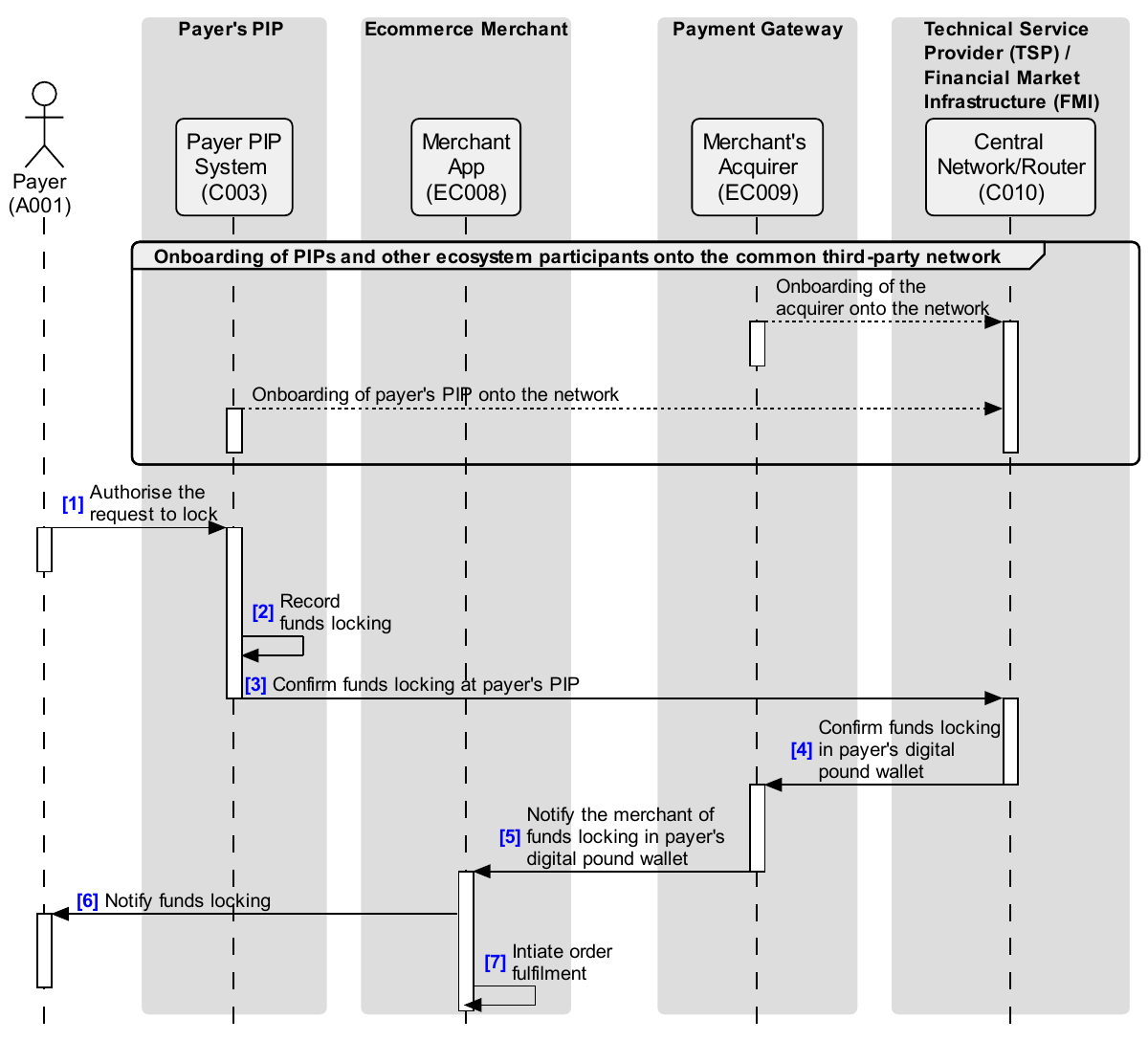}
\end{center}
\vspace{-4mm}
\caption{\footnotesize{Sequence diagram for design option 
`Locking at payer's PIP and confirming via a \changes{common third-party network}' (U3.S2.D4).}}
\end{figure}
\pagebreak

\begin{figure}[h!]
\label{fig:DesignOptionU3.S2.D5}
\begin{center}
    \includegraphics[width=1.0\paperwidth]{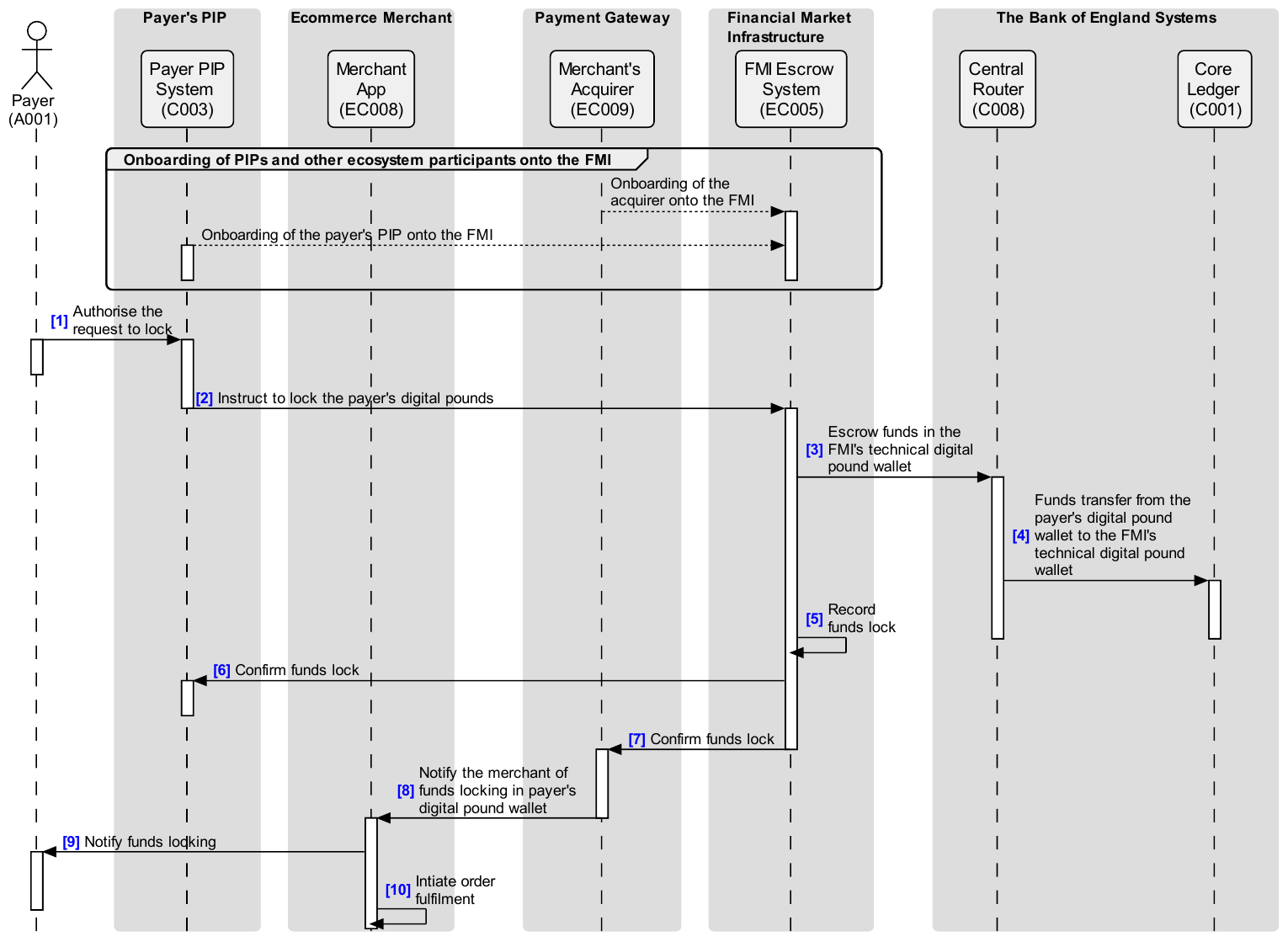}
\end{center}
\vspace{-4mm}
\caption{\footnotesize{Sequence diagram for design option 
`Locking and confirming using an FMI' (U3.S2.D5).}}
\end{figure}
\pagebreak

\subsection{Design options for `Releasing locked funds from digital
pound wallets and initiating funds transfer' (U3.S3)}
\label{app:uc3-seq-dig-rel-lock-topology}
\begin{figure}[h!]
\label{fig:DesignOptionU3.S3.D1}
\begin{center}
    \includegraphics[width=1.1\paperwidth]{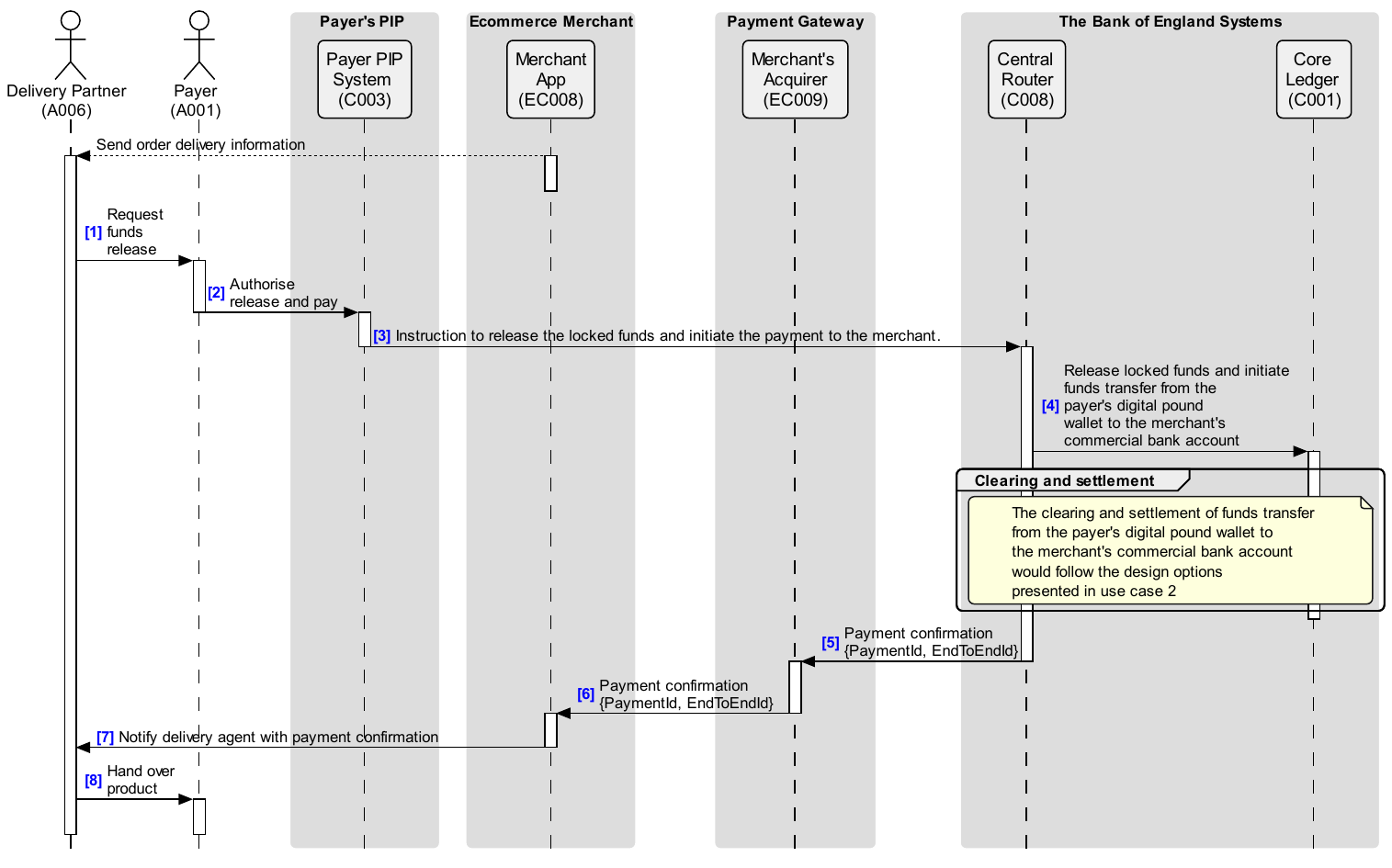}
\end{center}
\vspace{-4mm}
\caption{\footnotesize{Sequence diagram for design option 
`Release locked funds and initiate funds transfer using the CBDC system' (U3.S3.D1).}}
\end{figure}
\pagebreak

\begin{figure}[h!]
\label{fig:DesignOptionU3.S3.D2}
\begin{center}
    \includegraphics[width=1.1\paperwidth]{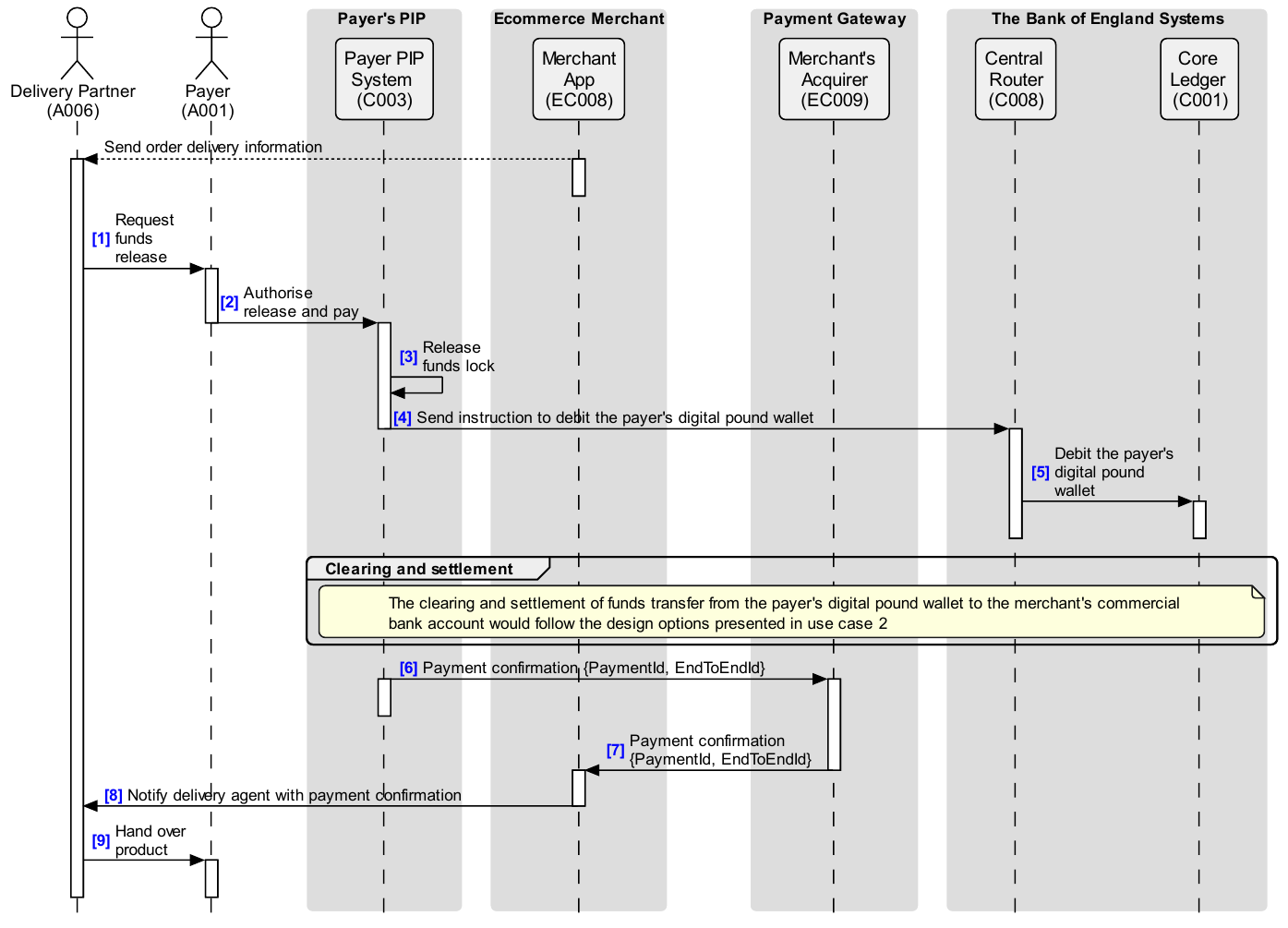}
\end{center}
\vspace{-4mm}
\caption{\footnotesize{Sequence diagram for design option 
`Release locked funds and initiate funds transfer using payer's PIP (U3.S3.D2).}}
\end{figure}
\pagebreak

\begin{figure}[h!]
\label{fig:DesignOptionU3.S3.D3}
\begin{center}
    \includegraphics[width=1.1\paperwidth]{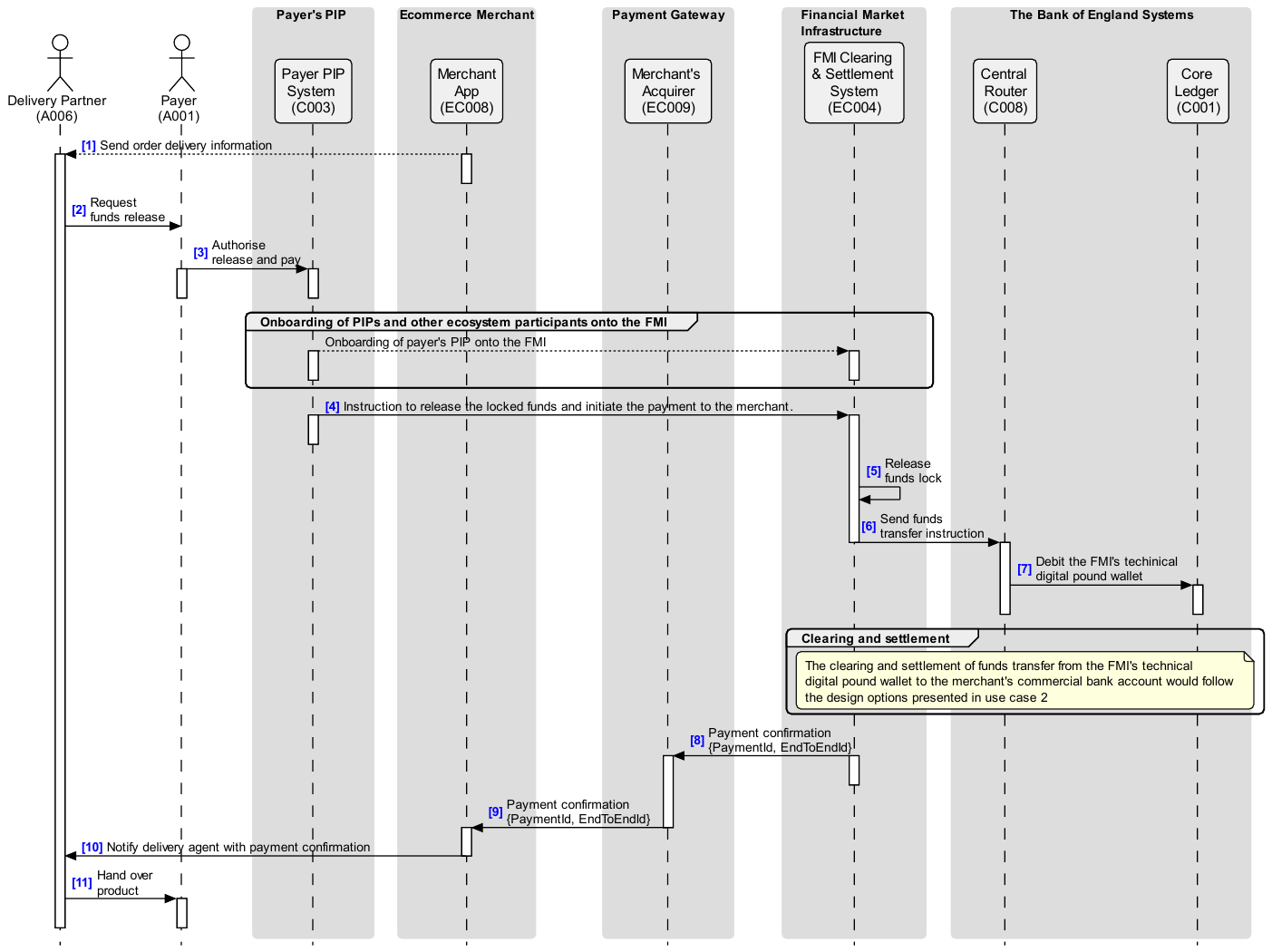}
\end{center}
\vspace{-4mm}
\caption{\footnotesize{Sequence diagram for design option 
`Release locked funds and initiate funds transfer using an FMI' (U3.S3.D3).}}
\end{figure}
\pagebreak
\end{landscape}
\end{document}